\documentclass[longbibliography,showpacs,floatfix,superscriptaddress,twocolumn,aps,prl]{revtex4-2}
\usepackage[figuresright]{rotating}  
\usepackage{amssymb}
\usepackage{bm}
\usepackage{amsmath}
\usepackage{mathtools}
\usepackage{psfrag}
\usepackage{multirow}
\usepackage{tabularx}
\usepackage{textcomp}
\usepackage{units}
\usepackage{lipsum}
\usepackage{wasysym} 
\usepackage{graphicx}

\usepackage{physics}
\usepackage{soul}
\usepackage{titlesec}
\usepackage{times}
\usepackage[colorlinks=true,citecolor=magenta,linkcolor=magenta,hypertexnames=false]{hyperref}
\usepackage{enumitem}

\usepackage[caption=false]{subfig}

\usepackage{tikz}

\definecolor{RubineRed}{cmyk}{0,1,0.13,0}

\renewcommand{\selectlanguage}[1]{}

\newcommand{\uvec}[1]{\hat{\mathbf{#1}}}

\begin{document}
\title{ Non-Abelian Anyon Condensation: a Path-Integral Monte Carlo Approach }
\newcommand{\TUM}{\affiliation{Technical University of Munich, TUM School of Natural Sciences, Physics Department, 85748 Garching, Germany}}
\newcommand{\MCQST}{\affiliation{Munich Center for Quantum Science and Technology (MCQST), Schellingstr. 4, 80799 M{\"u}nchen, Germany}}

\author{Rafael Flores-Calder\'on}
\TUM \MCQST

\author{Frank Pollmann}
\TUM \MCQST

\author{Michael Knap} 
\TUM \MCQST

\date{\today}

\begin{abstract}

Transitions out of non-Abelian topological order are difficult to describe in microscopic quantum models with numerical methods that remain tractable at large scales. We develop a sign-free path-integral framework for Kitaev quantum doubles $\mathcal D(G)$ by organizing single-link perturbations in terms of non-invertible electric and magnetic 1-form symmetries that proliferate distinct anyon species. For any finite group $G$, an exact \textit{matterization} isometry introduces vertex degrees of freedom and maps the link-only model onto a $G$ gauge--Higgs theory. Furthermore, the 1-form symmetries of the fixed point allow us to define generalized Fredenhagen--Marcu order parameters that become finite when the corresponding anyons condense. For $G = S_3$, quantum Monte Carlo simulations show that proliferating a non-Abelian electric anyon drives a first-order transition in which all nontrivial electric anyons condense. In the purely magnetic limit, the model reduces to a  $(2+1)$D pure $G$ gauge theory; for $G=S_3$, it exhibits a first-order confinement transition, diagnosed by the onset of a Wilson-loop area law and the restoration of an emergent magnetic 1-form symmetry. These results provide a unified numerical framework for non-Abelian anyon condensation, confinement, and generalized symmetry breaking.
 \end{abstract}
\maketitle
\paragraph{Introduction.--}

Topologically ordered phases of matter evade a description in terms of local order parameters. Their defining information is instead encoded nonlocally in long-range entanglement, topological ground-state degeneracies, and anyonic excitations
\cite{Wen1990,kitaev_fault-tolerant_2003,NayakSimonSternFreedmanDasSarma2008,KitaevTopoEntan,LevinWen2006}.
Transitions out of these phases are nevertheless generated by concrete microscopic processes. In Abelian topological orders, the proliferation and condensation of bosonic anyons provide a well-established route to confinement~\cite{Wegner1971Duality,FradkinShenker1979, Kogut1979,SenthilFisher2000,Trebst2007,Vidal2009,TupitsynKitaevProkofevStamp2010, WuDengProkofev2012, SomozaSernaNahum2021}, although other mechanisms based on tuning the anyon content are possible as well
\cite{Liu2024SimulatingTopologicalTransitions, SkeletonAbelian}. Extending this picture to non-Abelian phases is substantially more challenging: non-Abelian anyons possess internal fusion spaces, and their fusion can produce a sum of distinct topological sectors rather than a single excitation~\cite{BaisSchroersSlingerland2002,BaisSlingerland2009,Barkeshli2010,Burnell2011, Burnell2012, Kong2014AnyonCondensation,Burnell2018}. 
Recent realizations of non-abelian topological order and demonstrations of non-Abelian braiding on programmable quantum hardware further motivate the study of transitions out of non-Abelian phases \cite{Andersen2023,Iqbal2024NonAbelianTO,XuSun,Lo2026UniversalS3}.

Generalized symmetries provide a natural language for organizing these transitions. In two spatial dimensions, transporting an anyon around a closed contour defines a topological line operator whose action on another line records their mutual braiding.
For Abelian anyons, such lines form a group and generate an invertible 1-form symmetry. Noncontractible lines distinguish topological flux sectors in the deconfined phase, providing a higher-form analogue of spontaneous symmetry breaking \cite{Nussinov, GaiottoKapustinSeibergWillett2015, McGreevy2023GeneralizedSymmetries, Wen2019EmergentHigherSymmetries, XuRakovszkyKnapPollmann2025, xuFM, SernaSomozaNahum2024, liu_1form_2025}. Proliferating magnetic fluxes erases this sector dependence, confines electric strings, and restores the magnetic 1-form symmetry. For non-Abelian anyons, by contrast, line operators obey a fusion algebra and therefore generate non-invertible 1-form symmetries
\cite{BhardwajBottiniSchaferNamekiTiwari2023,Shao2023,ChoiSanghaviShaoZheng2025, BhardwajBottiniSchaferNamekiTiwari2025, noninvertible1form}.
How this non-invertible structure controls  transitions out of non-Abelian topological order in microscopic models remains an open problem.

\begin{figure}[ht!]
 \centering
 \includegraphics[width=\columnwidth]{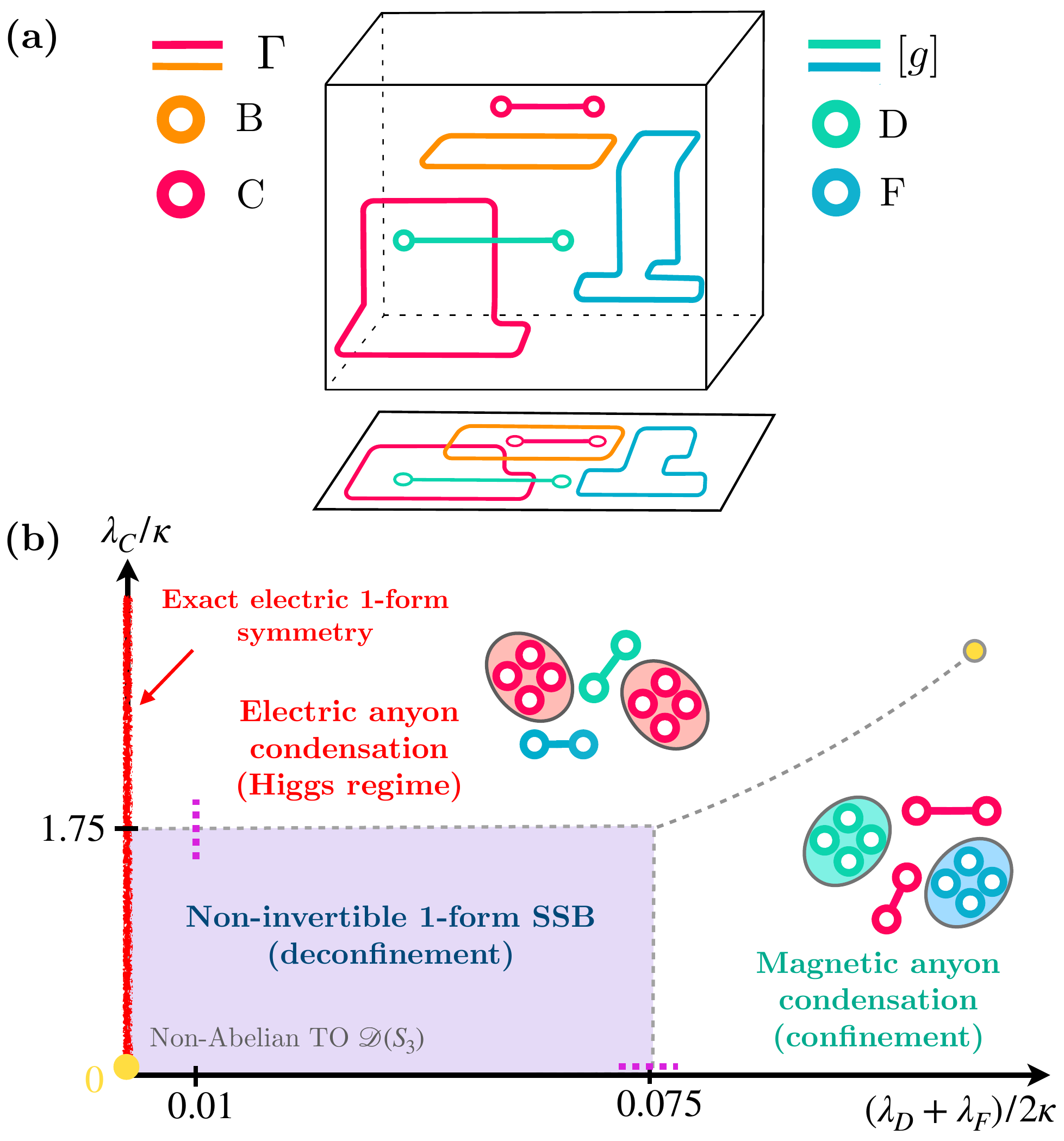}
 \caption{\textbf{From anyon condensation to a classical lattice gauge--Higgs
 theory.} (a) The lattice path integral lifts the two-dimensional quantum-double model in a field (bottom) to a three-dimensional classical ensemble of electric $\Gamma\in\operatorname{Rep}(G)$ and magnetic $[g]\in\operatorname{Class}(G)$ strings. (b) Phase diagram of $\mathcal D(S_3)$, where short open electric strings proliferate the $\mathrm C$ anyon ($\Gamma=E$), while magnetic strings proliferate the $\mathrm D$ and $\mathrm F$ anyons (fluxes $[s]$ and $[r]$, respectively). Here $\kappa$ is the plaquette energy scale. Along the vertical axis (red), the Hamiltonian preserves an exact non-invertible electric 1-form symmetry. Gray dashed lines denote first-order transitions, and purple dashed lines mark the cuts we study in detail. For the full numerical phase diagram, with $\lambda=\lambda_{\mathrm D}=\lambda_{\mathrm F}$ see End Matter.} \label{fig:2+1D_phasediag}
\end{figure}

In this work, we develop such a framework for non-Abelian anyon condensation in quantum doubles \(\mathcal D(G)\) of arbitrary finite groups, using local perturbations constructed from open 1-form symmetry operators.
Our first result is an exact \textit{matterization} isometry that introduces a $G$-valued degree of freedom at every vertex and maps the original link-only model onto a $G$ gauge--Higgs theory.
Using a path-integral representation, we map the perturbed quantum double to a sign-free $(2+1)$D classical gauge--Higgs theory, generalizing the Fradkin--Shenker construction~\cite{FradkinShenker1979,Kogut1979} to general finite groups. The resulting theory is accessible to large-scale Monte Carlo simulations, complementing previous studies based on deformed wavefunctions \cite{Fendley2008,Fidkowski2009,Marien2017Condensation,Schotte2019, XuSchuch2021NonAbelianTransitions,XuGarreRubioSchuch2022,WenTaoFibonacci}, exactly solvable models~\cite{Bombin2008NonAbelianKitaev,ChristianGreenHustonPenneys2023,LinBurnell2024AnyonCondensation,Zhao2025NonabelianCondensation} or mean-field and perturbative analyses \cite{Schulz2013,Schulz2014,Dusuel2015,RitzZwillingFuchsVidal2021}.

Our second result is a family of irrep-resolved Fredenhagen--Marcu (FM) order parameters~\cite{FredenhagenMarcu1983,FredenhagenMarcu1986,BaisRomers2009,GregorHuseMoessnerSondhi2011,xuFM} derived from the non-invertible electric 1-form symmetries of the quantum-double fixed point. These observables directly diagnose condensation in the corresponding electric sectors. We apply this construction to $\mathcal D(S_3)$; see Fig.~\ref{fig:2+1D_phasediag}. In the purely magnetic limit, the gauge constraint is preserved and the model reduces to a finite-group lattice gauge theory. Wilson-loop scaling reveals a first-order confinement transition accompanied by the restoration of an emergent magnetic 1-form symmetry. When instead the non-Abelian electric anyon $\mathrm C$ is proliferated, the FM order parameters in both nontrivial electric sectors become finite through a first-order transition, although the other nontrivial electric sector is not directly favored by the perturbation. Our results place confinement and Higgs transitions out of non-Abelian $\mathcal D(G)$ topological order within a unified symmetry-based framework that maps microscopic quantum models to sign-free classical gauge theories.

\begin{figure}[t]
 \centering
 \includegraphics[width=\columnwidth]{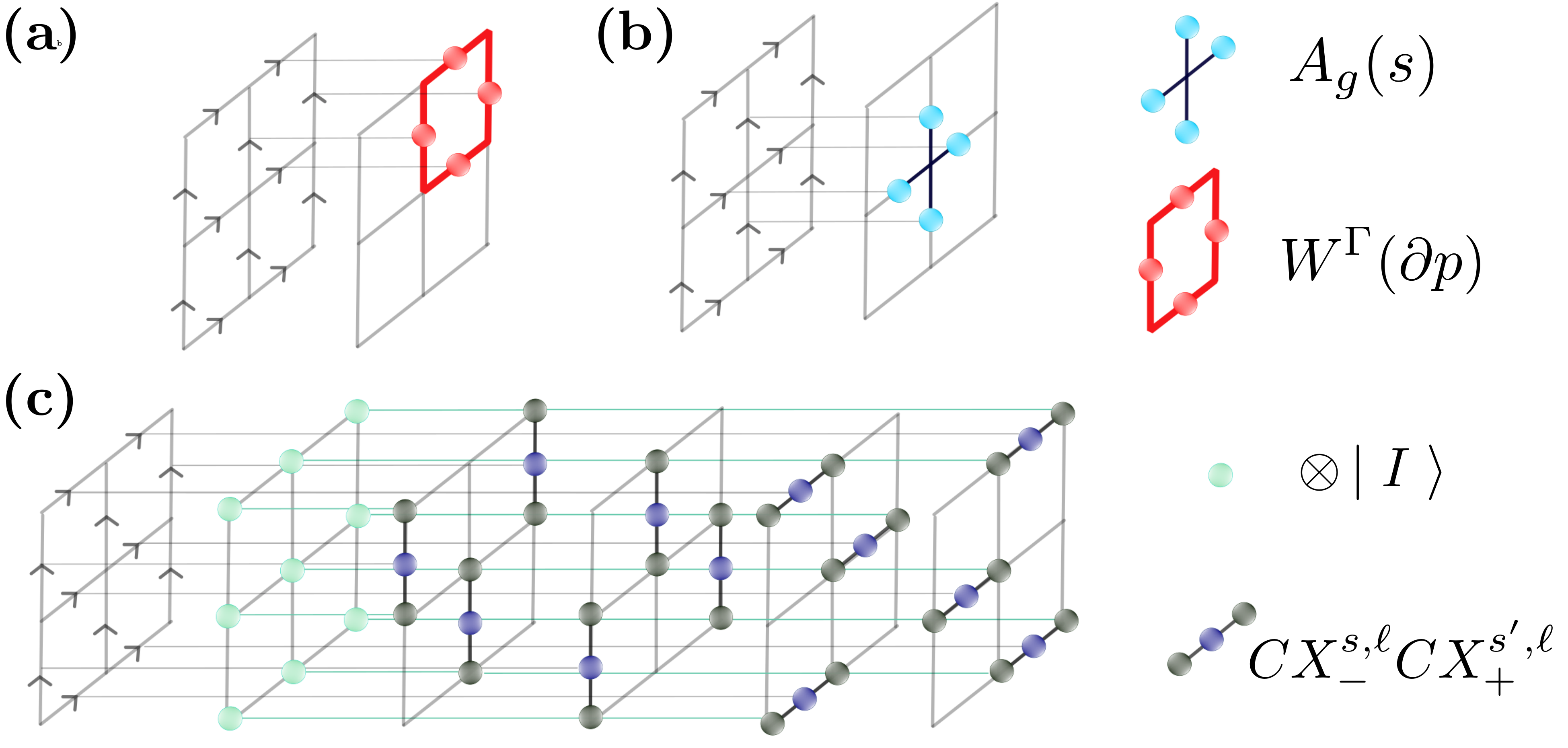}
 \caption{\textbf{Local operators and the matterization circuit.}
 (a) Qudits are placed on the links of the oriented lattice. The elementary Wilson loop
 $W^\Gamma(\partial p)$ measures the plaquette flux in irrep $\Gamma$.  We define $O_{\ell s}=-$ ($+$) when $\ell$ points away from (toward) a vertex $s$, and $O_{\ell p}=-$($+$) indicates that the conjugate $\bar{\Gamma}$ ($\Gamma$) representation is used if the orientation of the link is opposite (equal) to the plaquette path ordering $\mathcal{P}$ direction.  (b) The star operator $A_g(s)$ acts by group multiplication on the
 four links adjacent to $s$.  
 (c) The matterization isometry appends a vertex $G$-qudit (cyan dot) and uses
 controlled left and right multiplications, denoted $CX_-^{s,\ell}CX_+^{s',\ell}$, to rewrite every link relative to its endpoint variables.}
 \label{fig:sch-KW}
\end{figure}

\paragraph{Perturbed quantum double model.--}Let $G$ be a finite group and place one $G$-qudit with Hilbert space
$\mathcal H_\ell\simeq\mathbb C[G]=\operatorname{span}\{\ket{g}:g\in G\}$,
on every link $\ell$ of a square lattice (dark arrows in Fig.~\ref{fig:sch-KW}). All links point right or upward,
and $\bar g\equiv g^{-1}$. Generalized Pauli operators
$X_\pm^h$ implement left and right group multiplication $X_+^h\ket g=\ket{hg},
 X_-^h\ket g=\ket{g\bar h}$, whereas the diagonal operators $Z_\pm^h$ resolve the group-element basis
 $ Z_+^h\ket g=\delta_{h,g}\ket g,
 Z_-^h\ket g=\delta_{h,\bar g}\ket g$. Details are given in
the Supplemental Material (SM), Sec.~\ref{sec:SM-Gqudit}. The quantum-double fixed-point Hamiltonian is then 
\begin{align}
 H_{\mathrm{QD}}=J\sum_s[1-A(s)]+\kappa\sum_p[1-B(p)].
 \label{QD}
\end{align}
The star projector at each vertex $s$ and the plaquette projector on a plaquette $p$ can be written as
\begin{align}
 A(s)=\frac{1}{\abs{G}}\sum_{g\in G}A_g(s), \; \;
 B(p) = \frac{1}{\abs{G}}\sum_\Gamma d_\Gamma W^\Gamma(\partial p), \label{ABops}
\end{align}
where we defined star operator $A_g(s)=\prod_{\ell\in\operatorname{star}(s)}X_{O_{\ell s}}^g(\ell)$ and the Wilson loop $W^\Gamma(\mathcal{C})=\text{tr} \Big[\mathcal{P}\prod_{\ell\in \mathcal{C}}(\bm{Z}^{\Gamma}(\ell) )^{O_{\ell p}}\Big]$ with $\mathcal{C}=\partial p$ as in Fig.~\ref{fig:sch-KW}. Here, $\Gamma$ is an irrep of the group with dimension $d_\Gamma$ and 
$\bm Z^\Gamma(\ell)\ket g=\Gamma(g)\ket g$.
The ground state wavefunction $\ket{\Psi_{\mathcal{D}(G)} }$ of Eq.~\eqref{QD} on the plane can be visualized as a superposition of all possible magnetic flux loops with distinct group elements, including loops that intersect. The flat-$G$ connection on the plane results in $\ket{\Psi_{\mathcal{D}(G)}} \propto \prod_{s}A(s) \bigotimes_\ell \ket{1}_\ell\propto \sum_{\{ g_\ell\}} \prod_p \delta_{U_p,1 }\ket{\{ g_\ell\}}$, $U_p\equiv \mathcal{P}\prod_{\ell\in \partial p}g_\ell$. Due to the topological order of the ground state, quasiparticle excitations fractionalize.

The low-energy quasiparticles of $\mathcal D(G)$ are anyons labelled by
$\mathfrak a=([g],R)$, where $[g]$ is a conjugacy class and $R$ is an
irreducible representation (irrep) of the centralizer $Z_g$. Their quantum
dimension is $d_{\mathfrak a}=\abs{[g]}\dim R$; hence
$d_{\mathfrak a}>1$ identifies a non-Abelian anyon
\cite{bais_quantum_1992,BaisVanDrielDeWildPropitius1993,kitaev_fault-tolerant_2003}. The 1-form symmetries of the fixed point correspond to these anyon worldlines and are generally supported on closed ribbons \cite{noninvertible1form}. An open ribbon operator  
$F_\rho^{\mathfrak a}$ instead creates $\mathfrak a$ and its antiparticle $\bar{\mathfrak a}$ at the two
ends of $\rho$ \cite{kitaev_fault-tolerant_2003,noninvertible1form}. We perturb the fixed point
by the shortest open 1-form symmetry operators supported on single links (arrows in Fig.~\ref{fig:sch-KW}). To probe non-Abelian anyon condensation, we choose to proliferate so-called bosonic anyons. A necessary condition for an anyon to be a boson is that its self-statistics (captured by the monodromy or, in turn, the twist factor $\theta_{\frak a}$) is trivial \cite{BaisSlingerland2009}: $\frak L =\{ \frak a | \ \theta_{\frak a}=1\}$.

The perturbed Hamiltonian is
\begin{align}
 H=H_{\mathrm{QD}}
 -\sum_{\mathfrak a\in\frak L}\lambda_{\mathfrak a}
 \sum_\ell\tr F_\ell^{\mathfrak a},
 \label{eq:main-ribbon-field}
\end{align}
which lowers the energy of local $\mathfrak a$--$\bar{\mathfrak a}$ pairs
and thereby drives their proliferation. We now focus on the anyons belonging to $\frak L$ which are purely electric or magnetic. Purely electric anyons are those with trivial flux $([1],\Gamma)$ for which $\tr F_\ell^{\mathfrak a}=\tr \mathbf{Z}^{\Gamma}_+(\ell)$ whereas pure magnetic anyons instead have trivial centralizer irrep 
$([g],I)$, then $\tr F_\ell^{\mathfrak a}=\sum_{h\in [g]}X^h_+(\ell)$ and the Hamiltonian reduces to
\begin{align}
 H=H_{\mathrm{QD}}
 -\sum_{\ell,g\ne 1}
\lambda_{[g]}X_+^g(\ell)-\sum_{\ell,\Gamma\ne I}\lambda_\Gamma
 \tr\bm Z^\Gamma(\ell).
 \label{eq:main-electric-magnetic-fields}
\end{align}
Thus, $H=H_{\mathrm{QD}}+H_X+H_Z$. The gauge-invariance $[H_X,A(s)]=0$ holds because the couplings are class functions. This form also makes the consistency conditions transparent,
$\lambda_{[\bar g]}=\lambda_{[g]}^*$ and
$\lambda_{\bar\Gamma}=\lambda_\Gamma^*$. Thus an individual real class
sum is Hermitian when $[g]=[\bar g]$; otherwise $[g]$ and $[\bar g]$ must be paired.
The perturbed Hamiltonian with $\lambda_{[g]}=0$ satisfies an exact non-invertible electric 1-form symmetry (red line in Fig.~\ref{fig:2+1D_phasediag}(b)) generated by Wilson loops $ W^\Gamma(\mathcal{C})$ with $\mathcal{C}$ a  lattice loop. This higher-form symmetry is also present at the quantum-double fixed point together with magnetic and dyonic 1-form symmetries as detailed in Ref.~\cite{noninvertible1form}. The topological order of Eq.~\eqref{QD} can be interpreted as SSB of an anomalous non-invertible 1-form symmetry. Indeed, the torus ground states are degenerate and can be constructed from $\ket{\Psi_{\mathcal{D}(G)}}$ by applying 1-form symmetry operators acting on the two noncontractible cycles \cite{noninvertible1form}. For $\lambda_{\Gamma}=0$, the fixed-point magnetic 1-form symmetry becomes emergent and is spontaneously broken for small values of $\lambda_{[g]}$, as we show by numerically computing the noncontractible Wilson loops of the torus in the End Matter. More generally, we expect the topological order to be robust to small perturbations~\cite{kitaev_fault-tolerant_2003,HastingsWen2005}. The nature of the transition and its effective action can be studied from a generalization of the Fradkin--Shenker result \cite{FradkinShenker1979} by first mapping the model to a quantum lattice gauge theory.

\paragraph{Isometric map to a quantum gauge theory with matter--}A useful way to expose the gauge-theory structure of the perturbed quantum double is to introduce a local reference frame at every vertex. For the $\mathcal D(\mathbb Z_2)$ toric code, this construction underlies the familiar relation between the two-field model and the $\mathbb Z_2$ gauge theory~\cite{kitaev_fault-tolerant_2003,TupitsynKitaevProkofevStamp2010,XuRakovszkyKnapPollmann2025}, and the underlying controlled-multiplication circuit is the one used to gauge finite-group symmetries~\cite{Haegeman2015,Tantivasadakarn2023}.
Here we apply the construction to perturbed $\mathcal D(G)$ for arbitrary finite $G$ and prove that it is sign-free. We attach an auxiliary $G$-qudit (\textit{matter}) to every vertex $s$, initialized in the product state composed of the electric vacuum (trivial irrep $I$), $\ket{I}_s=\tfrac{1}{\sqrt{\abs{G}}}\sum_g \ket{g}_s$, and apply a finite-depth circuit of controlled
group multiplications, as illustrated by the three-qudit gates and cyan wires in
Fig.~\ref{fig:sch-KW}(c). For an oriented link
$\ell=(s\rightarrow s')\equiv ss'$, this mapping is an isometry,
$V_{\mathrm{KW}}=U_{\mathrm{KW}}\bigotimes_s\ket I_s$. The unitary transformation acts as
\begin{align}
 U_{\mathrm{KW}}\ket{f_s,g_\ell,f_{s'}}
 &=\ket{f_s,f_{s'}g_\ell\bar f_s,f_{s'}},\ V_{\mathrm{KW}}^\dagger V_{\mathrm{KW}}=\mathbb I.
\end{align}
Therefore gauge transformations of the links adjacent to a vertex can be reabsorbed by a single-site vertex operation. Because $U_{\mathrm{KW}}$ is unitary and the vertex qudits are initialized in
a fixed product state, $V_{\mathrm{KW}}$ preserves all inner products. Its image
is precisely the subspace invariant under simultaneous gauge transformations
of links and matter,
\begin{align}
V_{\mathrm{KW}}V_{\mathrm{KW}}^\dagger=P_{\mathrm{phys}}\equiv \prod_s\frac{1}{\abs{G}}
 \sum_{g\in G}A_g(s)X_+^g(s).
\end{align}
The map therefore, provides an exact change of variables between the original link-only model and a gauge theory with explicit matter fields. The label $\mathrm{KW}$ emphasizes that, for $G=\mathbb Z_2$, this construction reduces to the gauge-theory formulation associated with the Kramers--Wannier duality of Wegner's $\mathbb Z_2$ model \cite{Wegner1971Duality}. The density matrices of both models are related by a quantum channel $\rho_{GGT}= \mathcal{N}_{\mathrm{KW}}[\rho_{\mathcal{D}(G)}]$, where $\rho_{GGT}$ corresponds to the  auxiliary-space model and $\rho_{\mathcal{D}(G)}$ to the perturbed quantum-double, and $\mathcal{N}_{\mathrm{KW}}$ implements the isometry $V_{\mathrm{KW}}$~\cite{XuRakovszkyKnapPollmann2025}. Enforcing $P_{\mathrm{phys}} \ket{\Psi}=\ket{\Psi}$ for physical states is equivalent to enforcing
\begin{align}
X^g_+(s) \ket{\Psi} =A_{\bar{g}}(s)\ket{\Psi} , \quad  \forall \ s, \ \forall g \in G.
\end{align}
This is the \textit{$G$-Gauss law} for a link--vertex theory. The complete circuit construction and proof are given in SM Sec.~\ref{sec:SM-matterization}.

Under this isometry the term $H_X$ proliferating magnetic anyons is unchanged, while each electric 1-form operator acquires matter operators at the endpoints of its link in accordance with gauge invariance. The general transformed Hamiltonian is
\begin{align}
 &H_G^{(2+1)D}=J\sum_s(1-\tfrac{1}{\abs{G}}\sum_gX_+^g(s))
 +\kappa\sum_p[1-B(p)]\notag\\[-2pt]
 &\qquad -\sum_{\ell=ss'}\sum_{\Gamma\ne I}\lambda_\Gamma
 \tr\!\left[\bm Z^{\bar\Gamma}(s')\bm Z^\Gamma(\ell)
 \bm Z^\Gamma(s)\right]+H_X.
 \label{eq:main-gauge-Higgs-H}
\end{align}
In the group basis, the
last trace is the character $\chi_\Gamma( \Phi_\ell)$ of the gauge-invariant object $ \Phi_\ell=\bar f_{s'}g_\ell f_s$. This is the $G$ group
analogue of a Higgs hopping term: a charge moving between adjacent vertices is accompanied by the gauge connection on the intervening link. 

\paragraph{From $G$-qudits to a classical $G$ gauge--Higgs theory--} The isometry allows us to write the partition function of the perturbed quantum double in terms of the projector to physical states satisfying the $G$-Gauss law,  $Z=\Tr_{\mathcal H_{\mathrm{aux}}} \!\left[P_{\mathrm{phys}}e^{-\beta H_G^{(2+1)D}}\right]$.
Following a lattice path-integral procedure, detailed in SM Sec.~\ref{sec:SM-gauge-matter}, we map the partition function to a classical $G$ gauge--Higgs theory in (2+1)D. Therefore, we prove that for any finite $G$ with real, nonnegative $\lambda_{[g]}$ and Hermitian $H_X$ the perturbed $\mathcal{D}(G)$ is sign-free and amenable to the Monte Carlo methods that we develop in this work. Up to a constant term in the action, the mapping to a classical theory in (2+1)D is captured by the partition function
\begin{align}
 &Z\propto 
 \sum_{\{f_s^n,g_\ell^n,h_s^n\}}e^{-S_{\text{cl}}}
 , \; S_{\text{cl}}=-K\sum_{n,p\parallel xy}\delta_{U_p^n,1} \label{GGT-action}\\
 &-\sum_{n,\ell=ss'}
 \sum_{\Gamma\ne I}Q_\Gamma\chi_\Gamma( \Phi_\ell^n)-Q_\tau\sum_{n,s}\delta_{\Phi_{s\tau}^n,1}+\sum_{n,\ell}K_\tau([\bar U_{\ell\tau}^n]),\notag
\end{align}
where the matter-charge couplings are $Q_\Gamma = \lambda_\Gamma \Delta \tau$ and $Q_\tau=\ln\!\left[(1+(\abs{G}-1)e^{-J\Delta\tau})/
(1-e^{-J\Delta\tau})\right]$. Here $\Delta \tau = \beta/N_\tau$, $\beta$ is the inverse temperature of the quantum system, and $N_\tau$ is the number of layers in the time direction. 
The spatial gauge field coupling is $K=\kappa\Delta\tau$, and the explicit formula for the time coupling $K_\tau([g])$ is given in SM Sec.~\ref{sec:SM-pure-gauge}. 
The path-integral method inserts group-element states between Trotter steps, yielding classical $G$ variables $g_\ell^n$ and
$f_s^n$ for the spatial links and matter fields on each time slice $n$. The gauge-invariant combination 
$\Phi_\ell^n=\bar f_{s'}^n g_\ell^n f_s^n$ hops matter across spatial links, $\Phi_{s\tau}^n=\bar f_s^{n+1}h_s^n f_s^n$ does the same across adjacent time slices, and $U_{\ell\tau}^n=\bar g_\ell^n\bar h_{s'}^ng_\ell^{n+1}h_s^n$ is the corresponding temporal plaquette holonomy. The partition function trace makes $f$ and $g$ periodic in imaginary time.
The action of Eq.~\eqref{GGT-action} generalizes the Fradkin--Shenker result \cite{FradkinShenker1979} to general finite groups $G$. 
Setting $\lambda_\Gamma=0$ and taking $J\rightarrow\infty$ gives $Q_\tau\rightarrow 0$ and projects the vertex qudits onto $\ket I_s$. Thus, the vertex variables decouple and contribute only an overall factor, leaving a pure \((2+1)\)D \(G\)-gauge theory, as described in the End Matter. We now illustrate the approach for the smallest non-Abelian group $S_3$.

\begin{figure}[t]
 \centering
 \includegraphics[width=0.95\columnwidth]{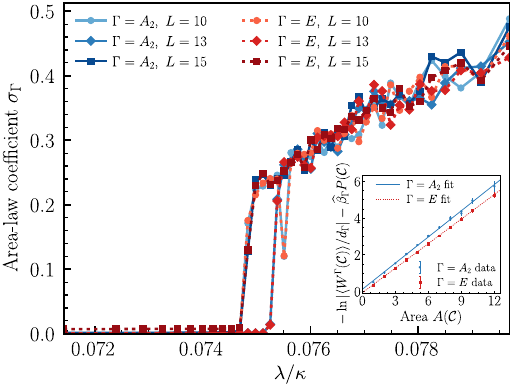}
 \caption{\textbf{Onset of the Wilson-loop area law at the confinement transition.}
 Effective area coefficient extracted from contractible $A_2$ (solid)
 and $E$ (dotted) Wilson loops for 
 $\beta=50$, and
 $\Delta\tau=0.3$. The fit includes both area and perimeter law contributions (inset $\lambda/\kappa=0.079$). The good overlap of the two irreps is consistent with the leading order
 representation-independent strong-field result. The sharp onset of
 the area coefficient marks confinement. Here, we set $\lambda_D=\lambda_F=\lambda$ and $\lambda_C=0$.
 } 
 \label{fig:main-area-law}
\end{figure}

\paragraph{Perturbed $\mathcal{D}(S_3)$ model and anyon proliferation.--} The group $S_3$ is the symmetry group of an equilateral triangle; see SM Sec.~\ref{sec:SM-S3} for details. It has three conjugacy classes and hence three irreducible
representations $A_1$, $A_2$, and $E$. Their dimensions  are
$d_{A_1}=d_{A_2}=1$, $d_E=2$. The eight
anyon types of $\mathcal{D}(S_3)$ \cite{kitaev_fault-tolerant_2003} are $\mathrm A=([1],I), \mathrm B=([1],A_2),
 \mathrm C=([1],E),
 \mathrm D=([s],I), \mathrm E=([s],A_2),
 \mathrm F=([r],I),
 \mathrm G=([r],\omega),\mathrm H=([r],\bar\omega)$.
Here $\omega=e^{2\pi i/3}$ denotes a $\mathbb{Z}_3$ irrep. Thus $\mathrm A$ is the vacuum;
$\mathrm B,\mathrm C$ are electric anyons, $\mathrm D,\mathrm F$ are magnetic anyons, and
$\mathrm E,\mathrm G,\mathrm H$ are dyons. The nontrivial bosons are
$\frak L=\{\mathrm B,\mathrm C,\mathrm D,\mathrm F\}$. We assign different proliferation coefficients labelled by their anyon types, which are magnetic
$\lambda_{\mathrm D}=\lambda_{[s]}$,
$\lambda_{\mathrm F}=\lambda_{[r]}$ or electric
$\lambda_{\mathrm B}=\lambda_{A_2}$,
$\lambda_{\mathrm C}=\lambda_E$. All $S_3$ conjugacy classes and irreps are
self-dual, so these four couplings may be chosen independently and taken real. To simplify the analysis, we  explore the phase diagram under symmetric proliferation of the magnetic anyons with average coupling $\lambda=(\lambda_{\mathrm D}+\lambda_{\mathrm F})/2$, and $\lambda=\lambda_{\mathrm D}=\lambda_{\mathrm F}$ in which case we have $H_X = -\lambda \sum_{\ell}\sum_{g\neq 1} X_+^g(\ell)$. 

\begin{figure}
 \centering
 \includegraphics[width=0.95\columnwidth]{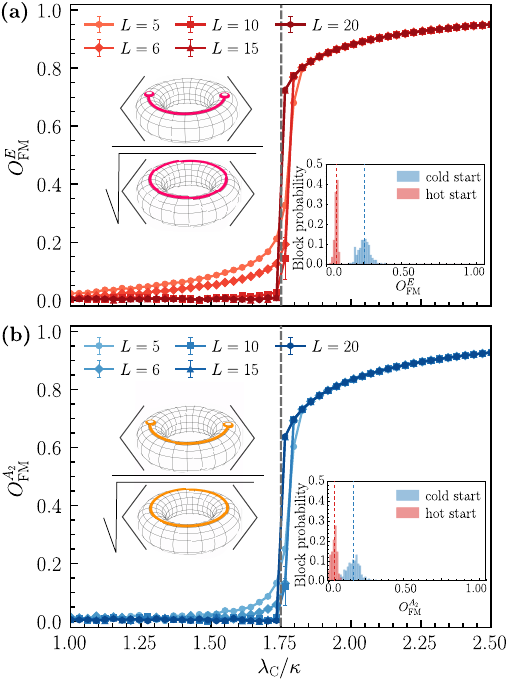}
 \caption{\textbf{Irrep-resolved electric-anyon condensation.}Fredenhagen--Marcu order parameters (a) $O_{\mathrm{FM}}^{E}$ and
 (b) $O_{\mathrm{FM}}^{A_2}$ versus $\lambda_{\mathrm C}/\kappa$ for the $S_3$
 gauge--Higgs theory at $J/\kappa=10$, $\lambda/\kappa=0.01$, and
 $\lambda_B=0$. We use $\beta=50$ and $\Delta\tau=0.1$. A simultaneous transition is observed at
 $\lambda_{\mathrm C}/\kappa\simeq 1.75$ in both FM order parameters (gray dashed line). Left insets show a schematic of the FM string order parameter associated with electric anyons $\mathrm{C}$ (pink) and $\mathrm{B}$ (orange). 
 Right insets show long-lived low- and high-FM order parameter branches at $\lambda_C/\kappa=1.75$ obtained from hot and cold starts, respectively, providing evidence for metastability consistent with a first-order transition. We present additional histograms in SM Sec.~\ref{sec:SM-numerics}.}\label{fig:FM}
\end{figure}

At fixed $J/\kappa\gg1$, no electric charges are allowed, the system is deconfined for $\lambda/\kappa \ll 1$, where contractible Wilson loops obey the perimeter law
\(\langle W^\Gamma(\mathcal C)\rangle\sim d_\Gamma e^{-\alpha_\Gamma P(\mathcal C)}\)$\expval{W^\Gamma(\mathcal C)} \sim d_\Gamma e^{-\alpha_\Gamma P(\mathcal C)}$. For $\lambda/\kappa\gg1$, the projector Hamiltonian $H_X$ instead selects the gapped electric vacuum $\ket{\mathbf I}=\bigotimes_\ell\ket I_\ell$. Its excitations are closed strings of length $l$ and energy $\lambda|G|l$, the leading area-law contribution is
$\expval{W^\Gamma(\mathcal C)} \sim d_\Gamma e^{-\sigma_\Gamma A(\mathcal C) -\beta_\Gamma P(\mathcal C)}$ with a sub-leading $\beta_\Gamma$ perimeter-law coefficient. These perturbative predictions (SM Sec.~\ref{sec:SM-confined-area-law}) persist at nonperturbative couplings, see Fig.~\ref{fig:main-area-law}, where the discontinuous onset of $\sigma_\Gamma$ signals a first-order transition. Through the mixed 't Hooft anomaly between the electric and magnetic 1-form symmetries~\cite{noninvertible1form}, Wilson loops also probe the ground-state sectors associated with spontaneous breaking of the magnetic symmetry; its restoration at confinement is discussed in the End Matter.

We now include dynamical matter and focus on the proliferation of the non-Abelian anyon $\mathrm C$, $\lambda_{\mathrm C}\neq 0$ setting $\lambda_{\mathrm B}=0$ and consider finite magnetic proliferation $\lambda/\kappa=0.01$.
Large Wilson loops are screened by dynamical electric charges, and therefore do not serve as a diagnostic of a phase transition toward the Higgs regime. 
We instead use generalized FM order parameters~\cite{FredenhagenMarcu1983,FredenhagenMarcu1986,BaisRomers2009,GregorHuseMoessnerSondhi2011,xuFM}. At the quantum-double fixed point, the closed operators $W^\Gamma(\mathcal C)$ generate a non-invertible electric 1-form symmetry~\cite{noninvertible1form}, while an open string $W^\Gamma(\mathcal S)$ creates electric anyons at its endpoints. Under the matterization isometry, these endpoints are dressed by matter fields, making the open string gauge invariant. The corresponding FM order parameter $O_{\mathrm{FM}}^\Gamma$ vanishes when the charges remain deconfined and becomes finite when they condense. The construction is illustrated in the left insets of Fig.~\ref{fig:FM}, where pink and orange lines denote the 1-form operators associated with the $\mathrm C$ and $\mathrm B$ anyons, respectively; its precise definition is given in the End Matter. 

We calculate the FM order parameter first for the $\mathrm C$ anyon, detected by the FM order parameter of the $E$ irrep. Figure~\ref{fig:FM}(a) shows a zoom-in near the transition point, where a jump in $O^E_{FM}$ is observed. Increasing
the linear length $L$ sharpens the discontinuity consistent with a first-order transition. The double peak structure observed histograms at $\lambda_C/\kappa=1.75$ (right insets) provide additional evidence for this observation. We interpret the finite value in the Higgs phase as a diagnostic of the condensation of the $\mathrm C$ anyon. Further values of $\lambda/\kappa$ are shown in the phase diagram of the End Matter. At the same transition point (gray dashed line in Fig.~\ref{fig:FM}), also the
$O_{\mathrm{FM}}^{A_2}$ order parameter jumps to a finite value. Their simultaneous onset is consistent with the fusion rule $\mathrm C\times \mathrm C = \mathrm A+\mathrm B+\mathrm C$, which states that even if $\mathrm B$ is not proliferated directly, it is affected by the $\mathrm C$ condensation. Following the branching theory of non-Abelian anyons \cite{BaisSchroersSlingerland2002,BaisSlingerland2009,KitaevKong2012, CongChengWang2016TQCGappedBoundaries}, we note that proliferating $\mathrm C$ may lead to $\mathrm C \rightarrow 1+e$, with $1$ denoting the vacuum. Here $e$ denotes the electric anyon of a $\mathcal{D}(\mathbb Z_2)$ toric-code phase. Further condensation of $e$ leads to a trivial phase. Our simulations are consistent with a direct transition into a trivial Higgs phase with the trivial ground state $\bigotimes_\ell \ket{1}_\ell$ in the $\lambda_{\mathrm C}\rightarrow \infty$ limit. 
A neighboring $\mathcal{D}(\mathbb Z_2)$ phase has been found along other deformation paths~\cite{XuGarreRubioSchuch2022,Lu2026SelfDualS3}; whether it appears in our model will be interesting to explore in future work.

\paragraph{Discussion \& outlook.--}
In this work, we developed a microscopic route from perturbed \(\mathcal D(G)\) quantum doubles to classical gauge theories. For any finite group $G$, an exact matterization isometry introduces vertex degrees of freedom and maps the original link-only model onto a $G$ gauge--Higgs theory. We show via a path-integral representation that the theory is sign-free and can be studied using large-scale Monte Carlo simulations. Within this formulation, the non-invertible electric 1-form symmetries of the quantum-double fixed point yield irrep-resolved FM order parameters that directly diagnose electric-anyon condensation. For $\mathcal D(S_3)$, this framework resolves two distinct routes out of non-Abelian topological order. In the purely magnetic limit, Wilson loops cross from perimeter- to area-law scaling while the distinct topological-flux sectors collapse. These results are consistent with a first-order confinement transition that restores the emergent magnetic 1-form symmetry. In our model, proliferating the non-Abelian electric anyon \(\mathrm C\) causes both the $E$ and $A_2$ FM responses to become finite through a first-order transition, even though no $\mathrm B$ anyon is directly proliferated. This demonstrates how non-Abelian fusion can induce condensation in a sector that is not directly perturbed.

Our construction turns generalized symmetry from a classification principle into a practical tool for studying non-Abelian anyon condensation. It connects microscopic quantum Hamiltonians, classical gauge--Higgs theories, and measurable string order parameters within a single scalable framework. Extending this approach to independently tunable electric and magnetic fields may reveal intermediate topological phases, self-dual critical points, and multicritical endpoints~\cite{selfdualhiggs,Lu2026SelfDualS3}. Our approach may also be useful for numerically exploring decohered states with non-abelian topological order~\cite{SalaVerresen2025LoopModels,SalaAliceaVerresen2025D4,SongZhang2025, Vadali2026NonAbelianProliferation}. More broadly, the same framework can guide the preparation and diagnosis of transitions out of non-Abelian topological order in programmable quantum simulators.

\paragraph{Acknowledgments}-- We acknowledge support from the Deutsche Forschungsgemeinschaft (DFG, German Research Foundation) under Germany’s Excellence Strategy--EXC--2111--390814868, TRR 360 – 492547816 and DFG grants No. KN1254/1-2, KN1254/2-1, the European Union (grant agreement No 101169765), as well as the Munich Quantum Valley, which is supported by the Bavarian state government with funds from the Hightech Agenda Bayern Plus.

\textit{{Data availability}}-- Numerical codes are available upon reasonable request on Zenodo~\cite{zenodo}.

\bibliography{manuscript_arxiv}

\clearpage

\appendix*
\setcounter{equation}{0}
\section{End Matter}\label{sec:EM}

\paragraph{Pure $G$-gauge theory and higher-form SSB.--}
Setting $\lambda_\Gamma=0$ removes electric charge proliferation terms.  Because the magnetic
hopping amplitudes are class functions, $H_X$ commutes with every $A(s)$, and
the limit $J\rightarrow\infty$ restricts the theory to
$A(s)\ket\psi=\ket\psi$.  On the maximally symmetric line, $\lambda_{[g]}=\lambda$,  every nonidentity
group element has the same amplitude and the
Hamiltonian becomes
\begin{align}
 H_G={}&\lambda\sum_\ell(1-\sum_{g\in G}X_+^g(\ell))
 +\kappa\sum_p[1-B(p)].\label{eq:main-pure-gauge-H}
\end{align}
The first term proliferates magnetic-flux pairs and competes with the
zero-flux projector.  General class-dependent amplitudes preserve the same
gauge structure and generate class-resolved temporal couplings, as derived in
SM Sec.~\ref{sec:SM-pure-gauge}.

To obtain the classical description, we resolve the quantum partition
function in the group basis at imaginary-time intervals $\Delta\tau$.  The
projector onto $A(s)=1$ supplies temporal links $h_s^n$, oriented from slice
$n$ to $n+1$.  Up to a Trotter error that vanishes as
$\Delta\tau\rightarrow 0$ we have $Z\propto\sum_{\{g_\mu(\mathbf x)\}}e^{-S_{\mathrm{cl}}[g]}$ with the action
\begin{align}
 S_{\mathrm{cl}}[g]
 &=K\sum_{p_s}(1-\delta_{U_{p_s},1})
 +K_\tau\sum_{p_\tau}(1-\delta_{U_{p_\tau},1}),
 \label{eq:main-anisotropic-LGT}
\end{align}
where $K=\kappa\Delta\tau$ and the normalized one-link transfer matrix gives $
 e^{-K_\tau}=\frac{e^{\abs{G}\lambda\Delta\tau}-1}
 {e^{\abs{G}\lambda\Delta\tau}+\abs{G}-1}.
 \label{eq:main-temporal-coupling}$
All holonomies use the right-to-left convention of Eq.~\eqref{ABops}, with
$g_\tau=h$.  Periodicity of the quantum trace requires summing over temporal
holonomies, so fixing temporal gauge must not remove sectors that wind around
imaginary time.  At the isotropic point $K_\tau=K$, the action may be
written, up to a constant, as
\begin{align}
 Z
 &=\sum_{\{g_\mu(\mathbf x)\}}e^{-S_{\text{cl}}[U_{\mu\nu}]},\quad
 U_{\mu\nu}(\mathbf x)
 =\mathcal P\!\prod_{\ell\in p_{\mu\nu}(s)}g_\ell^{O_{\ell p}},
 \notag\\[-2pt]
&S_{\text{cl}}[U_{\mu\nu}]=-\frac{K}{\abs{G}}
 \sum_{s,\mu<\nu}\sum_\Gamma d_\Gamma
 \tr\Gamma(U_{\mu\nu}(\mathbf x)).
 \label{eq:main-isotropic-LGT}
\end{align}
For a single irrep this has the Wilson form; here the group delta function
requires the sum over all irreps.
For $\lambda/\kappa\ll1$, the system is deconfined. From perturbation theory about the quantum-double fixed point we find that virtual flux pairs contribute to the Wilson-loop only at second order (SM Sec.~\ref{sec:SM-deconfined-perimeter-law}), yielding the perimeter law $
\frac{\langle W^\Gamma(\mathcal C)\rangle}{d_\Gamma}
\sim e^{-\alpha_\Gamma P(\mathcal C)},
\qquad
\alpha_\Gamma=
\left(\frac{\lambda}{2\kappa}\right)^2
\left[(\abs{G}-1)-\sum_{g\ne1}\chi_\Gamma(g)/d_\Gamma\right]
+O\!\left[(\lambda/\kappa)^4\right].
$
For an $x$-winding loop, the same decay is weighted by the character of the magnetic flux class $[g]$ threading the orthogonal cycle: $
\langle W_x^\Gamma\rangle
\sim\chi_\Gamma([g])e^{-\alpha_\Gamma L}.$
In the opposite large-$J$, large-$\lambda$ limit, $H_X$ is proportional to a sum of projectors whose ground state is the electric vacuum
$\ket{\mathbf I}=\bigotimes_\ell\ket I_\ell$
(SM Sec.~\ref{sec:SM-pure-gauge}). Within the $A(s)=1$ sector, excitations are closed electric strings of length $l$ and energy $\lambda\abs{G}l$. A contractible Wilson loop receives a nonzero perturbative contribution only after plaquette insertions cover its entire interior, giving
$\frac{\langle W^\Gamma(\mathcal C)\rangle}{d_\Gamma}
\propto e^{-\sigma_\Gamma A(\mathcal C)},
\qquad
\sigma_\Gamma\sim
\left|\ln\!\left[\frac{\kappa}{\lambda\abs{G}^2}\right]\right|. $
The leading area coefficient is independent of $\Gamma$. By contrast, a noncontractible loop cannot be assembled from any finite number of contractible plaquette insertions and therefore vanishes to every finite order in this expansion. 

\begin{figure}
 \centering
 \includegraphics[width=\columnwidth]{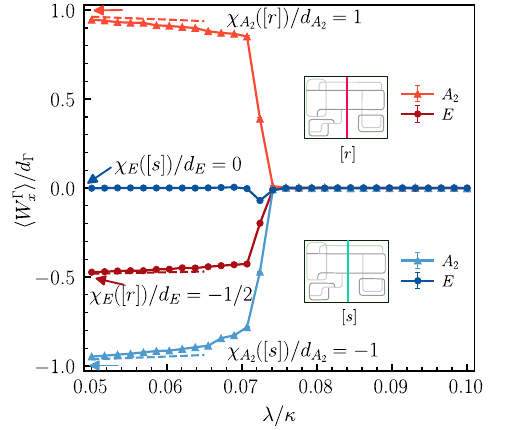}
 \caption{\textbf{Higher-form symmetry breaking and confinement.}
 Noncontractible $x$-cycle $\Gamma=\{A_2,E\}$ Wilson loops for $S_3$ at $L=10$, $\beta=60$ and
 $\Delta\tau=0.3$ for different initial random Monte Carlo runs, blue and red. In the deconfined phase the $[s]$ (blue) and $[r]$
 (red) flux sectors retain distinct expectation values approaching their respective characters as indicated by the arrows; the perturbative result in $\lambda/\kappa$ is shown as dashed lines. The values of the Wilson loops for each branch are consistent with fixed magnetic 1-form symmetry sectors indicating a broken emergent magnetic 1-form symmetry detected via electric 1-form symmetry operators. Across confinement all branches
 collapse to zero and become indistinguishable, signaling symmetry
 restoration.}
 \label{fig:main-higher-form}
\end{figure} 
\begin{figure*}[ht!]
    \centering
    \includegraphics[width=0.9\textwidth]{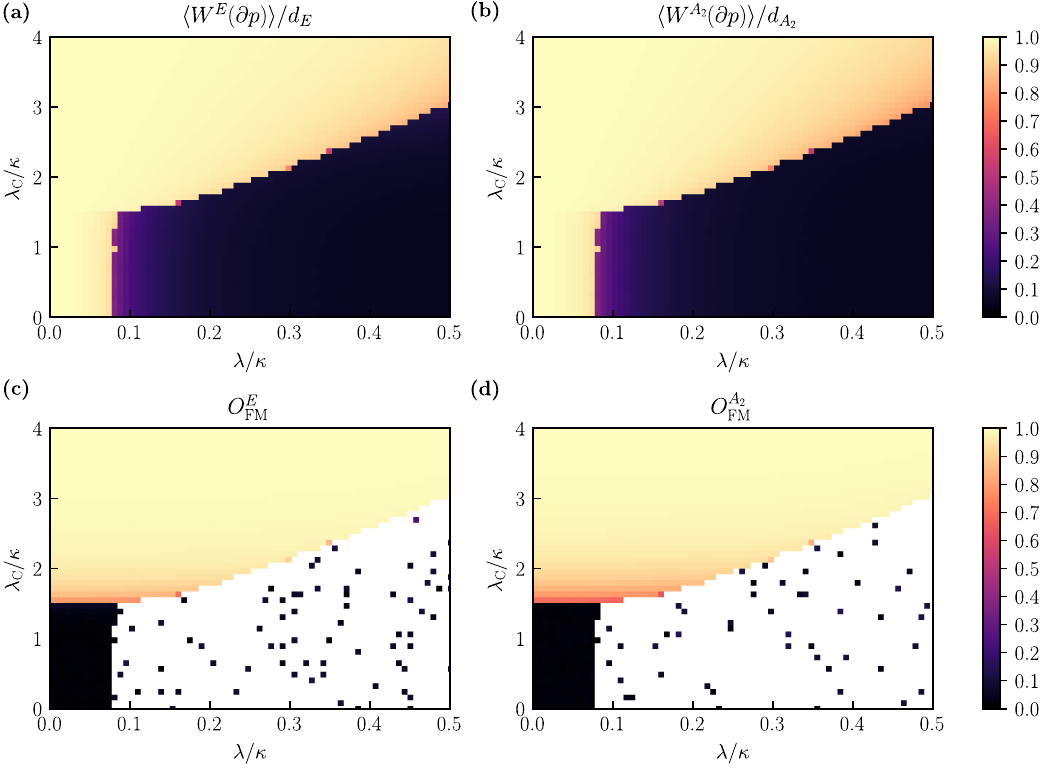}
    \caption{\textbf{Phase diagram of the $\mathcal{D}(S_3)$ quantum double under electric and magnetic anyon proliferation. }
    Heat maps obtained for $L=10$, $\beta=25$, $\kappa=1$, $J=10$,
    $\Delta\tau=0.25$, $\lambda_{A_2}=0$. The horizontal axis $\lambda=\lambda_{\mathrm D}= \lambda_{\mathrm F}$ is the strength of the coupling proliferating magnetic anyons, while the vertical axis  $\lambda_{\mathrm C}$ is the strength of proliferating the electric anyon $\mathrm C=([1],E)$. 
    (a) Normalized minimal $E$ Wilson plaquette
    $\langle W^E(\partial p)\rangle/d_E$,
    (b) the normalized $A_2$ Wilson plaquette
    $\langle W^{A_2}(\partial p)\rangle/d_{A_2}$, (c) $O_{\mathrm{FM}}^E$, and
    (d) $O_{\mathrm{FM}}^{A_2}$.  White points in the FM panels are not zeros: there the closed-loop
    denominator is nonpositive or lies within two blocking errors of zero, so
    the ratio is statistically unresolved and is masked. Black dots in the white region are numerical noise.}
    \label{fig:matter-phase-diagram}
\end{figure*}
The distinct plateaus in Fig.~\ref{fig:main-higher-form} for small $\lambda$ represent two states selected via spontaneous symmetry breaking (SSB) of the emergent magnetic 1-form symmetry discussed in Ref.~\cite{noninvertible1form}. Just like an Ising ferromagnet spontaneously breaks the $\mathbb{Z}_2$ symmetry by selecting an up or down state, in Fig.~\ref{fig:main-higher-form} a state with a nontrivial $[s]$ flux is selected in the blue realization and a $[r]$ flux state in the red realization. The SSB state is detected by Wilson loop operators that furnish the electric 1-form symmetry and obey
$\operatorname{Rep}(S_3)$ fusion. For $A_2$,
$\sum_{g\ne1}\chi_{A_2}(g)=-1$, giving
$\alpha_{A_2}=\tfrac32(\lambda/\kappa)^2+O[(\lambda/\kappa)^4]$.  For small $\lambda/\kappa$ the perturbative results (dashed lines in Fig.~\ref{fig:main-higher-form}) agree well with the exact numerics.  The opposite signs
come directly from the two character values.  Their common collapse agrees
with the confined-limit result that a noncontractible nontrivial Wilson loop
vanishes to every finite order in the strong-field expansion.
The collapse of the two sectors occurs at the same coupling at which
the area coefficient in Fig.~\ref{fig:main-area-law} turns on.  The isotropic
classical simulations in SM Sec.~\ref{sec:SM-numerics} additionally show an
energy discontinuity and long-lived phase coexistence near $\lambda/\kappa\simeq 0.075$,
consistent with a first-order confinement transition.

\paragraph{Gauge--matter phase diagram.--}
Figure~\ref{fig:matter-phase-diagram} extends the pure-gauge cut above by
varying both the magnetic coupling $\lambda$ and the electric coupling $\lambda_{\mathrm C}$.
The four panels separate the three regimes sketched in
Fig.~\ref{fig:2+1D_phasediag}.  In the lower-left region, both minimal Wilson
plaquettes remain near unity while both FM order parameters are small, as expected
for the deconfined $\mathcal D(S_3)$ phase with uncondensed electric charges.
At small $\lambda_{\mathrm C}$, increasing $\lambda/\kappa$ through approximately
$0.075$ suppresses both Wilson plaquettes.  In the same magnetically condensed
region the large closed Wilson loop falls below the statistical resolution
required for an FM ratio, producing the white domains in panels (c) and (d);
these points thus must not be interpreted as $O_{\mathrm{FM}}=0$, expected when the respective emergent 1-form symmetry is absent~\cite{xuFM}. Increasing $\lambda_{\mathrm C}/\kappa$ instead produces a light region in both FM
panels, with $O_{\mathrm{FM}}^E$ and $O_{\mathrm{FM}}^{A_2}$ approaching one.
This is the electric-charge-condensed Higgs regime.   The deconfined--Higgs boundary near
$\lambda_{\mathrm C}/\kappa\simeq1.75$ meets the pure-gauge confinement boundary near
$\lambda/\kappa\simeq0.075$ and continues diagonally between the electrically
and magnetically condensed regimes. The
finite-size cuts in Fig.~\ref{fig:FM} and the pure-gauge benchmarks in SM Sec.~\ref{sec:SM-numerics}
provide the corresponding sharpening and coexistence diagnostics.  The
simulation protocol and error analysis are detailed in
SM Sec.~\ref{sec:SM-numerics}.

\textit{FM order parameters from non-invertible 1-form symmetries.--} The FM order parameter removes
the bulk contribution using a closed loop of twice the length
\cite{FredenhagenMarcu1983,FredenhagenMarcu1986,BaisRomers2009,GregorHuseMoessnerSondhi2011,xuFM}. For an open string crossing half of the bonds in the horizontal direction $L_{1/2}$ ($|L_{1/2}|=|L|/2$) we define 
\begin{align}
&O_{FM}^\Gamma= \lim _{\left|L_{1 / 2}\right| \rightarrow \infty} \sqrt{\left|C_\Gamma\left(\left|L_{1 / 2}\right|\right)\right|},\notag\\
& C_\Gamma\left(\left|L_{1 / 2}\right|\right) =\dfrac{1}{\sqrt{d_\Gamma}} \dfrac{\ev{\text{tr} \Big[\mathcal{P}\prod_{\ell\in L_{1/2}}(\bm{Z}^{\Gamma}(\ell))\Big ]}}{ \sqrt{\ev{\text{tr} \Big[\mathcal{P}\prod_{\ell\in L}(\bm{Z}^{\Gamma}(\ell))\Big]}}}.
\end{align}

\clearpage
\newpage

\renewcommand{\t}[1]{\text{#1}}
\renewcommand{\theequation}{S\arabic{equation}}
\renewcommand{\selectlanguage}[1]{}
\renewcommand{\thefigure}{S\arabic{figure}}

\makeatletter
\@removefromreset{equation}{section}
\makeatother
\setcounter{equation}{0}
\setcounter{figure}{0}
\setcounter{secnumdepth}{1}

\renewcommand{\thesection}{S\arabic{section}}

\newpage
\onecolumngrid
\begin{center}
  \textbf{\large Supplemental material for `` Non-Abelian Anyon Condensation: a Path-Integral Monte Carlo Approach''}\\[.2cm]
  R. Flores-Calder\'on,$^{1,2,*}$ Frank Pollmann,$^{1,3}$ and Michael Knap$^{1}$\\[.1cm]
  {\itshape ${}^1$Technical University of Munich, TUM School of Natural Sciences, Physics Department, 85748 Garching, Germany\\
  ${}^2$Munich Center for Quantum Science and Technology (MCQST), Schellingstr. 4, 80799 M{\"u}nchen, Germany\\
(Dated: \today)\\[1cm]}
\end{center}

\makeatletter
\newif\ifSMtocactive
\begingroup
  \hypersetup{linkcolor=RubineRed}
  \color{RubineRed}
  \SMtocactivefalse
  \let\SM@contentsline\contentsline
  \def\SM@sectionlevel{section}
  \def\SM@subsectionlevel{subsection}
  \renewcommand{\contentsline}[4]{%
    \ifSMtocactive
      \def\SM@currentlevel{#1}%
      \ifx\SM@currentlevel\SM@sectionlevel
        \SM@contentsline{#1}{#2}{#3}{#4}%
      \else\ifx\SM@currentlevel\SM@subsectionlevel
        \SM@contentsline{#1}{#2}{#3}{#4}%
      \fi\fi
    \fi}
  \def\SMtocstart{\SMtocactivetrue}
  \renewcommand*\l@section[2]{%
    \@dottedtocline{1}{0em}{3.2em}{\small\bfseries #1}
      {\small\bfseries\color{RubineRed}#2}}
  \renewcommand*\l@subsection[2]{%
    \@dottedtocline{2}{1.8em}{2.8em}{\small #1}
      {\small\color{RubineRed}#2}}
  \let\SM@numberline\numberline
  \renewcommand{\numberline}[1]{\SM@numberline{#1.}}
  \setcounter{tocdepth}{2}
  \providecommand{\contentsname}{Contents}
  \tableofcontents
\endgroup
\addtocontents{toc}{\protect\SMtocstart}
\makeatother
\vspace{1cm}

This Supplemental Material begins with the numerical methods used for the
finite-size $S_3$ gauge--matter phase diagram in the End Matter.  We then fix the $G$-qudit,
quantum-double, and ribbon-operator conventions, derive the pure-gauge
classical mapping and its class-function generalization, and discuss Wilson
loops.  The final sections give the tilted-field matterization map, the complete
(2+1)D gauge--matter action, and the irrep-resolved Fredenhagen--Marcu order parameter
construction.

\section{Numerical methods}
\label{sec:SM-numerics}

\subsection{Gauge--matter simulations}

We simulate the group-element action derived in
Sec.~\ref{sec:SM-gauge-matter} on a periodic
$L\times L\times N_\tau$ lattice, without fixing temporal gauge.  A
configuration contains three positively oriented gauge links and one matter
variable $f$ per space--time site.  One systematic heat-bath sweep visits all
$3L^2N_\tau$ links and all $L^2N_\tau$ matter variables once.  At each visit
we enumerate the six possible $S_3$ elements and sample the exact conditional
distribution
\begin{align*}
 P(q\mid\text{rest})=
 \frac{e^{-S_{\mathrm{loc}}(q)}}
 {\sum_{q'\in S_3}e^{-S_{\mathrm{loc}}(q')}}.
\end{align*}
For a spatial link, $S_{\mathrm{loc}}$ contains the four adjacent plaquettes
and its spatial Higgs bond; for a temporal link it contains the four adjacent
temporal plaquettes and the corresponding temporal matter bond.  Updating a
matter variable changes its four incident spatial Higgs bonds and its two
temporal matter bonds.  Enumerating all six candidates makes each local step
rejection free and includes the non-Abelian multiplication order exactly.

The phase diagram in Fig.~\ref{fig:matter-phase-diagram} uses
$\kappa=1$, $J=10$, $\lambda_{A_2}=0$, $L=10$, $N_\tau=100$, and
$\Delta\tau=0.25$ ($\beta=25$).  We take equal magnetic class amplitudes
$\lambda_{[s]}=\lambda_{[r]}\equiv\lambda$ and use the exact temporal transfer
weights of Eq.~\eqref{eq:class-transfer-weight-irreps}, together with the
exact temporal matter coupling.  The grid contains 70 uniformly spaced
values $0.001\leq\lambda\leq0.5$ and 50 values
$0\leq\lambda_E\leq4$, for 3500 independently seeded points.  Every point
starts from the identity configuration, is thermalized for $10^4$ sweeps,
and is followed by 1000 measurements separated by four full heat-bath
sweeps.

The minimal Wilson observables are the normalized characters
$\langle W^\Gamma(\partial p)\rangle/d_\Gamma$ of a spatial plaquette,
averaged over all spatial translations and imaginary-time slices.  For the
FM observable we use a $5\times4$ spatial rectangle.  Its matter-dressed open
half-boundary has length $r=9$, while the closed denominator has perimeter
$2r=18$; both complementary halves and all translations are averaged.  The
FM ratio is always formed from ensemble means of the paired open and closed
histories.  Errors use a delete-one-block jackknife with ten consecutive
blocks of 100 measurements, preserving numerator--denominator correlations.
We report an FM value only when the closed-loop mean is positive, exceeds two
blocking errors, and remains positive in every jackknife replica.

Because the FM observable is a nonlinear ratio of expectation values, its
histogram cannot be formed from a configuration-by-configuration ratio.  We
instead divide each paired open- and closed-string history into consecutive
blocks and evaluate, for each block $b$,
\begin{align*}
 O_{{\rm FM},b}^{\Gamma}
 =\left|\frac{\overline w_{{\rm open},b}^{\Gamma}}
 {\sqrt{\overline w_{{\rm closed},b}^{\Gamma}}}\right|^{1/2}.
\end{align*}
Only blocks with a positive closed-loop mean enter the probability
distribution.  This construction probes the coarse-grained order parameter
on time scales longer than individual local fluctuations while retaining
information about metastable branches.

\begin{figure}[t]
    \centering
    \includegraphics[width=0.8\textwidth]{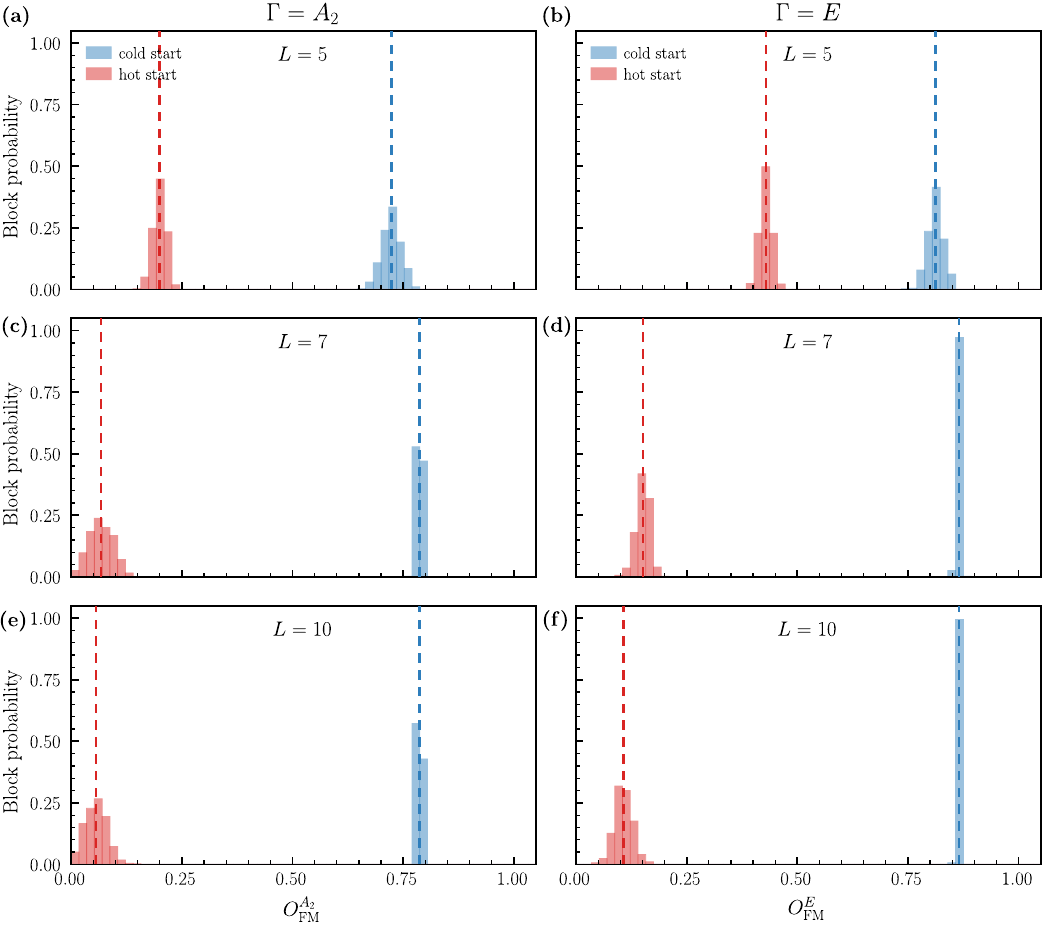}
    \caption{\textbf{Histograms of the irrep-resolved FM order
    parameter.}  Block-probability distributions of
    $O_{\mathrm{FM}}^{A_2}$ (left column) and $O_{\mathrm{FM}}^{E}$ (right
    column) for $L=5,7,10$ (top to bottom) near the gauge--matter transition.
    The simulations use $\lambda/\kappa=0.01$, $\kappa=1$, $J=10$,
    $\lambda_E/\kappa=1.5$, $\lambda_{A_2}=0$, $N_\tau=70$, and
    $\Delta\tau=0.25$ ($\beta=17.5$).  Each chain is thermalized for
    $2\times10^4$ heat-bath sweeps before $10^4$ measurements are collected.
    Blue and red histograms denote identity-configuration (cold) and random
    (hot) starts, respectively; dashed lines mark the corresponding branch
    means.  The cold histories remain in a high-$O_{\mathrm{FM}}$ Higgs
    branch, whereas the hot histories remain in a low-$O_{\mathrm{FM}}$
    uncondensed branch.  The narrowing distributions, depleted intermediate
    weight, and persistent branch separation with increasing $L$ are
    finite-size evidence for metastability and phase coexistence, supporting
    a first-order transition.  Both irreps display the same coexistence even
    though $\lambda_{A_2}=0$, consistent with
    $E\otimes E=A_1\oplus A_2\oplus E$.}
    \label{fig:Histograms}
\end{figure}

Figure~\ref{fig:Histograms} complements the smooth finite-size FM cuts in
Fig.~\ref{fig:FM} by resolving the two branches hidden by an ensemble mean.
Already at $L=5$, both irreps separate into a high-FM population reached from
the cold start and a low-FM population reached from the hot start.  The
distributions become narrower from $L=5$ to $L=10$, while their separation
does not decrease.  Thus the simultaneous onset of the $E$ and $A_2$
responses is not produced by a single broad crossover distribution: within
the simulated window, the Markov chains retain memory of two distinct
macroscopic states. Their robust separation instead diagnoses metastability.  In
combination with the sharpening of the FM curves with $L$, it supports a
first-order deconfined--Higgs transition near
$\lambda_E/\kappa\simeq1.75$ for the chosen finite-size parameters.  The
pure-gauge data below provide a separate benchmark for this hot--cold-start
coexistence criterion at a confinement transition.

Let us calculate the Trotter accuracy, which is relevant for the observables.  Write
$H=H_{\mathrm d}+H_{\mathrm o}$, with $H_{\mathrm d}$ diagonal in the group
basis, and set $N_\tau=\beta/\Delta\tau$.  Although the unsymmetrized product
has a generic operator error $O(\beta\Delta\tau)$, cyclicity of the trace gives
the exact identity
\begin{align}
 \Tr\!\left[(e^{-\Delta\tau H_{\mathrm d}}
 e^{-\Delta\tau H_{\mathrm o}})^{N_\tau}\right]
 =\Tr\!\left[(e^{-\Delta\tau H_{\mathrm d}/2}
 e^{-\Delta\tau H_{\mathrm o}}
 e^{-\Delta\tau H_{\mathrm d}/2})^{N_\tau}\right].
 \label{eq:trace-symmetric-trotter}
\end{align}
The right-hand side is the symmetric decomposition: its error per slice is
$O(\Delta\tau^3)$ and hence its accumulated error at fixed $\beta$ is
$O(\beta\Delta\tau^2)$.  The same identity holds with an observable inserted
when $[O,H_{\mathrm d}]=0$, as for the equal-time plaquette, Wilson-loop, and
FM string observables used here.  This estimate assumes that each block
exponential is evaluated exactly; approximating a transfer weight introduces
a separate finite-step error.

\subsection{Pure gauge theory simulations}

For the pure-gauge data we again impose periodic boundary conditions in all
three directions and retain all temporal-holonomy sectors.  The isotropic
benchmark uses an $L^3$ lattice, whereas the quantum-model scans use an
anisotropic $L\times L\times N_\tau$ lattice.  All plaquette and Wilson
holonomies follow the right-to-left convention of Eq.~\eqref{ABops}.

The anisotropic runs use a rejection-free single-link heat bath.  For each
visited link, all six $S_3$ elements are assigned their local spatial- and
temporal-plaquette actions and sampled with their exact Boltzmann weights.
The isotropic comparison instead uses a single-link Metropolis update: a
uniformly chosen $S_3$ element is accepted with probability
$\min(1,e^{-\Delta S})$.  For every stored configuration, loop observables
are averaged over translations before entering the Monte Carlo history.  The
$A_2$ and $E$ characters are evaluated on the same holonomy, with the latter
divided by $d_E=2$.

Statistical errors of ordinary observables are obtained from consecutive
block means,
\begin{align*}
 \delta O=\frac{\operatorname{std}(\overline O_1,\ldots,\overline O_{N_b})}
 {\sqrt{N_b}}.
\end{align*}
We use blocks of 128 measurements for the isotropic scan and ten stored
measurements for the anisotropic scans; anisotropic configurations are
separated by eight full heat-bath sweeps.  Near phase coexistence these error
bars quantify fluctuations within a sampled branch and do not include the
slower uncertainty associated with tunnelling, finite-size drift, or the
Trotter discretization.

\begin{figure}[t!]
\centering
        \includegraphics[scale=0.9]{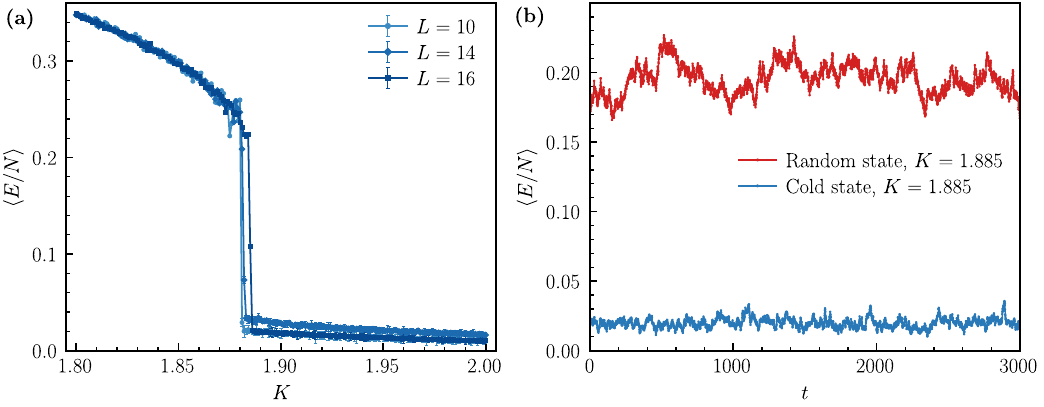}
        \label{fig:Ek}
    \caption{\textbf{Isotropic classical pure $G$ gauge theory transition.}  (a) Plaquette-defect density of Eq.~\eqref{CGGT} from
    single-link Metropolis simulations.  The discontinuity sharpens with
    increasing $L$.  (b) Histories initialized from random (hot) and
    identity-link (cold) configurations at $L=16$ and $K=1.885$.  Their
     long-lived separation is finite-size evidence for phase coexistence.}
    \label{fig:MC_CGGT}
\end{figure}

For Eq.~\eqref{CGGT} we measure the plaquette-defect density
$E=1-(3L^3)^{-1}\sum_p\delta_{U_p,1}$.  Figure~\ref{fig:MC_CGGT}(a) uses
$L=10,14,16$ and an ascending scan over $1.8\leq K\leq2.0$ in steps of
$10^{-3}$.  The first point is thermalized for $2^{11}$ sweeps and every
subsequent point for $2^{10}$ sweeps, using the final configuration at the
preceding coupling as its initial state; $2^{12}$ measurements are collected
at each point.  The increasingly sharp jump places the finite-size transition
near $K\simeq1.88$.  At $K=1.885$ and $L=16$, random and identity-link
initial states remain near $E\simeq0.20$ and $E\simeq0.02$, respectively,
for 3000 sweeps without observed tunnelling, Fig.~\ref{fig:MC_CGGT}(b).  This
supports a first-order transition. The anisotropic Wilson-loop scans use $\kappa=1$, $\beta=50$, and target
$\Delta\tau=0.3$, giving $N_\tau=167$ and the actual slice width
$\beta/N_\tau=0.2994$. At fixed $L$, the scan is
traversed from large to small $\lambda/\kappa$; the final configuration at one
field initializes the next, followed by rethermalization with an independent
random stream.

\begin{figure}[t!]
\centering
    \includegraphics[width=0.9\textwidth]{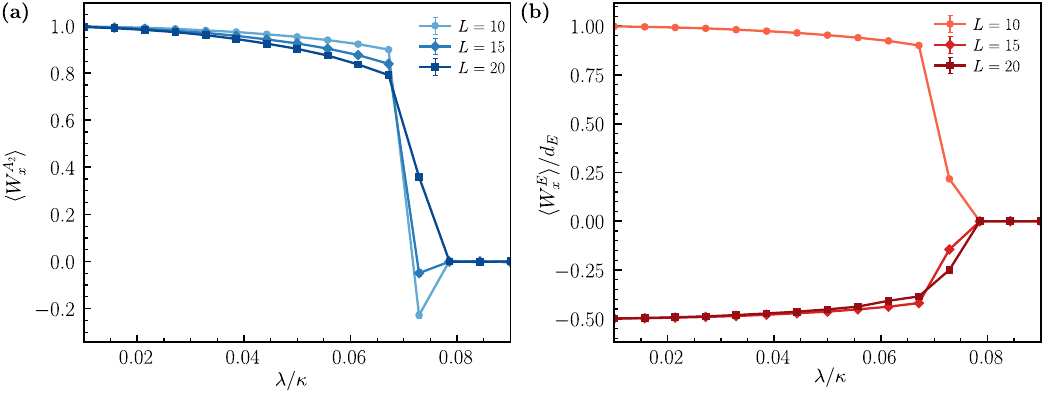}
    \caption{\textbf{Decay of noncontractible Wilson loops.} Results for (a)
    $\langle W_x^{A_2}\rangle$ and (b)
    $\langle W_x^E\rangle/d_E$ at  $\kappa=1$, $\beta=50$, and target
    $\Delta\tau=0.3$ for $L=10,15,20$. Deconfined-side plateaus depend on the flux
    sector retained by each annealed history, whereas both representations
    lose this sector dependence and approach zero across confinement.}
    \label{fig:non-contr}
\end{figure}

For the noncontractible loops in Fig.~\ref{fig:non-contr}, we use
$L=10,15,20$ and 15 equally spaced values
$0.01\leq\lambda/\kappa\leq0.09$.  Each point receives 5000 thermalization
sweeps followed by 800 measurements.  Below the transition, each annealed
history retains a selected flux sector over the simulation window.  The
sector-dependent signs and plateaus therefore should not be averaged across
sizes.  The relevant finite-size signature is the increasingly sharp loss of
sector memory: both irreps approach zero over the same narrow interval as
$L$ grows. The area-law analysis in Fig.~\ref{fig:main-area-law} uses $L=10,13,15$ and
40 field values concentrated in
$0.07\lesssim\lambda/\kappa\lesssim0.08$.  We perform $10^4$
thermalization sweeps and collect 1000 measurements per point.  On every
configuration we average 15 translated rectangles, including compact loops
up to $4\times3$ and elongated $m\times1$ loops up to $m=9$.  The effective
coefficient is obtained from an unweighted least-squares fit of
$-\ln|\langle W^\Gamma\rangle/d_\Gamma|$ versus area and perimeter, where we extract the area coefficient. 

\section{$G$-qudit and quantum-double conventions}
\label{sec:SM-Gqudit}

\subsection{$G$-qudit algebra}

For a finite group $G$, a $G$-qudit has Hilbert space
$\mathbb C[G]$ with orthonormal basis
$\{\ket g:g\in G\}$.  We write $\bar g=g^{-1}$ and use the generalized
shift and clock operators
\begin{align}
 X_+^h\ket g&=\ket{hg},&
 X_-^h\ket g&=\ket{g\bar h},\notag\\
 Z_+^h\ket g&=\delta_{h,g}\ket g,&
 Z_-^h\ket g&=\delta_{h,\bar g}\ket g.
 \label{eq:G-qudit-operators}
\end{align}
Thus $(X_\pm^h)^\dagger=X_\pm^{\bar h}$,
$Z_\pm^hZ_\pm^k=\delta_{h,k}Z_\pm^h$, and
$\sum_hZ_\pm^h=\mathbb I$.  Left and right shifts commute.  For a unitary
irrep $\Gamma$ of dimension $d_\Gamma$, it is convenient to collect the
diagonal operators into the matrix
\begin{align}
 [\bm Z^\Gamma]_{\alpha\beta}
 &=\sum_{g\in G}\Gamma(g)_{\alpha\beta}Z_+^g,
 &
 [\bm Z^{\bar\Gamma}]_{\alpha\beta}
 &=\sum_{g\in G}\Gamma(\bar g)_{\alpha\beta}Z_+^g.
 \label{eq:matrix-Z-definition}
\end{align}
In particular,
$\tr\bm Z^\Gamma\ket g=\chi_\Gamma(g)\ket g$.  The Fourier basis used
below is
\begin{align}
 \ket{\Gamma_{\alpha\beta}}
 =\sqrt{\frac{d_\Gamma}{\abs{G}}}
 \sum_{g\in G}\Gamma(g)_{\alpha\beta}\ket g,
 \label{eq:G-qudit-Fourier-basis}
\end{align}
and is orthonormal by Schur orthogonality.

On the square lattice, every positive-$x$ link points right and every
positive-$y$ link points upward.  Define $O_{\ell s}=-$ when $\ell$ points
away from a vertex $s$ and $O_{\ell s}=+$ when it points toward $s$.  The
local gauge transformation is
\begin{align}
 A_h(s)=\prod_{\ell\in\operatorname{star}(s)}X_{O_{\ell s}}^h(\ell),
 \qquad A(s)=\frac{1}{\abs{G}}\sum_{h\in G}A_h(s).
 \label{eq:SM-star-definition}
\end{align}
Label the plaquette links $\ell_1,\ell_2,\ell_3,\ell_4$ counterclockwise,
starting from the lower edge.  Thus the first two link arrows agree with the
traversal and the last two oppose it.  With
$O_{\ell p}=+$ in the first case and $O_{\ell p}=-$ in the second, define
\begin{align}
 B_k(p)=
 \sum_{h_4h_3h_2h_1=k}
 \prod_{m=1}^4Z_{O_{\ell_m p}}^{h_m}(\ell_m),
 \qquad B(p)=B_1(p).
 \label{eq:SM-plaquette-definition}
\end{align}
On a group-basis configuration this constraint is
$U_p=\bar g_4\bar g_3g_2g_1=k$.  All ordered boundary products below are
understood to accumulate factors from right to left.  In coordinate notation,
$g_1=g_\nu(\textbf{x})$, $g_2=g_\mu(\textbf{x}+\uvec{e}_\nu)$,
$g_3=g_\nu(\textbf{x}+\uvec{e}_\mu)$, and
$g_4=g_\mu(\textbf{x})$ give precisely the holonomy $\dd g_{\mu\nu}$ used in
Eq.~\eqref{CGGT}.  The ordered constraint is essential for non-Abelian $G$.  Equations
\eqref{eq:SM-star-definition} and \eqref{eq:SM-plaquette-definition} give
commuting projectors, and hence the quantum-double Hamiltonian in
Eq.~\eqref{QD}.

\subsection{Ribbon operators and the elementary condensation fields}
\label{sec:SM-ribbons}

A ribbon $\rho$ joins two sites of the quantum-double lattice, where a site
is a vertex together with an adjacent plaquette.  Kitaev's operators
$F_\rho^{h,g}$ form a basis of the ribbon algebra
\cite{kitaev_fault-tolerant_2003}.  Fourier transforming this algebra into
irreducible representations of $\mathcal D(G)$ gives matrix-valued operators
$F_\rho^{\mathfrak a;\mu\nu}$ labelled by
$\mathfrak a=([g],R)$.  Acting on a ground state, an open ribbon commutes with
all star and plaquette projectors away from its endpoints and creates
$\mathfrak a$ at one end and $\bar{\mathfrak a}$ at the other.  We use
\begin{align}
 \tr F_\rho^{\mathfrak a}
 \equiv\sum_\mu F_\rho^{\mathfrak a;\mu\mu}
 \label{eq:ribbon-trace-definition},
\end{align}
for the internal trace appearing in Eq.~\eqref{eq:main-ribbon-field}.

Only the shortest direct and dual ribbons are required in this work.  For an
oriented link $\ell$, we fix their normalization by
\begin{align}
 \tr F_\ell^{([1],\Gamma)}
 &\equiv\tr\bm Z^\Gamma(\ell),&
 \tr F_\ell^{([g],I)}
 &\equiv\sum_{h\in[g]}X_+^h(\ell).
 \label{eq:elementary-ribbon-fields}
\end{align}
The first operator creates a pure electric pair
$([1],\Gamma)$--$([1],\bar\Gamma)$ at the two vertices adjacent to $\ell$.
The second creates a pure magnetic pair $([g],I)$--$([\bar g],I)$ on the two
plaquettes adjacent to $\ell$.  Reversing the ribbon replaces
$\Gamma\mapsto\bar\Gamma$ and $[g]\mapsto[\bar g]$.  Consequently,
\begin{align}
 [\tr\bm Z^\Gamma]^\dagger&=\tr\bm Z^{\bar\Gamma},&
 \left[\sum_{h\in[g]}X_+^h\right]^\dagger
 &=\sum_{h\in[\bar g]}X_+^h,
 \label{eq:elementary-ribbon-adjoints}
\end{align}
which yields the Hermiticity conditions stated below
Eq.~\eqref{eq:main-electric-magnetic-fields}.  Closed versions of the first
operator are the Wilson loops used later in this supplement.

\subsection{The group $S_3$}
\label{sec:SM-S3}

The group $S_3$ is the permutation group of three objects and
has order $|S_3|=6$. Equivalently, it is the dihedral group $D_3$ of symmetries of an equilateral triangle. We use the presentation

\begin{figure}
\includegraphics[width=0.5\linewidth]{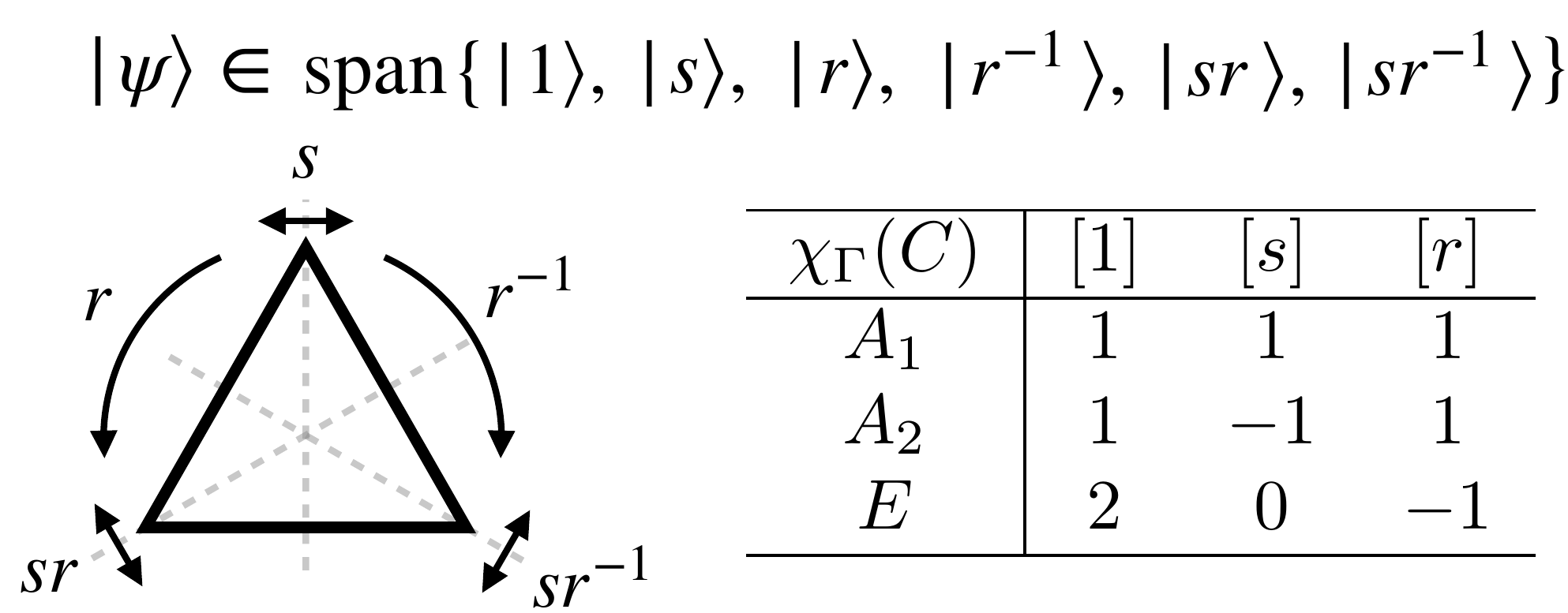}
\caption{\textbf{$S_3$-qudit and character table.} Elements of the group $S_3$ are the symmetry operations of an equilateral triangle (bottom left) and the properties of the group are summarized in its character table (bottom right). A general $S_3$-qudit $\ket{\psi}$ is given by a linear combination of all group elements (top).} \label{fig:s3_triangle}
\end{figure}
\begin{align}
S_3=\langle r,s \mid r^3=e,\ s^2=e,\ srs=r^{-1}\rangle ,
\end{align}
where $r$ is a rotation by $2\pi/3$ and $s$ is a reflection. Its
elements are $ S_3=\{e,r,r^2,s,sr,sr^2\}$, which label the basis of the local Hilbert space.
The conjugacy classes are given by
\begin{align}
[1]&=\{e\},&
[s]&=\{s,sr,sr^2\},&
[r]&=\{r,r^2\}.
\end{align}
Thus $S_3$ has three conjugacy classes and hence three irreducible
representations $\Gamma$, conventionally denoted by $A_1$, $A_2$, and $E$. Their dimensions $d_\Gamma\equiv \text{dim}(\Gamma)$ are
$d_{A_1}=1,\  d_{A_2}=1, \ d_E=2$,
satisfying
$\sum_{\Gamma} d_\Gamma^2
=
1^2+1^2+2^2
=
|S_3|$.
Here $A_1$ is the trivial representation, $A_2$ is the sign
representation, and $E$ is the two-dimensional representation. The characters of $S_3$ are calculated from the definition $\chi_\Gamma(g)=\tr [\Gamma(g)]$ for $g\in S_3$ and $\Gamma \in \text{Rep}(S_3)$. Since each character is a function of class, we may instead write $\chi_\Gamma([g])$, $[g]\in \operatorname{Class}(S_3)$. The resulting character table is presented in Fig.~\ref{fig:s3_triangle}. 

We also summarize the centralizer subgroups $Z_h$ of $S_3$. Recall that they are defined as the group elements that commute with a given $h\in G$; therefore,
\begin{align}
Z_{1}=S_3,\ Z_{s}=\{1,s\}\simeq\mathbb{Z}_2,\ Z_{r}=\{1,r,r^2\}\simeq\mathbb{Z}_3. \label{eq:S3_central}
\end{align}
All other group elements have a centralizer subgroup that is isomorphic to one from Eq.~\eqref{eq:S3_central}, which is why we may denote the relevant centralizer by $Z_g$ for a representative $g\in[g]$. 

In the following we explore deformations of the quantum double fixed point Hamiltonian which lead to confinement of the non-Abelian anyons. There are many possibilities and we choose to guide this pursuit by dualities to lattice gauge theory (LGT).
\subsection{Example: $G=\mathbb{Z}_2$, Ising gauge theory}
In this section, we recall the duality between gauge theory and Ising model for the $\mathbb{Z}_2$ toric code. In this theory $G=\mathbb{Z}_2$ and the degrees of freedom are Ising variables, the vector potential is $\sigma^z(\textbf{x},\uvec{e}_j)\equiv \sigma^z_j(\textbf{x})$ where $j=1,2$, the electric field is then $\sigma^x_j(\textbf{x})$ the Hamiltonian is then
\begin{align}
    H_{IGT} = -\lambda\sum_{\textbf{x},j} \sigma^x_{j}(\textbf{x})-\kappa \sum_{\textbf{x}} \sigma^z_1(\textbf{x})\sigma^z_2(\textbf{x}+\uvec{e}_1)\sigma^z_1(\textbf{x}+\uvec{e}_2)\sigma^z_2(\textbf{x}) \label{QIGT}.
\end{align}
This is the Toric Code Hamiltonian without the "electric matter" term $(1-A(s))$ if $\lambda=0$ and $\kappa=1$. It is also a "pure gauge theory" as can be seen by using the path integral approach to write $Z_{IGT}= \Tr e^{-\beta H_{IGT}}$ in terms of transfer matrices and noting it maps to a $(2+1)D$ classical lattice gauge theory with just the plaquette term now in 3D
\begin{align}
    \mathcal{H}_{\text{IGT}}= -K \sum_{x,\ \mu<\nu} \sigma^z_{\mu}(x)\sigma^z_\nu(x+\uvec{e}_\mu)\sigma^z_\mu(x+\uvec{e}_\nu)\sigma^z_\nu(x),\label{CIGT}
\end{align}
where $\mu,\nu=0,1,2$ and $x$ now has three components. There are some subtleties in the mapping. First, the quantum temperature $\beta$ corresponds to the extent in the imaginary time direction of the classical model $\beta\propto L_\tau$. This means that the infinite classical system corresponds to the ground state of the quantum model. Additionally, $\beta_{cl}K=a_\tau \kappa = \beta \kappa/N_\tau$ where $N_\tau$ is the discretized Trotter step and $1/\beta_{cl}$ is a classical temperature which also goes to zero if $\beta\rightarrow \infty $. Furthermore, the correspondence also requires a ferromagnetic  Ising exchange between the layers in the time direction with some coupling $K_\tau$. The correspondence also singles out particular values for the classical coupling, which are the only ones that map to the quantum theory. The ferromagnetic exchange can be exchanged for a plaquette coupling in time only for the model with $A(s)\equiv 1$.

Let us then define these sectors in terms of the generator of gauge transformations $A_s$, which we rewrite as
\begin{align}
    Q(\textbf{x})\ket{\psi} = \sigma_1^x(\textbf{x}) \sigma_1^x(\textbf{x}-\uvec{e}_1)\sigma_2^x(\textbf{x})\sigma_2^x(\textbf{x}-\uvec{e}_2)\ket{\psi}=\pm\ket{\psi}.
\end{align}
The gauge-invariant states with $Q(\textbf{x})\equiv +1$ that satisfy this condition are precisely the ones that belong in the ground-state subspace. It is easy to verify that $[Q(\textbf{x}),H_{\text{IGT}}]=0$ for all $\textbf{x}$, since $[A_s,B_p]=0$ and $\sigma^x$ operators commute. Furthermore, $[Q(\textbf{x}),Q(\textbf{y})]=0$ just as for $A_s$. Hence, the eigenstates of the Hamiltonian are also eigenstates of the generators $\{Q(\textbf{x})\}$. The plus sign corresponds to no charges in the ground state, while the minus sign leads to an odd gauge theory, where there are an odd number of edges with a minus sign. This is relevant to frustrated lattices since not all bonds can be minimized. We specialize to the even sector of the gauge theory $A_s\equiv 1$. The previous model displays a phase transition as the couplings are varied. One way to see and characterize the transition is to map the model to a transverse-field Ising model. We can do that by going to the dual lattice, which sits at the faces of the square plaquettes. We define new spin variables $S_p^z$ such that, for adjacent plaquettes $p,p'$ with bond $ij$ between them,
\begin{align}
    S_p^zS_q^z = \sigma^x_{ij}.
\end{align}
We then note that the second term of eq.~\eqref{QIGT} is implemented automatically if one plaquette variable is flipped. This means if we apply an operator $S_p^x$ to plaquette $p$ then this amounts to flipping all the $\sigma^x_{ij}$ along the boundary. Now since $\sigma^z_{ij}$ also flips the $\sigma^x_{ij}$ eigenstates then we conclude the relation
\begin{align}
    S_p^x= B_p.
\end{align}
We find then a mapping to the transverse field Ising model in (2+1)D, note that we must also implement $A_s\equiv1$ for this mapping to work
\begin{align}
    H_{IGT}= -\lambda\sum_{\left<p,p'\right>} S_p^zS_{p'}^z-\kappa \sum_p S_p^x.
\end{align}

We will show that the first quantum-to-classical mapping for arbitrary $G$ goes through and generalizes the Ising gauge theory. The second mapping to the Ising model does not generalize clearly because the isometry we describe always involves vertex and link degrees of freedom.
\section{Pure $G$ gauge theory}
\label{sec:SM-pure-gauge}

\subsection{Hamiltonian and gauge-invariant sector}

The pure-gauge limit of Eq.~\eqref{eq:main-electric-magnetic-fields} has
$\lambda_\Gamma=0$ and
\begin{align}
 H_G=H_X+H_B,\qquad
 H_B=\kappa\sum_p[1-B(p)],\qquad
 H_X=\sum_\ell H_X(\ell),
 \label{eq:SM-pure-gauge-general}
\end{align}
where the general class-function field $H_X(\ell)$ is given explicitly in
Eq.~\eqref{eq:class-electric-field}.  To verify its gauge invariance
explicitly, define the class-sum operator
\begin{align}
 T^{[g]}(\ell)=\sum_{h\in[g]}X_+^h(\ell),\qquad
 H_X(\ell)=c_X\mathbb I-\sum_{[g]}\lambda_{[g]}T^{[g]}(\ell).
 \label{eq:SM-class-sum-operator}
\end{align}
Let $\ell=(s\to s')$.  With the orientation convention of
Eq.~\eqref{eq:SM-star-definition}, then
\begin{align}
    \sum_{g} \lambda_{[g]}[A(s),X^g_+(j_1)] &= \dfrac{1}{\abs{G}}  \sum_h \sum_{g} \lambda_{[g]} (X_-^{hg}-X_-^{gh})X_-^hX_+^hX_+^h \notag\\
    &= \dfrac{1}{\abs{G}}  \sum_h \sum_g \lambda_{[g]} X_-^{hg}X_-^hX_+^hX_+^h -  \sum_g \lambda_{[g]} X_-^{gh}X_-^hX_+^hX_+^h  \notag\\
    &= \dfrac{1}{\abs{G}}  \sum_h \sum_{\tilde{g}=hgh^{-1}} \lambda_{[hgh^{-1}]}X_-^{\tilde{g}h}X_-^hX_+^hX_+^- -\sum_{g} \lambda_{[g]}X_-^{\tilde{g}h}X_-^hX_+^hX_+^h=0.
\end{align}
Interestingly, we see that only class-function coefficients ensure $[H_X,A(s)]=0$ under $\tilde{g}=hgh^{-1}$. Hermiticity is a separate condition,
$\lambda_{[\bar g]}=\lambda_{[g]}^*$, derived in
Eq.~\eqref{eq:class-hermiticity}; in particular, self-inverse conjugacy
classes are required only when a single real class is included by itself.

The mutually commuting star projectors define the gauge-invariant subspace,
\begin{align}
 P_{\mathrm{phys}}=\prod_s A(s),\qquad
 A(s)\ket{\psi_{\mathrm{phys}}}=\ket{\psi_{\mathrm{phys}}}\quad\forall s,
 \qquad
 Z_G^Q=\Tr\!\left[P_{\mathrm{phys}}e^{-\beta H_G}\right].
 \label{eq:SM-pure-gauge-projector}
\end{align}
On the maximally symmetric line used for the pure-$S_3$ simulations,
$\lambda_{[g]}=\lambda$ for every nonidentity class.  After absorbing the
identity contribution into an additive constant, Eq.~\eqref{eq:SM-pure-gauge-general}
reduces to the main-text Hamiltonian
\begin{align}
 H_G=H_X+H_B={}&\lambda\sum_\ell\left(1-\sum_{g\in G}X_+^g(\ell)\right)
 +\kappa\sum_p[1-B(p)],
 \qquad A(s)\ket{\psi_{\mathrm{phys}}}=\ket{\psi_{\mathrm{phys}}},
 \label{QGGT}
\end{align}
which becomes the Ising gauge theory for $G=\mathbb Z_2$, up to an energy
shift.  Equivalently, the physical projector may be enforced energetically
using the same star coupling $J$ as in Eq.~\eqref{QD}:
\begin{align}
 H_G(J)=H_G+J\sum_s[1-A(s)],\qquad
 \lim_{J\to\infty}\Tr e^{-\beta H_G(J)}
 =\Tr\!\left[P_{\mathrm{phys}}e^{-\beta H_G}\right].
 \label{QGGT_soft}
\end{align}

\subsection{Pure $G$-gauge theory via (2+1)D path integral}

As a warm up we treat first the easier case of $\lambda_{[g]}=\lambda>0$ for every nonidentity conjugacy class, so that all nonidentity group elements have the same hopping amplitude. We will discuss the more general case in a later section. We map the projected partition function in
Eq.~\eqref{eq:SM-pure-gauge-projector} to a classical model in one extra
dimension.  Since $P_{\mathrm{phys}}$ commutes with $H_X+H_B$, inserting
group-basis resolutions of the identity gives
\begin{align}
Z^{Q}_{G}
&=\Tr\!\left[P_{\mathrm{phys}}e^{-\beta(H_X+H_B)}\right]\notag\\
&=\sum_{\{g_i^0\},\{g_i^{N_\tau}\}}
\bra{\{g_i^0\}}e^{-\Delta\tau(H_X+H_B)}\cdots
e^{-\Delta\tau(H_X+H_B)}\ket{\{g_i^{N_\tau}\}}
\bra{\{g_i^{N_\tau}\}}P_{\mathrm{phys}}\ket{\{g_i^0\}}\notag\\
&=\sum_{\{g_i^0\},\{g_i^{N_\tau}\}}
Z^Q_{GGT}(\{g_i^0\},\{g_i^{N_\tau}\})
Z_{\mathrm{gauge}}(\{g_i^{N_\tau}\},\{g_i^0\}).
\end{align}
Here $\{g_i^n\}$ denotes all spatial-link variables on slice $n$,
$N_\tau\Delta\tau=\beta$, and the last factor imposes the Gauss law.  Its
group-basis matrix element is
\begin{align}
    Z_{\mathrm{gauge}}(\{g_i^{N_\tau}\},\{g_i^0\})
    &=\bra{\{g_i^{N_\tau}\}}P_{\mathrm{phys}}\ket{\{g_i^0\}}
    =\bra{\{g_i^{N_\tau}\}}\prod_s A(s)\ket{\{g_i^0\}}\notag\\
    &= \frac{1}{\abs{G}^{N_v}}\sum_{\{\Omega_s\in G\}}
    \prod_{\ell=(s\to s')}
    \delta_{g_\ell^{N_\tau},\,\Omega_{s'}g_\ell^0\Omega_s^{-1}}.
\end{align}
Here $N_v$ is the number of spatial vertices. Each $\Omega_s$ is shared by all links incident on $s$. This kernel closes the
imaginary-time evolution with a gauge twist connecting the two time slices at $\tau=0$ and $\tau=N_\tau$. 
Let us insert identities of the whole lattice of edges at each intermediate slice $n=1,\ldots,N_\tau-1$, keeping both endpoints fixed:

\begin{align}
    Z^Q_{GGT}(\{g_i^0\},\{g_i^{N_\tau}\}) &= \sum_{\{g_i^1\},\ldots,\{g_i^{N_\tau-1}\}}\prod_{n=0}^{N_\tau-1}\bra{\{g_i^{n}\}}e^{-\Delta \tau (H_X+H_B)}\ket{\{g_{i}^{n+1}\}}\notag\\
    &\approx  \sum_{\{g_i^1\},\ldots,\{g_i^{N_\tau-1}\}}\prod_{n=0}^{N_\tau-1}\bra{\{g_i^n\}}e^{-\Delta \tau H_B}e^{-\Delta \tau H_X}\ket{\{g_i^{n+1}\}} \\
    &=  \sum_{\{g_i^1\},\ldots,\{g_i^{N_\tau-1}\}}\prod_{n=0}^{N_\tau-1}e^{-\Delta \tau H_B(\{g^n_i\})}\bra{\{g^n_i\}}e^{-\Delta \tau H_X}\ket{\{g_j^{n+1}\}}.
\end{align}
All boundary products below are ordered from right to left and include inverses for links traversed against their orientation, in the same convention as $A(s)$. By acting on the configuration of group elements, we find 
\begin{align}
    e^{-\Delta \tau H_B(\{g_i^n\})} = \prod_p \exp \{{-\Delta \tau \ \kappa (1- \delta_{\prod_{l\in \partial p}g^n(l),1})}\}.
\end{align}
Thus, if $\kappa \rightarrow \infty $ at fixed $\Delta\tau>0$, then $e^{-\Delta \tau H_B(\{g_i^n\})} \rightarrow \prod_p\delta_{\prod_{l\in \partial p}g^n(l),1}$, enforcing plaquette flatness (zero magnetic flux) for every plaquette. The kinetic term instead acts like
\begin{align}
    \bra{\{g_i^n\}}H_X\ket{\{g^{n+1}_j\}} = \bra{\{g_i^n\}}\sum_{j} \lambda(1- \sum_{h}X^h_+(j))\ket{\{g^{n+1}_j\}}\notag\\
    = \sum_{j}\lambda \bra{\{g_i^n\}} (1- \sum_{h}X^h_+(j))\ket{\{g^{n+1}_j\}}.
\end{align}
    The action of the X operator on an arbitrary basis state is 
\begin{align}
    \mel{g}{\sum_{h\in G}\lambda X_+^h}{f} = \lambda\sum_{h\in G}\delta_{g,hf}=\lambda.
\end{align}
This implies the result
\begin{align}
    \sum_{j}\lambda \bra{\{g_i^n\}} (1- \sum_{h}X^h_+(j))\ket{\{g_i^{n+1}\}} =&  \sum_{j} \lambda \delta_{\{g_i^n\},\{g_i^{n+1}\}}-\sum_{j}\left(\sum_{h} \lambda \delta_{g_j^n\bar{g}_j^{n+1},h}\right) \prod_{l \neq j}\delta_{g_l^n,g_l^{n+1}}\notag\\
    =&-\sum_{j}\left(\sum_{h\neq 1} \lambda \delta_{g_j^n\bar{g}_j^{n+1},h}\right)\prod_{l \neq j}\delta_{g_l^n,g_l^{n+1}}.
\end{align}
We have
\begin{align}
     \sum_{j}\lambda \bra{\{g_i^n\}} (1- \sum_{g}X^g_+(j))\ket{\{g_i^{n+1}\}}  = -\lambda \sum_{j}(1-\delta_{g_j^n,g_j^{n+1}})\prod_{l \neq j}\delta_{g_l^n,g_l^{n+1}} = - \lambda \delta^{(1)}_{\{g_i^n\},\{g_i^{n+1}\}},
\end{align}
where the last delta equals one when the two configurations differ on exactly one edge and equals zero otherwise. We have the lowest order result
\begin{align}
     Z^Q_{GGT}(\{g_i^0\},\{g_i^{N_\tau}\}) &\simeq\sum_{\{g_i^1\},\ldots,\{g_i^{N_\tau-1}\}}\prod_{n=0}^{N_\tau-1}\exp \{\sum_p {-\Delta \tau \ \kappa (1- \delta_{\prod_{l\in \partial p}g^n(l),1})}\}\left(\delta_{\{g_i^n\},\{g_i^{n+1}\}}+ \Delta \tau \lambda  \delta^{(1)}_{\{g_i^n\},\{g_i^{n+1}\}} +O(\Delta \tau^2)\right)\notag\\
     &\simeq\sum_{\{g_i^1\},\ldots,\{g_i^{N_\tau-1}\}}\prod_{n=0}^{N_\tau-1} \exp\left\{ -\beta_{cl}\left(J_V\sum_p {(1- \delta_{\prod_{l\in \partial p}g^n(l),1})}-J_\tau \sum_l (\delta_{g_l^n,g_l^{n+1}}-1)\right)\right\},
\end{align}
We retain the transfer matrix only through first order in $\Delta\tau$. The following classical discretization reproduces it up to $O(\Delta\tau^2)$ per time step at fixed couplings. We define
\begin{align}
    K_\tau\equiv\beta_{cl}J_\tau\equiv-\ln(\lambda\Delta\tau),
    \qquad K_s\equiv\beta_{cl}J_V\equiv\kappa\Delta\tau.
\end{align}
Continuous time evolution is given by $\Delta \tau\rightarrow 0$ which means $\beta_{cl}J_\tau\rightarrow \infty,\  \beta_{cl}J_V\rightarrow 0 $ while keeping constant 
\begin{align}
   \lambda = e^{-\beta_{cl}J_\tau }/ \Delta \tau , \quad \dfrac{\kappa}{\lambda} =  \beta_{cl}J_V e^{\beta_{cl}J_\tau }.
\end{align}
The form of the partition function is that of a classical system in one more dimension with a Hamiltonian
\begin{align}
    \mathcal{H}^{cl}_{GGT} = J_V\sum_{p,n} {(1- \delta_{\prod_{l\in \partial p}g^n(l),1})}-J_\tau \sum_{l,n} (\delta_{g_l^n,g_l^{n+1}}-1),
\end{align}
the extent of the extra dimension is $N_\tau = \beta/\Delta \tau \rightarrow \infty $. We can now map this to a pure lattice gauge theory in 2+1 dimensions with anisotropic couplings. To do so, we define
\begin{align}
\mathcal{H}^{2+1}_{GGT} = J_V\sum_{p} (1-\delta_{\prod_{l\in \partial p}g(l),1})+J_\tau \sum_{p_\tau} (1-\delta_{\prod_{l\in \partial p_\tau}g(l),1}),
\end{align}
with temporal-link variables $h_s^n$ oriented from slice $n$ to slice $n+1$.
For $\ell=(s\to s')$, the temporal plaquette holonomy is
\begin{align}
U_{\ell\tau}^n=(g_\ell^n)^{-1}(h_{s'}^n)^{-1}g_\ell^{n+1}h_s^n.
\end{align}
When $h_s^n=1$, its identity constraint reduces to
$\delta_{g_\ell^n,g_\ell^{n+1}}$.
This Hamiltonian is gauge invariant under the action of multiplication by an
arbitrary group element $h$ on the legs of every spacetime vertex, in the same
orientation convention as $A(s)$.  We may use this redundancy to impose
\textit{temporal gauge}.  With periodic imaginary time, however, all temporal
links can be fixed to the identity only when the ordered holonomy
$\Omega_s=h_s^{N_\tau-1}\cdots h_s^0$ is trivial.  In general one temporal
boundary link remains and carries $\Omega_s$.  We must therefore sum over
all temporal-holonomy sectors rather than retaining only
$\Omega_s=1$.
Accordingly, the final spacetime sum includes every temporal link, including the unfixed temporal boundary links.

Since $P_{\mathrm{phys}}^2=P_{\mathrm{phys}}$ and it commutes separately with $H_X$ and $H_B$, we may insert it at every time step. Its vertex variables become the temporal links. For ordinary, unnormalized sums over group variables on all links of spacetime, we obtain
\begin{align}
    Z^Q_G \simeq \abs{G}^{-N_vN_\tau}\sum_{\{g(l)\}} \exp\left\{-K_s\sum_{p} (1-\delta_{\prod_{l\in \partial p}g(l),1})-K_\tau \sum_{p_\tau} (1-\delta_{\prod_{l\in \partial p_\tau}g(l),1})\right\}, \quad 
    \beta_{cl}J_\tau = -\ln(\dfrac{\beta \lambda }{N_\tau}), \quad     \beta_{cl}J_V=\dfrac{\beta \kappa }{N_\tau}.
\end{align}
Within this lowest-order discretization, imposing isotropy  gives $\lambda\Delta\tau=e^{-\kappa\Delta\tau}$.  In the following, we study the isotropic classical model as a separate specialization of the classical action.
The partition function then becomes (we use now the LGT notation, $K=\beta_{cl}J_V=\beta_{cl}J_\tau$), denote the set of all plaquettes $\Lambda^2$ and links $\Lambda^1$ of the lattice, with each plaquette counted once using $\mu<\nu$, then 
\begin{align}
    Z^{2+1D}_{G}= \sum_{\{g_\mu(\textbf{x})\}} e^{-S[\dd g]},\qquad S[\dd g]= K \sum_{(\textbf{x},\mu,\nu)\in \Lambda^2}(1-\delta_{\dd g_{\mu\nu}(\textbf{x}),1}), \quad \dd g_{\mu\nu}(\textbf{x}) = \bar{g}_\mu(\textbf{x})\bar{g}_\nu(\textbf{x}+\uvec{e}_\mu)g_\mu(\textbf{x}+\uvec{e}_\nu)g_\nu(\textbf{x}) \label{CGGT}.
\end{align}
Let us use the decomposition into characters to write  $f(g)=\delta_{g,1}$ as
\begin{align}
    f(g)= \sum_\Gamma \chi_\Gamma(g) \tilde{f}_\Gamma , \qquad \tilde{f}_\Gamma =\dfrac{1}{\abs{G}}\sum_g \bar{\chi}_\Gamma(g) f(g).
\end{align}
Then since $\chi_\Gamma(1)=d_\Gamma$ we obtain $\delta_{g,1} = \tfrac{1}{\abs{G}}\sum_\Gamma d_\Gamma\chi_\Gamma(g)$. Using this to rewrite the action and using properties of representations, it becomes a generalization of Wilson's action which sums over all irreps of the group:
\begin{align}
         S[\dd g]&= K \sum_{(\textbf{x},\mu,\nu)\in \Lambda^2}(1-\tfrac{1}{\abs{G}} \sum_\Gamma d_\Gamma \tr{\Gamma(\dd g_{\mu\nu}(\textbf{x}))}) \notag\\
     &=  K \sum_{(\textbf{x},\mu,\nu)\in \Lambda^2}(1-\tfrac{1}{\abs{G}} \sum_\Gamma d_\Gamma\tr{\Gamma(g_\mu(\textbf{x}))^{-1}\Gamma(g_\nu(\textbf{x}+\uvec{e}_\mu))^{-1}\Gamma(g_\mu(\textbf{x}+\uvec{e}_\nu))\Gamma(g_\nu(\textbf{x}))}).
\end{align}

\subsection{Class-function $\lambda_{[g]}$ and exact temporal weights}

The preceding mapping used the same hopping amplitude for all nonidentity
group elements. We now allow for an arbitrary class
function $\lambda_{[g]}$. Denoting the conjugacy classes of $G$ by $[g]\in\operatorname{Class}(G)$,
we write the electric part of the Hamiltonian as a sum of one-link terms,
\begin{align}
 H_X=\sum_\ell H_X(\ell), \quad
 H_X(\ell)=c_X\mathbb{I}
 -\sum_{[g]\in\operatorname{Class}(G)}\lambda_{[g]}
 \sum_{h\in[g]}X_+^h(\ell),
 \label{eq:class-electric-field}
\end{align}
where $\lambda_{[g]}$ is the amplitude per element of $[g]$ and $c_X\in\mathbb{R}$.
The constant $c_X$ is
the coefficient of the identity operator on each link, and therefore only
fixes the zero of energy. For example, if the field is written as
\begin{align}
 H_X(\ell)=\sum_{[g]\in\mathcal S}\lambda_{[g]}
 \left(\mathbb{I}-\sum_{h\in[g]}X_+^h(\ell)\right),
\end{align}
then $c_X=\sum_{[g]\in\mathcal S}\lambda_{[g]}$. If instead the Hamiltonian
is written with a separate identity shift for every group-element hop,
\begin{align}
 H_X(\ell)=\sum_{[g]}\lambda_{[g]}\sum_{h\in[g]}
 \left(\mathbb{I}-X_+^h(\ell)\right),
\end{align}
then $c_X=\sum_{[g]}\abs{[g]}\lambda_{[g]}$. If $H_X(\ell)$ contains only the X terms, then $c_X=0$. A coefficient $\lambda_{[1]}$ multiplying
$X_+^1=\mathbb{I}$ can likewise be absorbed through
$c_X\rightarrow c_X-\lambda_{[1]}$. Thus $c_X$ is not a
temporal coupling and will cancel from all normalized temporal weights below.

Gauge invariance and Hermiticity impose distinct conditions. Since
$(X_+^h)^\dagger=X_+^{\bar h}$, one finds
\begin{align}
 H_X(\ell)^\dagger=H_X(\ell)
 \quad\Longleftrightarrow\quad
\lambda_{[\bar g]}=\lambda_{[g]}^*,
 \qquad [\bar g]=\{\bar h:h\in[g]\}.
 \label{eq:class-hermiticity}
\end{align}
Thus $[g]=[\bar g]$ is sufficient, but not necessary. If $[g]\neq[\bar g]$, a
Hermitian perturbation contains both class sums with conjugate coefficients.
For real couplings this reduces to $\lambda_{[\bar g]}=\lambda_{[g]}$. Only when one
insists on including a single class independently must that class be
self-inverse.

The electric transfer matrix can be evaluated exactly within each Trotter
step without introducing irreducible representations. For two group-basis
states $\ket{u}$ and $\ket{v}$, let $r=u\bar v$. Since the electric terms on
distinct links commute,
\begin{align}
 w_{\Delta\tau}(r)
 &\equiv\mel{u}{e^{-\Delta\tau H_X(\ell)}}{v}=e^{-\Delta\tau c_X}\sum_{n=0}^{\infty}
 \frac{(\Delta\tau)^n}{n!}
 \sum_{h_1,\ldots,h_n\in G}
 \left(\prod_{j=1}^n\lambda_{[h_j]}\right)
 \delta_{r,h_1h_2\cdots h_n}.
 \label{eq:class-transfer-weight}
\end{align}
The $n=0$ term is understood as $\delta_{r,1}$. Simultaneously conjugating
$r$ and every $h_j$ in Eq.~\eqref{eq:class-transfer-weight} proves that
$w_{\Delta\tau}(kr\bar k)=w_{\Delta\tau}(r)$. Hence the transfer weight itself
depends only on the conjugacy class $[r]$.

Equivalently, this result may be written entirely in terms of class
multiplication data. For a representative $g_0\in[g]$, define
\begin{align}
 N^{[g]}_{[g_1]\cdots[g_n]}
 =\#\{(h_1,\ldots,h_n):h_j\in[g_j],\ h_1\cdots h_n=g_0\}.
\end{align}
This number is independent of the chosen representative. Equation
\eqref{eq:class-transfer-weight} then becomes
\begin{align}
 w_{\Delta\tau}([g])=e^{-\Delta\tau c_X}
 \sum_{n=0}^{\infty}\frac{(\Delta\tau)^n}{n!}
 \sum_{[g_1],\ldots,[g_n]}
 N^{[g]}_{[g_1]\cdots[g_n]}\prod_{j=1}^n\lambda_{[g_j]}.
 \label{eq:class-transfer-weight-classes}
\end{align}
This expression resums all sequences of electric hops that produce a given
temporal flux class.

For comparison, the same convolution can be diagonalized in irreducible
representations. Let $\Gamma$ denote a unitary irrep of dimension $d_\Gamma$
and character $\chi_\Gamma$. The central projectors in the left regular
representation are
\begin{align}
 P_\Gamma=\frac{d_\Gamma}{\abs{G}}
 \sum_{h\in G}\chi_\Gamma(\bar h)X_+^h,
 \qquad
 P_\Gamma P_{\Gamma'}=\delta_{\Gamma,\Gamma'}P_\Gamma,
 \qquad
 \sum_\Gamma P_\Gamma=\mathbb{I}.
 \label{eq:class-central-projectors}
\end{align}
The one-link Hamiltonian is a scalar on each irrep sector,
\begin{align}
 H_X(\ell)P_\Gamma=\varepsilon_\Gamma P_\Gamma,
 \qquad
\varepsilon_\Gamma=c_X-\frac{1}{d_\Gamma}
 \sum_{[g]\in\operatorname{Class}(G)}\abs{[g]}\lambda_{[g]}\chi_\Gamma([g]).
 \label{eq:class-irrep-eigenvalue}
\end{align}
Consequently,
\begin{align}
 e^{-\Delta\tau H_X(\ell)}
 &=\sum_\Gamma e^{-\Delta\tau\varepsilon_\Gamma}P_\Gamma, \quad
 w_{\Delta\tau}([g])
 =\frac{1}{\abs{G}}
 \sum_\Gamma d_\Gamma e^{-\Delta\tau\varepsilon_\Gamma}
 \chi_\Gamma([\bar g]).
 \label{eq:class-transfer-weight-irreps}
\end{align}
The inverse class in the last line follows from our convention
$r=u\bar v$, for which
$\mel{u}{P_\Gamma}{v}=d_\Gamma\chi_\Gamma(r^{-1})/\abs{G}$.
Condition~\eqref{eq:class-hermiticity} guarantees that every
$\varepsilon_\Gamma$ is
real, although the group-basis weights can still be complex when distinct
inverse classes carry complex-conjugate amplitudes.

We now derive explicitly how this one-link kernel produces the classical
action. Let
$\ket{\bm{g}^n}=\bigotimes_\ell\ket{g_\ell^n}$ denote the spatial-link
configuration on time slice $n$, and define
\begin{align}
 H_B(\bm{g}^n)=\kappa\sum_{p\parallel xy}
 \left(1-\delta_{U_p(\bm{g}^n),1}\right).
\end{align}
Inserting a complete group basis after a first-order Trotter decomposition
gives
\begin{align}
 Z&\simeq
 \prod_{n=0}^{N_\tau-1}\sum_{\{g_\ell^n\}}
 e^{-\Delta\tau H_B(\bm{g}^n)}
 \mel{\bm{g}^n}{e^{-\Delta\tau H_X}}
 {\bm{g}^{n+1}},
 \qquad N_\tau=\frac{\beta}{\Delta\tau}.
 \label{eq:class-trotter-partition}
\end{align}
Here $\bm{g}^{N_\tau}=\bm{g}^0$ implements the trace.
The equality becomes exact as $\Delta\tau\rightarrow0$; at finite step the
first-order decomposition has a total Trotter error of order
$\beta\Delta\tau^2$. Since the terms $H_X(\ell)$ act on distinct links and
commute, the many-link matrix element factorizes exactly,
\begin{align}
 \mel{\bm{g}^n}{e^{-\Delta\tau H_X}}{\bm{g}^{n+1}}
 =\prod_\ell w_{\Delta\tau}(r_\ell^n),
 \qquad r_\ell^n=g_\ell^n\bar g_\ell^{n+1}.
 \label{eq:class-many-link-factorization}
\end{align}
To make contact with the single-link-change expression used above, assume
that the identity-class contribution has been absorbed into $c_X$. Expanding
Eq.~\eqref{eq:class-many-link-factorization} to first order yields
\begin{align}
 &\mel{\bm{g}^n}{e^{-\Delta\tau H_X}}{\bm{g}^{n+1}}
 =e^{-\Delta\tau c_XN_\ell}
 \left[
 \delta_{\bm{g}^n,\bm{g}^{n+1}}
 +\Delta\tau\,\delta_\lambda^{(1)}
 (\bm{g}^n,\bm{g}^{n+1})+O(\Delta\tau^2)
 \right],\notag\\
 &\delta_\lambda^{(1)}(\bm{g}^n,\bm{g}^{n+1})
 =\sum_\ell\sum_{[g]\neq[1]}\lambda_{[g]}
 \delta_{[r_\ell^n],[g]}
 \prod_{\ell'\neq\ell}
 \delta_{g_{\ell'}^n,g_{\ell'}^{n+1}},
 \qquad
 \delta_{\bm{g}^n,\bm{g}^{n+1}}
 =\prod_\ell\delta_{g_\ell^n,g_\ell^{n+1}}.
 \label{eq:class-weighted-single-link-delta}
\end{align}
The weighted delta is nonzero only when the two configurations differ on one
link, and it assigns that change the amplitude $\lambda_{[g]}$ determined by the
conjugacy class of the relative group element. If all nonidentity classes
have a common amplitude $\lambda$, then
$\delta_\lambda^{(1)}=\lambda\delta^{(1)}$, reproducing the earlier
single-link result.

For finite $\Delta\tau$ it is preferable to retain the exact link weight. We
normalize it by the identity-class weight and define
\begin{align}
 K_\tau([1])=0,\qquad
 e^{-K_\tau([g])}=\frac{w_{\Delta\tau}([g])}
 {w_{\Delta\tau}([1])}.
 \label{eq:class-temporal-coupling}
\end{align}
Using Eq.~\eqref{eq:class-transfer-weight-irreps}, the exact temporal coupling
has the equivalent representation-space form
\begin{align}
 K_\tau([g])=-\ln\left[
 \frac{\displaystyle\sum_\Gamma d_\Gamma
 e^{-\Delta\tau\varepsilon_\Gamma}\chi_\Gamma([\bar g])}
 {\displaystyle\sum_\Gamma d_\Gamma^2
 e^{-\Delta\tau\varepsilon_\Gamma}}
 \right],
 \qquad K_\tau([1])=0.
 \label{eq:class-temporal-coupling-irreps}
\end{align}
The common contribution $e^{-\Delta\tau c_X}$ contained in every
$e^{-\Delta\tau\varepsilon_\Gamma}$ cancels between numerator and denominator.
Equation~\eqref{eq:class-temporal-coupling-irreps} is exactly
equivalent to the group- and class-based expressions
in Eqs.~\eqref{eq:class-transfer-weight} and
\eqref{eq:class-transfer-weight-classes}; it is not an additional
approximation. On every pair of adjacent time slices,
\begin{align}
 \prod_\ell w_{\Delta\tau}([r_\ell^n])
 =w_{\Delta\tau}([1])^{N_\ell}
 \exp\left[-\sum_\ell K_\tau([r_\ell^n])\right].
 \label{eq:class-product-to-action}
\end{align}
Substitution in Eq.~\eqref{eq:class-trotter-partition} therefore gives, in
temporal gauge and up to the configuration-independent normalization
$\mathcal N=w_{\Delta\tau}([1])^{N_\ell N_\tau}$,
\begin{align}
 Z&\simeq\mathcal N\sum_{\{\bm{g}^n\}}
 \exp\left[-K\sum_{n,p\parallel xy}
 (1-\delta_{U_p(\bm{g}^n),1})
 -\sum_{n,\ell}K_\tau([r_\ell^n])\right],
 \qquad K=\Delta\tau\kappa.
 \label{eq:class-temporal-gauge-action}
\end{align}
In particular, the factor $e^{-\Delta\tau c_X}$ contained in every link
weight contributes only the overall energy-shift factor
$e^{-\beta c_XN_\ell}$ to $\mathcal N$ and never enters $K_\tau([g])$.

Finally, restore temporal links $h_s^n\in G$, all oriented from vertex $s$ on
slice $n$ to the same vertex on slice $n+1$.  As in the spatial plane, all
positive coordinate links therefore point to the right, upward, or forward in
time.  For a spatial link $\ell=(s\rightarrow s')$, use the same plaquette
orientation as in Eq.~\eqref{CGGT},
\begin{align}
 U_{\ell\tau}^n
 =\bar g_\ell^n\bar h_{s'}^n g_\ell^{n+1}h_s^n.
 \label{eq:class-temporal-plaquette}
\end{align}
In temporal gauge $h_s^n=1$, this is
$U_{\ell\tau}^n=\bar r_\ell^n$.  More generally, $r_\ell^n$ is conjugate to
$\bar U_{\ell\tau}^n$, so the transfer weight is
$w_{\Delta\tau}([\bar U_{\ell\tau}^n])$. Under a local gauge transformation
$U_{\ell\tau}^n$ changes by conjugation, and hence its conjugacy class is
gauge invariant. The anisotropic classical action can thus be written
entirely in terms of group elements and conjugacy classes as
\begin{align}
 S[g]=K\sum_{p\parallel xy}(1-\delta_{U_p,1})
 +\sum_{p\ni\tau}K_\tau([\bar U_p])
 =K\sum_{p\parallel xy}(1-\delta_{U_p,1})
 +\sum_{p\ni\tau}\sum_{[g]\neq[1]}K_\tau([g])\delta_{[\bar U_p],[g]}.
 \label{eq:class-function-action}
\end{align}
Here $\delta_{[U_p],[g]}$ is one when the plaquette holonomy belongs to $[g]$ and
zero otherwise, making gauge invariance manifest. For $[g]\neq[1]$ with
$\lambda_{[g]}>0$, the continuous-time limit
gives
\begin{align}
 w_{\Delta\tau}([g])=\Delta\tau\,\lambda_{[g]}+O(\Delta\tau^2),
 \qquad
 K_\tau([g])=-\ln(\Delta\tau\,\lambda_{[g]})+O(\Delta\tau).
 \label{eq:class-continuous-limit}
\end{align}
If $\lambda_{[g]}=0$, the first nonzero contribution may occur at higher order
through products of allowed classes; if the support of the electric field
cannot generate $[g]$, then $w_{\Delta\tau}([g])=0$ and $K_\tau([g])=+\infty$.

As a check, suppose that every nonidentity group element has the same
amplitude $\lambda$. The exact normalized temporal weight
then reduces to a single coupling,
\begin{align}
 e^{-K_\tau}
 =\frac{e^{\abs{G}\lambda\Delta\tau}-1}
 {e^{\abs{G}\lambda\Delta\tau}+\abs{G}-1}.
 \label{eq:uniform-exact-temporal-coupling}
\end{align}
Its continuous-time expansion is
$e^{-K_\tau}=\lambda\Delta\tau+O(\Delta\tau^2)$, reproducing the coupling used
in the preceding path-integral derivation.

For the case relevant to our numerical analysis, the nonidentity classes of
$S_3$ are $[s]$ and $[r]$, of sizes three and two, respectively. Taking
\begin{align}
 H_X(\ell)=c_X\mathbb{I}
 -\lambda_{[s]}\sum_{h\in[s]}X_+^h(\ell)
 -\lambda_{[r]}\sum_{h\in[r]}X_+^h(\ell),
\end{align}
Eq.~\eqref{eq:class-transfer-weight} gives the three class weights
\begin{align}
 w_{\Delta\tau}([1])
 &=\frac{e^{-\Delta\tau c_X}}{3}
 \left[e^{2\lambda_{[r]}\Delta\tau}
 \cosh(3\lambda_{[s]}\Delta\tau)+2e^{-\lambda_{[r]}\Delta\tau}\right],\notag\\
 w_{\Delta\tau}([s])
 &=\frac{e^{-\Delta\tau c_X}}{3}
 e^{2\lambda_{[r]}\Delta\tau}\sinh(3\lambda_{[s]}\Delta\tau),\notag\\
 w_{\Delta\tau}([r])
 &=\frac{e^{-\Delta\tau c_X}}{3}
 \left[e^{2\lambda_{[r]}\Delta\tau}
 \cosh(3\lambda_{[s]}\Delta\tau)-e^{-\lambda_{[r]}\Delta\tau}\right].
 \label{eq:S3-exact-class-weights}
\end{align}
The two temporal couplings follow directly as
$K_\tau([s])=-\ln[w_{\Delta\tau}([s])/w_{\Delta\tau}([1])]$ and
$K_\tau([r])=-\ln[w_{\Delta\tau}([r])/w_{\Delta\tau}([1])]$. If
$\lambda_{[s]}=\lambda_{[r]}=\lambda$, the two weights coincide and
Eq.~\eqref{eq:uniform-exact-temporal-coupling} is recovered with
$\abs{G}=6$.

For Hermitian $H_X$ with real, nonnegative coefficients all
$w_{\Delta\tau}([g])$ are nonnegative and
$K_\tau([g])=K_\tau([\bar g])$, so Eq.~\eqref{eq:class-function-action} is an
ordinary real classical action, hence we have shown the perturbed $\mathcal{D}(G)$ model of the main text is sign-free. Finally, the result above is exact for the electric
transfer matrix at fixed $\Delta\tau$. The mapping of the full Hamiltonian at
finite step still has the Trotter error from separating $H_X$ and the
plaquette Hamiltonian, which
vanishes as $\Delta\tau\rightarrow0$.

\subsection{Duality to string-nets and confined excitations}
Let us consider the Hamiltonian of Eq. \eqref{QGGT}. Instead of considering as a basis the magnetic degrees of freedom let us consider the electric picture, which will lead us to string-nets 
\cite{LevinWen2005,PhysRevB.80.155136}. We start by analyzing the confined phase of our theory. This is accomplished by setting $\kappa=0$, and for simplicity choosing $\lambda=\lambda_{[g]}$. The ground state can then be solved exactly, and it is given by a product state of all links in the following way
\begin{align}
 H \xrightarrow[\kappa=0]{} \lambda\sum_{j} (1- \sum_{g}X^g_+(j))    , \qquad  \ket{\Psi_{\kappa=0}} =\bigotimes_{j} \ket{I}_j \equiv \ket{\textbf{I}}.
\end{align}
Where we defined the equal-weight superposition of all group element states as
\begin{align}
    \ket{I}_j = \tfrac{1}{\sqrt{\abs{G}}}\sum_g \ket{g}_j.
\end{align}
The identity notation is intentional: in the electric picture, we use the representation basis, and the identity representation with $\Gamma(g)=1$ and $d_I=1$ has the associated state $\ket{I}$. We rewrite the confined Hamiltonian as
\begin{align}
    H_{\kappa=0}=\lambda\sum_{j} (1- \sum_{g}X^g_+(j)) = \lambda \abs{G}\sum_j (1-\ketbra{I}{I}_j)-\lambda \sum_j(\abs{G}-1).
\end{align}
We see that, up to a constant energy term, this is a sum of local projectors $\mathcal{P}_I(j) = 1-\ketbra{I}{I}_j$. The ground state is the product state from the trivial irrep $\ket{I}$ with energy $\lambda \sum_j(1-\abs{G})$. To consider excited states, we must also consider the gauge constraint $A(s)\ket{\psi_{phy}}= \ket{\psi_{phy}}$. As we shall see, a single edge in the state $\ket{\Gamma_{\alpha\beta}}$, rather than in the trivial irrep, is not allowed. Nevertheless, such an excitation would cost $\lambda \abs{G}$. Each violated link contributes an additional $\lambda \abs{G}$, leading to a gapped spectrum in which a string of length $\ell$ has an energy above the ground state of
\begin{align}
    \Delta E = \lambda\abs{G}\ell.
\end{align}
Let us analyze what the gauge constraint implies for a state with two edges in two different irreps $\Gamma,\Gamma'$, let us put them on opposite sides, then 
\begin{align}
    A(s)\ket{\Gamma'_{\alpha\beta}}\ket{I}\ket{\Gamma_{\lambda\delta}} \ket{I} &=\tfrac{1}{\abs{G}}\sum_gX^g_-X_-^gX_+^gX_+^g\ket{\Gamma'_{\alpha\beta}}\ket{I}\ket{\Gamma_{\lambda\delta}} \ket{I} = \tfrac{1}{\abs{G}}\sum_g \sum_{\beta',\lambda'}\ket{\Gamma'_{\alpha\beta'}}\Gamma'_{\beta'\beta}(g)\ket{I}\Gamma_{\lambda\lambda'}(g^{-1})\ket{\Gamma_{\lambda'\delta}} \ket{I}\notag\\
    &=\tfrac{1}{\abs{G}}\sum_{\beta',\lambda'}\sum_g \Gamma^{-1}_{\lambda\lambda'}(g)\Gamma'_{\beta'\beta}(g) \ket{\Gamma'_{\alpha\beta'}}\ket{I}\ket{\Gamma_{\lambda'\delta}} \ket{I}\notag\\
    &= \sum_{\beta',\lambda'}\tfrac{1}{\abs{G}} \tfrac{\abs{G}}{d_\Gamma} \delta_{\Gamma,\Gamma'} \delta_{\lambda \beta} \delta_{\lambda',\beta'}\ket{\Gamma'_{\alpha\beta'}}\ket{I}\ket{\Gamma_{\lambda'\delta}} \ket{I} = \delta_{\lambda \beta}\tfrac{1}{d_\Gamma}\delta_{\Gamma,\Gamma'}\sum_{\beta'}\ket{\Gamma_{\alpha\beta'}}\ket{I}\ket{\Gamma_{\beta'\delta}} \ket{I}.
\end{align}
From this action we conclude that if we impose $A(s)\equiv 1$ on physical states then an excitation with two irreps must have the same irrep incoming and outgoing, locally for a vertex of the lattice the form of the state is
\begin{align}
\sum_\beta\ket{\Gamma_{\alpha\beta}}\ket{I}\ket{\Gamma_{\beta\delta}} \ket{I}.
\end{align}
Since we impose the gauge constraint for all states this means then that the eigenstate with the smallest energy has four edges occupied by an irrep $\Gamma$ and they form a closed loop 
\begin{align}
    \ket{\Psi_\Gamma(\square)} = (\otimes_{j\notin \square} \ket{I}_j )\otimes  \tr{\ket{\bm\Gamma}\ket{\bm\Gamma}\ket{\bm\Gamma} \ket{\bm \Gamma}},  
    \qquad  \Delta E = 4\lambda\abs{G},
\end{align}
we defined the matrix state $\ket{\bm\Gamma}$ as the matrix of states $\ket{\bm\Gamma}_{\alpha\beta}\equiv\ket{\Gamma_{\alpha\beta}}$ and the trace acts on the irrep vector space. For a cylinder there are $(\abs{\operatorname{Class}(G)}-1) L^2$ distinct states corresponding to $L^2$ total plaquettes and $\abs{\operatorname{Class}(G)}-1$ conjugacy classes which is equal to the number of irreps not being the trivial one. Implied in the notation of the previous state is that to write the state in this simple form the lattice must be oriented such that the plaquette that is occupied forms a loop with the same orientation of the arrows.

\subsection{Wilson loop area law from perturbation theory in the confined limit}
\label{sec:SM-confined-area-law}
 
Physically, all excited states are gapped, and to satisfy the gauge constraint, they must be closed electric strings labelled by irreps $\Gamma$. The $\lambda$ term acts to give a string tension or electric energy to loops or strings proportional to their length. The ground state has no strings (identity irrep), and for a small $\kappa$ perturbation in eq. \eqref{QGGT}, it will contain only the smallest irrep loops. We can see that from perturbation theory as follows.  We begin by noting that the ground state for $\kappa=0$ is nondegenerate, $\ket{\Psi_{\kappa=0}}= \ket{\textbf{I}}$. We can then write the correction to the ground-state energy and wave function as
\begin{align}
\ket{\Psi_\kappa}=\ket{\textbf{I}}+\dfrac{Q_0}{E_0-H_0}V\ket{\textbf{I}}+O(\kappa^2),
\end{align}
where $Q_0=1-\ketbra{\textbf{I}}$. Since the V term acts by changing the irrep from $I$ to $\Gamma$ we have no contribution to $\ev{V}_0$ except for $\Gamma=I$.
Let us calculate the value of the Wilson loop operators in this electric vacuum.
\begin{align}
\ev{\tfrac{1}{\abs{G}}\sum_\Gamma W^\Gamma(\mathcal{C})d_\Gamma}{\Psi_{\kappa=0}}= \sum_{\{g_\ell\}}\delta_{\prod_{\ell\in\mathcal{C}} g_\ell,1} = \abs{G}^{N-L_\sigma}\dfrac{1}{\abs{G}^N}\sum_{g_1,\dots,g_{L_\sigma}}\delta_{\prod_{\ell\in\sigma} g_\ell,1}= \abs{G}^{-L_{\sigma}}\abs{G}^{L_\sigma -1} = \dfrac{1}{\abs{G}}.
\end{align}

It is easy to see then since the vacuum is composed of trivial irreps that
\begin{align}
\ev{W^\Gamma(\mathcal{C})}{\textbf{I}}=\delta_{\Gamma,I}.
\end{align}
Now we write $H_0=\lambda \abs{G} \sum_\ell \mathcal{P}_I(\ell)$ up to a constant energy term. Furthermore, we note that 
\begin{align}
\textbf{Z}^\Gamma \ket{I} = \dfrac{1}{\sqrt{d_\Gamma}}\ket{\mathbf{\Gamma}}.
\end{align}

Again up to a constant energy term $V= \kappa \sum_p (1-B(p)) $ can be written as $V= \kappa N_p-\kappa \sum_p\tfrac{1}{\abs{G}}\sum_{\Gamma\neq I} W^\Gamma(\partial p) d_\Gamma$, here the orientation of the edges follows the up-right convention so that the Wilson loop is defined with dual representations for counter links. The $H_0$  difference in energy of this excitation with respect to the electric vacuum is given by the previous confinement equation so $E_0-H_0$ leads to $-4 \lambda \abs{G}$. The electric vacuum gets dressed to lowest order in perturbation theory by elementary electric loops

\begin{align}
\ket{\Psi_\kappa}=\ket{\textbf{I}}+\dfrac{\kappa}{4\lambda \abs{G}^2}\sum_{\Gamma\neq I}\sum_p \sqrt{d_\Gamma} \ket{\Psi_\Gamma(\square_p)} +O(\kappa^2) = \ket{\textbf{I}}+\dfrac{\kappa}{4\lambda \abs{G}^2}\sum_{\Gamma\neq I}\sum_p  d_\Gamma W^\Gamma(\partial p) \ket{\textbf{I}} +O(\kappa^2).
\end{align}

The correction to the Wilson loop expectation value for $\Gamma\neq I$ will be given by

\begin{align}
\expval{W^\Gamma(\mathcal{C})}{\Psi_\kappa} = \left(  \bra{\textbf{I}}+\dfrac{\kappa}{4\lambda \abs{G}^2}\sum_{\Gamma'\neq I}\sum_{p'}  d_{\Gamma'} \bra{\textbf{I}}W^{\bar{\Gamma ' }}(\partial p')  \right) W^\Gamma(\mathcal{C}) \left( \ket{\textbf{I}}+\dfrac{\kappa}{4\lambda \abs{G}^2}\sum_{\Gamma'\neq I}\sum_{p'} d_{\Gamma'} W^{\Gamma'}(\partial p')\ket{\textbf{I}}\right)+\dots.
\end{align}
Since the $\ket{\Gamma_{\alpha \beta}}$ states are an orthonormal basis we only need to consider processes that return to the electric vacuum by erasing the elementary loops. Therefore, unless $\mathcal{C}=\partial p$ (we assume $\mathcal{C}$ has orientation of the lattice loops ) for some plaquette the process is not allowed. The only contributions in that case are
\begin{align}
\expval{W^\Gamma(\mathcal{C})}{\Psi_\kappa} =  2\dfrac{\kappa}{4\lambda \abs{G}^2} \bra{\textbf{I}}  W^\Gamma(\mathcal{C})\sum_{\Gamma'\neq I}\sum_{p} \delta_{\mathcal{C},\partial p } d_{\Gamma'} W^{\Gamma'}(\partial p)\ket{\textbf{I}}+O(\kappa^2).
\end{align}
We use now the fusion property of the non-invertible Wilson loops to obtain
\begin{align}
\expval{W^\Gamma(\mathcal{C})}{\Psi_\kappa} =  2\dfrac{\kappa}{4\lambda \abs{G}^2} \bra{\textbf{I}} \sum_{\Gamma'\neq I}\sum_{p} \delta_{\mathcal{C},\partial p }d_{\Gamma'} \sum_{\tilde{\Gamma}}N^{\tilde{\Gamma}}_{\Gamma\Gamma'}W^{\tilde{\Gamma}}(\partial p)\ket{\textbf{I}}+O(\kappa^2).
\end{align}
The only nonzero matrix elements come from fusion into the electric vacuum, meaning $\tilde{\Gamma}=I$. Therefore, we obtain
\begin{align}
\expval{W^\Gamma(\mathcal{C})}{\Psi_\kappa} =  2\delta_{\mathcal{C},\partial p }\dfrac{\kappa}{4\lambda \abs{G}^2} \sum_{\Gamma'\neq I}N^{I}_{\Gamma\Gamma'} d_{\Gamma'} +O(\kappa^2).
\end{align}
Therefore, to obtain a nonzero contribution to the Wilson loop, we must go to higher orders in perturbation theory. In particular, for a Wilson loop consisting of two plaquettes, we need the second-order wave function, which is given by 
\begin{align}
\ket{\Psi_\kappa}=\ket{\textbf{I}}+\dfrac{\kappa}{4\lambda \abs{G}^2}\sum_{\Gamma\neq I}\sum_p  d_\Gamma W^\Gamma(\partial p) \ket{\textbf{I}}+\mathcal{P}_{phy}\dfrac{Q_0}{E_0-H_0}V\mathcal{P}_{phy}\dfrac{Q_0}{E_0-H_0}V\ket{\textbf{I}} +O(\kappa^3).
\end{align}
We defined the pure-gauge projector $\mathcal{P}_{phy}=\prod_s A(s)$. We insert it in the expansion because the only allowed states are pure gauge. A Wilson loop consisting of two plaquettes therefore receives a modification only at second order in perturbation theory. The fusion of smaller Wilson loops within the physical subspace leads to a larger Wilson loop only if the number of plaquettes $A(\mathcal{C})$ inside $\mathcal{C}$ matches the number of applications of $V$. Thus, for an arbitrary loop, we have
\begin{align}
\expval{W^\Gamma(\mathcal{C})}{\Psi_\kappa} \propto \left(\dfrac{\kappa}{\lambda {\abs{G}}^2}\right)^{A(\mathcal{C})}=e^{-\sigma_\Gamma A(\mathcal{C})},\quad \sigma_\Gamma \propto \abs{\ln(\tfrac{\kappa}{\lambda {\abs{G}}^2})},\quad  \dfrac{\kappa}{\lambda}\ll 1
\end{align}
We find the area law is satisfied. Interestingly, this argument predicts that the area law coefficient is to leading order independent of the irreducible representation $\Gamma$. This is indeed found numerically.

An important consequence of the perturbative analysis is that noncontractible loops receive no corrections at any order in perturbation theory. Therefore, within the confined regime,

\begin{align}
\expval{W_x^\Gamma(\mathcal{C})}{\Psi_\kappa} = \expval{W_y^\Gamma(\mathcal{C})}{\Psi_\kappa} = 0 , \quad \Gamma\neq I.
\end{align}

\subsection{Wilson loop perimeter law from perturbation theory in the deconfined limit}
\label{sec:SM-deconfined-perimeter-law}

Let us analyze the behavior of the Wilson loops within perturbation theory in the deconfined limit. For simplicity, we take the state $\ket{\Psi_0}=\ket{\Psi_{\mathcal{D}}(G)}$. We do not need to consider hybridization with other ground states because topological order implies that the hybridization is exponentially suppressed with system size. The Wilson loop we consider is the electric $W^\Gamma(\mathcal{C})$, which in the ground state has the value
\begin{align}
\expval{W^\Gamma(\mathcal{C})}{\Psi_0} = d_\Gamma \braket{\Psi_0}=d_\Gamma.
\end{align}
The modified ground state wave function to first order in the perturbation $H_X= \lambda(\sum_\ell(1-\sum_g X^g_+(\ell)))$, or equivalently up to a constant energy term $H_X=\lambda \abs{G}\sum_\ell (1-P_I(\ell))$, is
\begin{align}
\ket{\Psi_\lambda} =  \ket{\Psi_0} + \dfrac{Q_0}{E_0-H_0}H_X\ket{\Psi_0}+O(\lambda^2).
\end{align}
Since $H_0= \kappa (\sum_p (1-B(p)))$, we have $E_0=0$. Additionally, each $X_+^g$ violates two plaquette constraints, which implies that such a state has energy $E_0-H_0\rightarrow -2\kappa$. Because the X operators are elementary ribbon operators, their action directly gives the overlap with the excited eigenstates. We then have
\begin{align}
\ket{\Psi_\lambda} =  \ket{\Psi_0} +\dfrac{\lambda}{2\kappa}(\sum_\ell\sum_{g\neq 1} X^g_+(\ell))\ket{\Psi_0}+O(\lambda^2).
\end{align}
Now using that the Wilson loop is diagonal in group basis and overlap of excited state with the ground states is zero we have
\begin{align}
\expval{W^\Gamma(\mathcal{C})}{\Psi_\lambda}=d_\Gamma+ (\tfrac{\lambda}{2\kappa})^2\bra{\Psi_0}\left(\sum_{\ell'}\sum_{h \neq 1} X^h_+(\ell')\right)W^\Gamma(\mathcal{C})\left(\sum_\ell\sum_{g \neq 1} X^g_+(\ell)\right)\ket{\Psi_0}+O(\lambda^3).
\end{align}
We write this explicitly and use that we designed the confining term such that $[A(s),H_X]=0$ and $[W^\Gamma(\mathcal{C}),A(s)]=0$ for closed loops $\mathcal{C}$ so that we have
\begin{align}
\expval{W^\Gamma(\mathcal{C})}{\Psi_\lambda}&=d_\Gamma+ (\tfrac{\lambda}{2\kappa})^2\sum_{g \neq 1}\sum_{h \neq 1}\sum_{\ell'}\sum_\ell \bra{\textbf{1}}\prod_sA(s) X^h_+(\ell')W^\Gamma(\mathcal{C})X^g_+(\ell)\prod_sA(s)\ket{\textbf{1}}+O(\lambda^3)\notag\\
&=d_\Gamma+ (\tfrac{\lambda}{2\kappa})^2\sum_{g \neq 1}\sum_{h \neq 1}\sum_{\ell'}\sum_\ell \bra{\textbf{1}}\prod_sA(s) X^h_+(\ell')W^\Gamma(\mathcal{C})X^g_+(\ell)\ket{\textbf{1}}+O(\lambda^3).
\end{align}
Now there are two cases $\ell \in \mathcal{C}$ or $\ell \not \in \mathcal{C}$ for the second case the operator $W^\Gamma(\mathcal{C})$ can commute through the X and give $d_\Gamma$ on the all identity state. For the links not in the loop we have 
\begin{align}
\sum_{\ell \not \in \mathcal{C}}\sum_{g\neq 1} W^\Gamma(\mathcal{C})X^g_+(\ell)\ket{\textbf{1}} =  \sum_{\ell \not \in \mathcal{C}}\sum_{g\neq 1} X^g_+(\ell)W^\Gamma(\mathcal{C})\ket{\textbf{1}}=d_\Gamma \sum_{\ell \not \in \mathcal{C}}\sum_{g\neq 1} X^g_+(\ell)\ket{\textbf{1}}.
\end{align} 
The matrix element for these links involves another $X^h_+(\ell')$, which creates a defect with zero overlap with $\ket{\Psi_0}$ unless $h=\bar{g}$ and $\ell'=\ell$, so that
\begin{align}
\sum_{g \neq 1,h \neq 1}\sum_{\ell'  \not \in \mathcal{C}}\sum_{\ell \not \in \mathcal{C}} \bra{\textbf{1}}\prod_sA(s) X^h_+(\ell')W^\Gamma(\mathcal{C})X^g_+(\ell)\ket{\textbf{1}} =\sum_{\ell'  \not \in \mathcal{C}} \sum_{\ell \not \in \mathcal{C}}\sum_{g\neq 1}  d_\Gamma \bra{\textbf{1}}\prod_s A(s) X^{\bar{g}}_+(\ell')X^g_+(\ell)\ket{\textbf{1}}  = (\abs{G}-1)d_\Gamma (N-P(\mathcal{C})).
\end{align}
We denote by $N$ the number of links in the lattice and by $P(\mathcal{C})$ the perimeter (number of links) of $\mathcal{C}$.
Let us now calculate the matrix element of links in the loop. In that case, we write 
\begin{align}
\sum_{\ell \in \mathcal{C}}\sum_{g\neq 1} W^\Gamma(\mathcal{C})X^g_+(\ell)\ket{\textbf{1}}  &= \sum_{\ell \in \mathcal{C}}\sum_{g\neq 1} \tr{\cdots \textbf{Z}^\Gamma(\ell) \cdots }X^g_+(\ell)\ket{\textbf{1}} =  \sum_{\ell \in \mathcal{C}}\sum_{g\neq 1} X^g_+(\ell)\tr{\cdots \mathbf{\Gamma}(g)\textbf{Z}^\Gamma(\ell) \cdots }\ket{\textbf{1}} \notag\\
&= \sum_{\ell \in \mathcal{C}}\sum_{g\neq 1} X^g_+(\ell)\tr{\mathbf{\Gamma}(g)}\ket{\textbf{1}} = \sum_{\ell \in \mathcal{C}}\sum_{g\neq 1}\chi_\Gamma(g) X^g_+(\ell)\ket{\textbf{1}}.
\end{align}
Now similarly to the previous case only $\ell'$ that fall in $\ell$ will cancel the defect and allow for finite overlap with $\ket{\Psi_0}$ which means the loop sum reduces to 
 \begin{align}
 \sum_{g \neq 1,h \neq 1}\sum_{\ell' \in \mathcal{C}}\sum_{\ell \in \mathcal{C}} \bra{\textbf{1}}\prod_sA(s) X^h_+(\ell')W^\Gamma(\mathcal{C})X^g_+(\ell)\ket{\textbf{1}}& = \sum_{\ell \in \mathcal{C}} \sum_{\ell' \in \mathcal{C}} \sum_{g\neq 1}  \chi_\Gamma(g)\bra{\textbf{1}}\prod_s A(s) X^{\bar{g}}_+(\ell')X^g_+(\ell)\ket{\textbf{1}} \notag\\
 &=\sum_{\ell \in \mathcal{C}} \sum_{g\neq 1} \chi_\Gamma(g)\bra{\textbf{1}}\prod_s A(s) X^{\bar{g}}_+(\ell)X^g_+(\ell)\ket{\textbf{1}} =\sum_{g\neq 1}  \chi_\Gamma(g)P(\mathcal{C}).
 \end{align}
 Putting all the pieces together we get
 \begin{align}
 \expval{W^\Gamma(\mathcal{C})}{\Psi_\lambda}=d_\Gamma+ (\tfrac{\lambda}{2\kappa})^2\left[(\abs{G}-1)d_\Gamma (N-P(\mathcal{C}))+\sum_{g\neq 1}\chi_\Gamma(g)P(\mathcal{C})\right]+O(\lambda^3).
 \end{align}
Finally let us note that by similar arguments the normalization of the perturbed wave function becomes
\begin{align}
\braket{\Psi_\lambda}&= 1 + (\tfrac{\lambda}{2\kappa})^2 \sum_{g \neq 1}\sum_{h \neq 1}\sum_{\ell'}\sum_\ell \bra{\textbf{1}}\prod_sA(s) X^h_+(\ell')X^g_+(\ell)\ket{\textbf{1}}+O(\lambda^3)\notag\\
&=1 + (\tfrac{\lambda}{2\kappa})^2 (\abs{G}-1)N+O(\lambda^3).
\end{align}
 The Wilson loop to lowest order in perturbation theory is then
 \begin{align}
 \dfrac{\expval{W^\Gamma(\mathcal{C})}{\Psi_\lambda}}{\braket{\Psi_\lambda}} = d_\Gamma\left(1-\left(\dfrac{\lambda}{2\kappa}\right)^2P(\mathcal{C})\left[(\abs{G}-1)-\sum_{g\neq 1}\chi_\Gamma(g)/d_\Gamma\right]\right)+O(\lambda^3).
 \end{align}
 This expansion has the perimeter law form 
 \begin{align}
\dfrac{\expval{W^\Gamma(\mathcal{C})}{\Psi_\lambda}}{\braket{\Psi_\lambda}} \propto d_\Gamma e^{-\alpha P(\mathcal{C})} ,\quad \alpha \propto \left(\dfrac{\lambda}{2\kappa}\right)^2\left[(\abs{G}-1)-\sum_{g\neq 1}\chi_\Gamma(g)/d_\Gamma\right]+O\left(\left(\tfrac{\lambda}{2\kappa}\right)^3\right).
 \end{align}
 Another important difference from the confined case is that noncontractible loops also receive perimeter-law corrections. For a ground state with magnetic flux class $[g]$ traversing the $y$ direction, this implies
\begin{align}
 \dfrac{\expval{W^\Gamma_x}{\Psi_\lambda([g])}}{\braket{\Psi_\lambda([g])}}\propto \chi_\Gamma([g])e^{-\alpha L}.
\end{align}

\section{Tilted-field quantum double and matterization}
\label{sec:SM-matterization}

In this section we consider the more general model with both magnetic and
electric fields, using the same irrep normalization as in the main text:

\begin{align}
H_{\text{TFQD}}&= \kappa \sum_{p}(1-B(p))+ J\sum_s (1-A(s))+H_X+H_Z,\notag\\
H_X&=\sum_\ell H_X(\ell),\qquad
H_X(\ell)=c_X\mathbb I-
\sum_{[g]\in\operatorname{Class}(G)}\lambda_{[g]}\sum_{h\in[g]}X_+^h(\ell),
\qquad
H_Z=-\sum_\ell\sum_{\Gamma\ne I}\lambda_\Gamma
\tr\bm Z^\Gamma(\ell).
\end{align}
Hermiticity requires $\lambda_{\bar\Gamma}=\lambda_\Gamma^*$.  For $S_3$
all irreps are self-conjugate, so the independent $\lambda_\Gamma$ may be
taken real.  In particular, no factor $d_\Gamma/\abs{G}$ is absorbed into
these physical field strengths.

\subsection{Matterization: construction of the isometry}

In this section, we map the tilted-field $G$-qudit model with link degrees of freedom to a theory with both gauge (link) and matter (vertex) degrees of freedom. Although this may seem overly complicated, it allows us to formulate a path-integral QMC algorithm in the presence of both fields. To accomplish this, we define an auxiliary Hilbert space. Let $\mathcal{H}\equiv\mathcal{H}_{gauge}$ be the original link Hilbert space of the $G$ quantum double. As described above, we can write it as a tensor product of local Hilbert spaces $\mathcal{H}=\bigotimes_\ell\mathcal{H}_\ell$ with $\mathcal{H}_\ell\simeq \mathbb{C}[G]$. Let us denote the vertices of the lattice by $s$ or $i$. We then define an auxiliary Hilbert space by
\begin{align}
\mathcal{H}_{aux}=\mathcal{H}_{gauge}\bigotimes \mathcal{H}_{matter},
\end{align} 
where $\mathcal{H}_{matter}=\otimes_{s}\mathcal{H}_s$ and $\mathcal{H}_s\simeq \mathbb{C}[G]$. Effectively, the full space lives on the Lieb lattice. Since we are interested in the original problem we must enforce a projection on this additional degrees of freedom so as to freeze them. We choose the vertex state to be a simple product state composed of the \textit{electric vacuum} $\ket{I}$. Concretely let the matter state be fixed to 
\begin{align}
\ket{\psi_{matt}} = \bigotimes_s \ket{I}_s.
\end{align} 
The projector to this state is for each vertex $s$ given by
\begin{align}
\mathcal{P}^I_{s} \ket{\psi_{matt}} = \ket{\psi_{matt}} , \qquad \mathcal{P}^I_{s} = \dfrac{1}{\abs{G}}\sum_g X^g_+(s), \quad \mathcal{P}^I=\prod_s \mathcal{P}^I_{s}.
\end{align}
Moreover, since alternatively the matter state is a magnetic condensate we also have $
X^g_+(s) \ket{\psi_{matt}} = \ket{\psi_{matt}} $, $ X^g_-(s) \ket{\psi_{matt}} = \ket{\psi_{matt}}, \quad \forall g\in G$. Enforcing the first equation for all group elements fixes the state to be $\ket{\psi_{matt}}$. By expressing the auxiliary wavefunction as $\ket{\psi}= \ket{\psi_{gauge}}\otimes \ket{\psi_{matter}}$ we enforce the projector on the auxiliary wavefunction as
\begin{align}
X^g_+(s) \ket{\psi} = \ket{\psi}, \quad X^g_-(s) \ket{\psi} = \ket{\psi}, \quad \forall g\in G \label{gauge-constraint}.
\end{align} 
The procedure is now to define a unitary transformation on the auxiliary Hilbert space which mixes vertex and link dofs in a way that allows us to treat the $A(s)$ term fully without requiring $U\rightarrow \infty$.

We define the unitary transformation $U_{KW}$ as implementing 
\begin{align}
U_{KW}: \quad \ket{f_s}\ket{g}_{ss'}\ket{f_{s'}}
\longrightarrow
\ket{f_s}\ket{f_{s'}g\bar f_s}_{ss'}\ket{f_{s'}},
\end{align}
where $\ket{g}_{ss'}\equiv \ket{g}_{\ell}$ denotes a link qudit with orientation $s$ to $s'$. Alternatively we can write in terms of control X operators
\begin{align}
U_{KW} = \prod_{s} \left(\prod_{\ell \rightarrow s }CX_-^{s,\ell}\prod_{\ell \leftarrow s }CX_+^{s,\ell}\right).
\end{align}

The inverse transformation is then
\begin{align}
U_{KW}^\dagger : \quad
\ket{f_s}\ket{g}_{ss'}\ket{f_{s'}}
\longrightarrow
\ket{f_s}\ket{\bar f_{s'}g f_s}_{ss'}\ket{f_{s'}}.
\end{align}
 We now want to find the projection relation of Eq.~\eqref{gauge-constraint} in the new basis. We find  by defining $\ket{\Psi} = U_{KW} \ket{\psi}$ that
\begin{align}
U_{KW} X^g_+(s) U_{KW}^\dagger U_{KW} \ket{\psi} = U_{KW} \ket{\psi}\notag\\
U_{KW} X^g_+(s) U_{KW}^\dagger \ket{\Psi} = \ket{\Psi}.
\end{align}
We may calculate the operator $U_{KW} X^g_+(s) U_{KW}^\dagger $ by applying it over a basis state in the group representation. The result is 
\begin{align}
U_{KW} X^g_+(s) U_{KW}^\dagger \ket{\Psi} = A_g(s)X^g_+(s) \ket{\Psi} = \ket{\Psi}, \qquad   A(s) = \dfrac{1}{\abs{G}}\sum_g A_g(s).
\end{align}
Remarkably, in the transformed basis, the four-link star term $A(s)$ becomes a single-body operator. This is the gauge constraint with matter:
\begin{align}
X^g_+(s) \ket{\Psi} =A_{\bar{g}}(s)\ket{\Psi} , \quad  \forall \ s, \ \forall g \in G.
\end{align}
Transforming the projector leads similarly to 
\begin{align}
P_{phys}(s)=U_{KW}\mathcal{P}_s^I U_{KW}^\dagger =\dfrac{1}{\abs{G}} \sum_g A_g(s)X^g_+(s), \quad P_{phys}=\prod_s P_{phys}(s).
\end{align}

Let us clarify the mathematical mappings we are considering here. Consider an expectation value of some observable $\mathcal{O}_{gauge}$ in the original theory $\mathcal{H}_{gauge}$ for some wavefunction $\ket{\psi_{gauge}}$
\begin{align}
&\ev{\mathcal{O}_{gauge}}{\psi_{gauge}}= \left(\otimes_s\bra{I}_s\otimes \bra{\psi_{gauge}}\right)\mathcal{O}_{gauge}\otimes 1_{matt}\left(\ket{\psi_{gauge}}\otimes_s\ket{I}_s\right)\notag\\
&= \left(\otimes_s\bra{I}_s\otimes \bra{\psi_{gauge}}\right)\mathcal{O}_{gauge}\otimes \mathcal{P}^I\left(\ket{\psi_{gauge}}\otimes_s\ket{I}_s\right) \notag \\
&=\left(\otimes_s\bra{I}_s\otimes \bra{\psi_{gauge}}\right)U_{\text{KW}}^\dagger U_{\text{KW}}\mathcal{O}_{gauge}\mathcal{P}^IU_{\text{KW}}^\dagger U_{\text{KW}}\left(\ket{\psi_{gauge}}\otimes_s\ket{I}_s\right) \notag\\
&= \bra{\Psi}(U_{\text{KW}}\mathcal{O}_{gauge}U_{\text{KW}}^\dagger) (U_{\text{KW}}\mathcal{P}^IU_{\text{KW}}^\dagger)\ket{\Psi}
\notag\\
&= \bra{\Psi}(U_{\text{KW}}\mathcal{O}_{gauge}U_{\text{KW}}^\dagger) P_{phys}\ket{\Psi}, 
\end{align}
where we defined the state
\begin{align}
&\ket{\Psi} = U_{\text{KW}}\left(\ket{\psi_{gauge}}\otimes_s\ket{I}_s\right) = V_{\text{KW}}\ket{\psi_{gauge}}.\notag\\
 &V_{\text{KW}}: \mathcal{H}_{gauge}\rightarrow\mathcal{H}_{gauge}\bigotimes \mathcal{H}_{matt} , \quad V_{\text{KW}} = U_{KW} \bigotimes_s \ket{I}_s. \quad 
\end{align}
In this language $V_{\text{KW}}$ is an isometry of the link space $\mathcal{H}_{gauge}$. Indeed, we can write at the level of density matrices 
\begin{align}
\rho_{GGT}= \ket{\Psi}\bra{\Psi} = V_{\text{KW}} (\rho_{\mathcal{D}(G)})V_{\text{KW}}^\dagger, \quad \rho_{\mathcal{D}(G)} = \ket{\psi_{\mathcal{D}(G)}}\bra{\psi_{\mathcal{D}(G)}}\notag\\
V_{\text{KW}}^\dagger V_{\text{KW}} = \mathbb{I}, \quad V_{\text{KW}}V_{\text{KW}}^\dagger  = U_{KW}\mathcal{P}^I U_{KW}^\dagger = P_{phys} = \prod_s\dfrac{1}{\abs{G}} \sum_g A_g(s)X^g_+(s).
\end{align}
Therefore, there exists a \textit{Kramers--Wannier quantum channel} (KWQC) $\mathcal{N}_{KW}$ with a rank-1 Kraus operator $K=V_{\text{KW}}$ between the $G$-gauge-theory model with density matrix $\rho_{GGT}$ and the $G$ quantum-double model.

\begin{align}
\rho_{GGT}= \mathcal{N}_{KW}[\rho_{\mathcal{D}(G)}].
\end{align}

For the multiplication operators we have 
\begin{align}
&U_{\text{KW}}X_+^g(\ell)U_{\text{KW}}^\dagger
\ket{f_s}\otimes \ket{h} \otimes \ket{f_{s'}}
=U_{\text{KW}}X_+^g(\ell)\ket{f_s}\otimes
\ket{\bar f_{s'} h f_s}\otimes \ket{f_{s'}}\notag \\
&=U_{\text{KW}}\ket{f_s}\otimes
\ket{g\bar f_{s'} h f_s}\otimes \ket{f_{s'}}
=\ket{f_s}\otimes \ket{f_{s'}g\bar f_{s'}h}\otimes \ket{f_{s'}}.
\end{align}
So that the sum becomes
\begin{align}
&\sum_gX_+^g(\ell)U_{\text{KW}}^\dagger
\ket{f_s}\otimes \ket{h} \otimes \ket{f_{s'}}
=\sum_gX_+^{f_{s'}g\bar f_{s'}}(\ell)
\ket{f_s}\otimes \ket{h} \otimes \ket{f_{s'}}\notag \\
&=\sum_{\tilde g}X_+^{\tilde g}(\ell)
\ket{f_s}\otimes \ket{h} \otimes \ket{f_{s'}},
\end{align}
where $\tilde g=f_{s'}g\bar f_{s'}$. Similar result holds for
$\sum_g X_-^g(\ell)$.

\subsubsection{Class-function $H_X$ under matterization}

The preceding calculation extends from the uniform group sum to arbitrary
class-function coefficients. Consider the electric link Hamiltonian introduced
in Eq.~\eqref{eq:class-electric-field},
\begin{align}
 H_X=\sum_\ell H_X(\ell),\qquad
 H_X(\ell)=c_X\mathbb{I}
 -\sum_{[g]\in\operatorname{Class}(G)}\lambda_{[g]}
 \sum_{h\in[g]}X_+^h(\ell).
 \label{eq:matterization-class-X-field}
\end{align}
Let $\ell=(s\rightarrow s')$, and denote the matter group elements at its
source and target by $f_s$ and $f_{s'}$, respectively. The basis-state
calculation above is equivalent to the operator identity
\begin{align}
 U_{\mathrm{KW}}X_+^g(\ell)U_{\mathrm{KW}}^\dagger
 =\sum_{f_{s'}\in G}Z_+^{f_{s'}}(s')
 X_+^{f_{s'}g\bar f_{s'}}(\ell),
 \label{eq:matterization-single-X-plus}
\end{align}
where $Z_+^{f_{s'}}(s')=\ketbra{f_{s'}}{f_{s'}}_{s'}$ projects onto the
target matter state.
It follows directly that
\begin{align}
 U_{\mathrm{KW}}H_X(\ell)U_{\mathrm{KW}}^\dagger
 &=c_X\mathbb{I}-\sum_{f_{s'}\in G}Z_+^{f_{s'}}(s')
 \sum_{[g]}\lambda_{[g]}\sum_{h\in[g]}X_+^{f_{s'}h\bar f_{s'}}(\ell)\notag\\
 &=c_X\mathbb{I}-\sum_{f_{s'}\in G}Z_+^{f_{s'}}(s')
 \sum_{[g]}\lambda_{[g]}\sum_{\tilde g\in[g]}X_+^{\tilde g}(\ell)
 =H_X(\ell).
 \label{eq:matterization-class-X-invariant}
\end{align}
In the second line we used the fact that conjugation by $f_{s'}$ maps every
conjugacy class $[g]$ bijectively onto itself. The matter projectors then sum to
the identity. Consequently, neither the class-dependent hopping amplitudes nor
the offset $c_X$ acquire a matter dressing:
\begin{align}
 U_{\mathrm{KW}}
 (H_X\otimes\mathbb{I}_{\mathrm{matter}})
 U_{\mathrm{KW}}^\dagger
 =H_X\otimes\mathbb{I}_{\mathrm{matter}}.
 \label{eq:matterization-class-HX-invariant}
\end{align}
Uniformity over the full group is therefore not required; invariance on each
conjugacy class is sufficient. Hermiticity remains the separate condition
$\lambda_{[\bar g]}=\lambda_{[g]}^*$ from Eq.~\eqref{eq:class-hermiticity}.

For completeness, a right-multiplication operator is conjugated by the matter
element at the source endpoint,
\begin{align}
 U_{\mathrm{KW}}X_-^g(\ell)U_{\mathrm{KW}}^\dagger
 =\sum_{f_s\in G}Z_+^{f_s}(s)X_-^{f_s g\bar f_s}(\ell).
\end{align}
Thus class sums of $X_-^g$ are invariant for the same reason.

The same statement can be expressed directly through the isometry
$V_{\mathrm{KW}}=U_{\mathrm{KW}}\bigotimes_s\ket{I}_s$. Since
$V_{\mathrm{KW}}V_{\mathrm{KW}}^\dagger=P_{\mathrm{phys}}$, one obtains
\begin{align}
 V_{\mathrm{KW}}H_XV_{\mathrm{KW}}^\dagger
 &=(H_X\otimes\mathbb{I}_{\mathrm{matter}})P_{\mathrm{phys}},\notag\\
 V_{\mathrm{KW}}^\dagger
 (H_X\otimes\mathbb{I}_{\mathrm{matter}})V_{\mathrm{KW}}
 &=H_X.
 \label{eq:isometry-class-HX}
\end{align}
Hence, on the physical image of the isometry, the electric term has exactly
the same action and spectrum as in the original link Hilbert space. In
particular, its irrep energies $\varepsilon_\Gamma$ and the temporal couplings
$K_\tau([g])$ derived above are unchanged by matterization.

This invariance fails for coefficients that are not class functions. If
$H_X(\ell)=c_X\mathbb{I}-\sum_g\lambda_gX_+^g(\ell)$, changing variables to
$\tilde g=f_{s'}g\bar f_{s'}$ gives
\begin{align}
 U_{\mathrm{KW}}H_X(\ell)U_{\mathrm{KW}}^\dagger
 =c_X\mathbb{I}-\sum_{f_{s'},\tilde g}
 \lambda_{\bar f_{s'}\tilde g f_{s'}}
 Z_+^{f_{s'}}(s')X_+^{\tilde g}(\ell).
 \label{eq:matterization-nonclass-X}
\end{align}
The hopping amplitude now depends on the target matter state and represents a
matter-controlled electric transition. It reduces to the undressed link term
precisely when
$\lambda_{\bar f_{s'}\tilde g f_{s'}}=\lambda_{\tilde g}$, namely when
$\lambda_g$ is a class function. Therefore, in the final transformed
Hamiltonian the uniform link term may be generalized simply by the
replacement
\begin{align}
 \lambda\sum_\ell\left(1-\sum_gX_+^g(\ell)\right)
 \longrightarrow
 \sum_\ell\left[c_X\mathbb{I}
 -\sum_{[g]}\lambda_{[g]}\sum_{h\in[g]}X_+^h(\ell)\right],
 \label{eq:ungauged-general-class-X}
\end{align}
without modifying any of the matter or gauge--matter terms.

We now transform the diagonal electric field using exactly the irrep
normalization adopted in Eq.~\eqref{eq:main-electric-magnetic-fields},
\begin{align}
 H_Z=-\sum_\ell\sum_{\Gamma\ne I}\lambda_\Gamma
 \tr\bm Z^\Gamma(\ell),
 \qquad \lambda_{\bar\Gamma}=\lambda_\Gamma^*.
 \label{eq:SM-irrep-electric-field}
\end{align}
For an oriented link $\ell=(s\rightarrow s')$, the matterization unitary acts on
each matrix element as
\begin{align}
&U_{\mathrm{KW}}Z^\Gamma_{\alpha\beta}(\ell)
 U_{\mathrm{KW}}^\dagger
 \ket{f_s,h,f_{s'}}
 =\Gamma(\bar f_{s'}hf_s)_{\alpha\beta}
 \ket{f_s,h,f_{s'}}\notag\\
&\hspace{1.8cm}=\left[
 \bm Z^{\bar\Gamma}(s')\bm Z^\Gamma(\ell)
 \bm Z^\Gamma(s)\right]_{\alpha\beta}
 \ket{f_s,h,f_{s'}}.
\end{align}
Taking the representation trace therefore gives directly
\begin{align}
U_{\mathrm{KW}}H_ZU_{\mathrm{KW}}^\dagger
 =-\sum_{\ell=(s\rightarrow s')}\sum_{\Gamma\ne I}
 \lambda_\Gamma
 \tr{\bm Z^{\bar\Gamma}(s')\bm Z^\Gamma(\ell)
 \bm Z^\Gamma(s)}.
 \label{eq:ungauged-irrep-electric-field}
\end{align}
Hermiticity follows from pairing $\Gamma$ and $\bar\Gamma$ with
$\lambda_{\bar\Gamma}=\lambda_\Gamma^*$.

\paragraph{Equivalent group-basis parametrization.}
For completeness, the same diagonal field can be written using the group
projectors $Z_+^g$. To avoid confusing their coefficients with the physical
$\lambda_\Gamma$ above, write
\begin{align}
 H_Z^{(g)}=\sum_\ell\sum_{g\in G}a^gZ_+^g(\ell),
 \qquad a^g\in\mathbb R.
 \label{eq:group-basis-diagonal-field}
\end{align}
Fourier completeness gives
\begin{align}
 Z_+^g(\ell)=\frac{1}{\abs{G}}\sum_\Gamma d_\Gamma
 \tr{\Gamma(g)^\dagger\bm Z^\Gamma(\ell)}.
 \label{eq:group-projector-irrep-expansion}
\end{align}
For $\ell=(s\rightarrow s')$, define the gauged matrix
\begin{align}
 \bm{\mathcal Z}_\ell^\Gamma
 \equiv\bm Z^{\bar\Gamma}(s')\bm Z^\Gamma(\ell)\bm Z^\Gamma(s).
\end{align}
Then Eq.~\eqref{eq:group-projector-irrep-expansion} and the transformation of
$\bm Z^\Gamma(\ell)$ above imply
\begin{align}
 U_{\mathrm{KW}}Z_+^g(\ell)U_{\mathrm{KW}}^\dagger
 =\frac{1}{\abs{G}}\sum_\Gamma d_\Gamma
 \tr{\Gamma(g)^\dagger\bm{\mathcal Z}_\ell^\Gamma}.
 \label{eq:ungauged-group-projector}
\end{align}

For arbitrary real $a^g$, introduce matrices
\begin{align}
 \widehat{\bm\Lambda}^\Gamma
 \equiv d_\Gamma\sum_{g\in G}a^g\Gamma(g).
 \label{eq:group-field-matrix-Fourier-transform}
\end{align}
Schur orthogonality gives the inverse transform
\begin{align}
 a^g=\frac{1}{\abs{G}}\sum_\Gamma
 \tr{\Gamma(g)^\dagger\widehat{\bm\Lambda}^\Gamma}.
 \label{eq:group-field-matrix-Fourier-inverse}
\end{align}
Substitution in Eq.~\eqref{eq:ungauged-group-projector} yields
\begin{align}
 U_{\mathrm{KW}}H_Z^{(g)}U_{\mathrm{KW}}^\dagger
 =\frac{1}{\abs{G}}
 \sum_{\ell=(s\rightarrow s')}\sum_\Gamma
 \tr{(\widehat{\bm\Lambda}^\Gamma)^\dagger
 \bm{\mathcal Z}_\ell^\Gamma}.
 \label{eq:ungauged-general-group-field}
\end{align}
We keep the dagger explicitly; replacing it by a barred representation would
require an additional matrix-index convention.

If $a^g=a^{[g]}$ is a class function, Schur's lemma gives
\begin{align}
 \widehat{\bm\Lambda}^\Gamma
 &=d_\Gamma\sum_{[h]}a^{[h]}\sum_{g\in[h]}\Gamma(g)\notag\\
 &=\sum_{[h]}\abs{[h]}a^{[h]}\chi_\Gamma([h])\,\bm I
 \equiv\widehat a_\Gamma\bm I,
 \qquad
 \widehat a_\Gamma=\sum_{[h]}\abs{[h]}a^{[h]}\chi_\Gamma([h]).
 \label{eq:class-group-field-Fourier-transform}
\end{align}
Hence
\begin{align}
 U_{\mathrm{KW}}H_Z^{(g)}U_{\mathrm{KW}}^\dagger
 =\frac{1}{\abs{G}}
 \sum_{\ell=(s\rightarrow s')}\sum_\Gamma
 \widehat a_\Gamma^*\tr{\bm{\mathcal Z}_\ell^\Gamma}.
 \label{eq:ungauged-class-group-field}
\end{align}
The trivial-irrep contribution is a constant. Matching every nontrivial
irrep term to the physical convention in
Eq.~\eqref{eq:SM-irrep-electric-field} requires
\begin{align}
 \lambda_\Gamma
 =-\frac{\widehat a_\Gamma^*}{\abs{G}}
 =-\frac{1}{\abs{G}}\sum_{[g]}\abs{[g]}a^{[g]}\chi_\Gamma([g])^*,
 \qquad \Gamma\ne I.
 \label{eq:physical-irrep-group-field-relation}
\end{align}
Conversely, with $c=\widehat a_I/\abs{G}$ physically arbitrary because it
only fixes the omitted trivial-irrep contribution,
\begin{align}
a^g=c-\sum_{\Gamma\ne I}\lambda_\Gamma\chi_\Gamma(g).
 \label{eq:group-field-from-physical-irreps}
\end{align}
The condition $\lambda_{\bar\Gamma}=\lambda_\Gamma^*$ makes $a^g$ real.
No condition $[g]=[\bar g]$ is needed in this diagonal parametrization: each
$Z_+^g$ is itself a Hermitian projector.  The inverse-class pairing condition
instead applies to the translation class sums entering $H_X$.
For $S_3$, Eq.~\eqref{eq:group-field-from-physical-irreps} reads
\begin{align}
 a^{[1]}&=c-\lambda_{A_2}-2\lambda_E,&
 a^{[r]}&=c-\lambda_{A_2}+\lambda_E,&
 a^{[s]}&=c+\lambda_{A_2}.
 \label{eq:S3-group-field-from-physical-irreps}
\end{align}
When $a^g=a$ is uniform, only the trivial irrep remains and
$H_Z^{(g)}=a\sum_\ell\mathbb I$ is an additive constant.

We use the fact that 
\begin{align}
\sum_p (1-B(p))=\sum_p \left(1-\dfrac{1}{\abs{G}} \sum_\Gamma d_\Gamma W^\Gamma (\partial p)\right).
\end{align}
So that
\begin{align}
U_{\text{KW}}\sum_p (1-B(p))U_{\text{KW}}^\dagger = U_{\text{KW}}\sum_p \left(1-\dfrac{1}{\abs{G}} \sum_\Gamma d_\Gamma W^\Gamma (\partial p)\right)U_{\text{KW}}^\dagger = \sum_p \left(1-\dfrac{1}{\abs{G}} \sum_\Gamma d_\Gamma U_{\text{KW}} W^\Gamma (\partial p) U_{\text{KW}}^\dagger \right)\notag\\
=\sum_p \left(1-\dfrac{1}{\abs{G}} \sum_\Gamma d_\Gamma U_{\text{KW}} \text{tr} \Big[\mathcal{P}\prod_{\ell\in \partial p}(\bm{Z}^{\Gamma}(\ell))^{O_\ell}\Big ] U_{\text{KW}}^\dagger \right)=\sum_p \left(1-\dfrac{1}{\abs{G}} \sum_\Gamma d_\Gamma  \text{tr} \Big[\mathcal{P}\prod_{\ell\in \partial p}(\bm{Z}^{\Gamma}(\ell))^{O_\ell}\Big ]  \right).
\end{align}

Only the action on the physical image of the isometry is relevant.  We choose
the following convenient extension to the whole auxiliary Hilbert space:
\begin{align}
H_G^{(2+1)D}={}&J\sum_s (1-\mathcal P_s^I)+H_X
+\kappa \sum_p (1-B(p))\notag\\
&-\sum_{\ell=(s\rightarrow s')}\sum_{\Gamma\ne I}\lambda_\Gamma
\tr{\bm{Z}^{\bar{\Gamma}}(s')\bm{Z}^\Gamma(\ell)
\bm{Z}^{\Gamma}(s)},
\qquad
\mathcal P_s^I=\frac{1}{\abs{G}}\sum_gX_+^g(s).
\label{eq:ungauged-HG}
\end{align}
It satisfies
\begin{align}
 [H_G^{(2+1)D},P_{phys}]&=0,\notag\\
 H_G^{(2+1)D}P_{phys}
 &=U_{\mathrm{KW}}(H_{\mathrm{TFQD}}\otimes\mathbb I_{\mathrm{matter}})
 U_{\mathrm{KW}}^\dagger P_{phys}.
 \label{eq:HG-on-physical-image}
\end{align}
The normalization in $\mathcal P_s^I$ is essential: on physical states
$A(s)P_{phys}=\mathcal P_s^I P_{phys}$. 
\section{Classical mapping with matter}
\label{sec:SM-gauge-matter}
We now use the isometry to derive the classical partition function.  This
also fixes an important point about the enlarged Hilbert space: its
unphysical states must not be included in the trace.  The intertwining
relation $H_G^{(2+1)D}V_{\mathrm{KW}}=V_{\mathrm{KW}}H_{\mathrm{TFQD}}$
and $V_{\mathrm{KW}}V_{\mathrm{KW}}^\dagger=P_{phys}$ give
\begin{align}
 Z&=\Tr_{\mathcal H_{\mathrm{gauge}}}
 e^{-\beta H_{\mathrm{TFQD}}}
 =\Tr_{\mathcal H_{\mathrm{gauge}}}
 \left[V_{\mathrm{KW}}^\dagger e^{-\beta H_G^{(2+1)D}}
 V_{\mathrm{KW}}\right]=\Tr_{\mathcal H_{\mathrm{aux}}}
 \left[P_{phys}e^{-\beta H_G^{(2+1)D}}\right].
 \label{eq:isometry-projected-partition}
\end{align}
Thus the projector cannot be dropped: replacing the last line by an
unprojected auxiliary-space trace would introduce spurious states.  Since
$[H_G^{(2+1)D},P_{phys}]=0$ and $P_{phys}^2=P_{phys}$, for
$N_\tau\Delta\tau=\beta$ one has the exact identity
\begin{align}
 P_{phys}e^{-\beta H_G^{(2+1)D}}
 =\left(P_{phys}e^{-\Delta\tau H_G^{(2+1)D}}\right)^{N_\tau}.
 \label{eq:projector-every-slice}
\end{align}
This form preserves gauge invariance at every Trotter step.

Denote a group-basis configuration on slice $n$ by
\begin{align}
 \ket{\bm f^n,\bm{g}^n}
 =\bigotimes_s\ket{f_s^n}\bigotimes_{\ell}\ket{g_\ell^n},
\end{align}
where $f_s^n$ is a matter variable and $g_\ell^n$ is a spatial gauge
variable.  For an oriented link $\ell=(s\rightarrow s')$, expansion of the
local projectors in $P_{phys}$ gives
\begin{align}
 P_{phys}\ket{\bm{f},\bm{g}}
 =\frac{1}{\abs{G}^{N_s}}\sum_{\{h_s\}}
 \bigotimes_s\ket{\bar h_sf_s}
 \bigotimes_{\ell=(s\rightarrow s')}
 \ket{\bar h_{s'}g_\ell h_s}.
 \label{eq:projector-basis-action}
\end{align}
The variables $h_s^n\in G$ are the gauge fields on temporal links and are all
oriented from $(s,n)$ to $(s,n+1)$.  The inverses in
Eq.~\eqref{eq:projector-basis-action} occur because the projector acts on the
ket at slice $n+1$ whereas $h_s^n$ points toward that slice.

For the Trotter decomposition, separate the diagonal terms from the two
off-diagonal terms $H_X$ and $J\sum_s(1-\mathcal P_s^I)$.  In the group
basis the diagonal energy is
\begin{align}
 E_{\mathrm d}(\bm{f}^n,\bm{g}^n)
 ={}&\kappa\sum_{p\parallel xy}
 \left(1-\delta_{U_p(\bm{g}^n),1}\right)
 -\sum_{\ell=(s\rightarrow s')}\sum_{\Gamma\ne I}\lambda_\Gamma
 \tr{\Gamma(\bar f_{s'}^n)\Gamma(g_\ell^n)\Gamma(f_s^n)},\quad  \Phi_\ell^n=\bar f_{s'}^n g_\ell^n f_s^n.
 \label{eq:diagonal-gauge-matter-energy}
\end{align}
Indeed, $ \Phi_\ell^n$ is precisely the original link group element obtained by
applying $U_{\mathrm{KW}}^\dagger$.  The second term in
Eq.~\eqref{eq:diagonal-gauge-matter-energy} is the group-basis evaluation on
slice $n$ of the explicit Higgs interaction derived above,
\begin{align}
 -\sum_{\ell=(s\rightarrow s')}\sum_{\Gamma\ne I}\lambda_\Gamma
 \tr{\bm Z^{\bar\Gamma}(s')\bm Z^\Gamma(\ell)\bm Z^\Gamma(s)}.
 \label{eq:explicit-Higgs-interaction}
\end{align}
Indeed,
$\tr{\Gamma(\bar f_{s'}^n)\Gamma(g_\ell^n)\Gamma(f_s^n)}
=\chi_\Gamma( \Phi_\ell^n)$.  
The one-link kernel of the general class-sum $H_X(\ell)$ is the class
function $w_{\Delta\tau}([g])$ derived in
Eqs.~\eqref{eq:class-transfer-weight}--\eqref{eq:class-transfer-weight-irreps}.
The matter kernel is also exact.  Since $\mathcal P_s^I$ is a projector,
\begin{align}
 e^{-\Delta\tau J(1-\mathcal P_s^I)}
 &=e^{-\Delta\tau J}\mathbb I
 +(1-e^{-\Delta\tau J})\mathcal P_s^I,\notag\\
 m_{\Delta\tau}(r)
 &\equiv\mel{u}{e^{-\Delta\tau J(1-\mathcal P_s^I)}}{v}
 =e^{-\Delta\tau J}\delta_{r,1}
 +\frac{1-e^{-\Delta\tau J}}{\abs{G}},
 \qquad r=u\bar v.
 \label{eq:exact-matter-kernel}
\end{align}
Using Eqs.~\eqref{eq:projector-basis-action} and
\eqref{eq:exact-matter-kernel}, one Trotter matrix element is
\begin{align}
 &\mel{\bm{f}^n,\bm{g}^n}
 {e^{-\Delta\tau H_G^{(2+1)D}}P_{phys}}
 {\bm{f}^{n+1},\bm{g}^{n+1}}\notag\\
 &\quad=\frac{e^{-\Delta\tau E_{\mathrm d}(\bm{f}^n,\bm{g}^n)}}
 {\abs{G}^{N_s}}\sum_{\{h_s^n\}}
 \prod_{\ell=(s\rightarrow s')}
 w_{\Delta\tau}([\bar U_{\ell\tau}^n])
 \prod_s m_{\Delta\tau}(\Phi_{s\tau}^n)
 +O(\Delta\tau^2),\notag\\
 U_{\ell\tau}^n
 &\equiv\bar g_\ell^n\bar h_{s'}^n g_\ell^{n+1}h_s^n,
 \qquad
 \Phi_{s\tau}^n\equiv\bar f_s^{n+1}h_s^n f_s^n.
 \label{eq:full-gauge-matter-transfer-matrix}
\end{align}
Equation~\eqref{eq:full-gauge-matter-transfer-matrix} is the desired
single-step derivation.  The $O(\Delta\tau^2)$ term comes only from separating
the diagonal and off-diagonal parts; both displayed local kernels have been
exponentiated exactly.  Explicitly, the relative link element entering
$w_{\Delta\tau}$ is
$g_\ell^n\bar h_s^n\bar g_\ell^{n+1}h_{s'}^n$, which is conjugate to
$\bar U_{\ell\tau}^n$.  Likewise, the relative matter element is conjugate to
$\Phi_{s\tau}^n$.  This proves the class arguments in
Eq.~\eqref{eq:full-gauge-matter-transfer-matrix} with the stated forward-link
orientation.

It is useful to verify gauge covariance before converting the weights to an
action.  Under an independent $u_s^n\in G$ at every spacetime vertex,
\begin{align}
 f_s^n&\longmapsto u_s^nf_s^n,\notag\\
 g_{ss'}^n&\longmapsto u_{s'}^ng_{ss'}^n\bar u_s^n,\notag\\
 h_s^n&\longmapsto u_s^{n+1}h_s^n\bar u_s^n.
 \label{eq:spacetime-gauge-transformation}
\end{align}
Consequently, both $ \Phi_\ell^n$ and $\Phi_{s\tau}^n$ are invariant, while
$U_{\ell\tau}^n\mapsto u_s^nU_{\ell\tau}^n\bar u_s^n$.  The class dependence
of $w_{\Delta\tau}$ is therefore exactly the condition needed for the
temporal plaquette weight to be gauge invariant.

Normalize the two kernels by their identity values.  For $H_X$ we retain
Eq.~\eqref{eq:class-temporal-coupling},
\begin{align}
 e^{-K_\tau([g])}=\frac{w_{\Delta\tau}([g])}
 {w_{\Delta\tau}([1])},\qquad K_\tau([1])=0.
\end{align}
For the matter kernel, all nonidentity elements have the same weight, and
\begin{align}
 m_{\Delta\tau}(r)
 &=m_{\Delta\tau}(1)
 \exp\left[-Q_\tau(1-\delta_{r,1})\right], \quad 
 m_{\Delta\tau}(1)=\frac{1+(\abs{G}-1)e^{-\Delta\tau J}}{\abs{G}},\notag\\
Q_\tau
 &=\ln\left[\frac{1+(\abs{G}-1)e^{-\Delta\tau J}}
 {1-e^{-\Delta\tau J}}\right].
 \label{eq:matter-temporal-coupling}
\end{align}
Substituting these expressions into all $N_\tau$ transfer steps gives
\begin{align}
 Z&=\mathcal N_{\Delta\tau}
 \sum_{\{f_s^n,g_\ell^n,h_s^n\}}e^{-S_{\text{GGT}}}
 +O(\beta\Delta\tau^2), \quad 
 \mathcal N_{\Delta\tau}
 =\left[\abs{G}^{-N_s}w_{\Delta\tau}([1])^{N_\ell}
 m_{\Delta\tau}(1)^{N_s}\right]^{N_\tau},\notag\\
 S_{\text{GGT}}={}&K\sum_{n,p\parallel xy}
 (1-\delta_{U_p^n,1})
 +\sum_{n,\ell}K_\tau([\bar U_{\ell\tau}^n])-\Delta\tau
 \sum_{n,\ell=(s\rightarrow s')}\sum_{\Gamma\ne I}\lambda_\Gamma
 \tr{\Gamma(\bar f_{s'}^n)\Gamma(g_\ell^n)\Gamma(f_s^n)}
 +Q_\tau\sum_{n,s}
 (1-\delta_{\Phi_{s\tau}^n,1}),
 \label{eq:general-class-gauge-matter-action}
\end{align}
Here $K=\Delta\tau\kappa$. The gauge part of this action contains only group elements and conjugacy
classes, while the spatial Higgs coupling is displayed explicitly in its
representation form.  
\subsection{Periodic imaginary time and temporal holonomy}

The trace in Eq.~\eqref{eq:isometry-projected-partition} imposes periodic
boundary conditions on the time-slice variables,
\begin{align}
 f_s^{N_\tau}=f_s^0,\qquad g_\ell^{N_\tau}=g_\ell^0,
 \label{eq:periodic-matter-gauge-fields}
\end{align}
and the allowed gauge transformations are periodic,
$u_s^{N_\tau}=u_s^0$.  It does \emph{not} constrain the temporal gauge-field
holonomy to be the identity.  With every temporal link oriented from $n$ to
$n+1$, define the ordered Polyakov holonomy
\begin{align}
 \Omega_s=h_s^{N_\tau-1}h_s^{N_\tau-2}\cdots h_s^0.
 \label{eq:temporal-Polyakov-holonomy}
\end{align}
Under a periodic gauge transformation,
$\Omega_s\mapsto u_s^0\Omega_s\bar u_s^0$, so its conjugacy class is
gauge invariant and must be summed over.

A periodic gauge transformation can set
$h_s^n=1$ for $n=0,\ldots,N_\tau-2$, but the last temporal link then remains
$h_s^{N_\tau-1}=\Omega_s$ up to conjugation.  For
$\ell=(s\rightarrow s')$, the only temporal factors modified by the boundary
are therefore
\begin{align}
 U_{\ell\tau}^{N_\tau-1}
 &=\bar g_\ell^{N_\tau-1}\bar\Omega_{s'}
 g_\ell^0\Omega_s,\notag\\
 \Phi_{s\tau}^{N_\tau-1}
 &=\bar f_s^0\Omega_s f_s^{N_\tau-1}.
 \label{eq:temporal-boundary-holonomy-factors}
\end{align}
Thus full temporal gauge, $h_s^n=1$ on every temporal link, is possible with
periodic gauge transformations only in the trivial-holonomy sector
$\Omega_s=1$.  Equivalently, one may set every temporal link to the
identity using a nonperiodic gauge transformation, but then the fields obey
boundary conditions twisted by $\Omega_s$.

The matter variables do not remove this sum over holonomy sectors.  Rather,
the temporal Higgs factors $\Phi_{s\tau}^n$ give those sectors a dynamical
weight.  Because $f_s^n$ itself is periodic, the periodic transformation
$u_s^n=\bar f_s^n$ is always allowed and fixes unitary gauge, $f_s^n=1$.
In this gauge
\begin{align}
  \Phi_\ell^n=g_\ell^n,\qquad \Phi_{s\tau}^n=h_s^n,
\end{align}
so the two matter interactions become bond potentials for the spatial and
temporal gauge links, as in the gauge-fixed $\mathbb Z_2$ gauge--Higgs model.
In the strict no-matter limit below, $H_Z=0$ and
$Q_\tau=0$, this direct holonomy weight disappears, while the
sum over the pure-gauge temporal holonomy sectors remains.

\subsection{Checks and limiting cases}

First consider the theory without matter.  In the present construction this
means setting $H_Z=0$ and imposing the charge-free sector selected by the star
projector, equivalently taking $J\rightarrow\infty$ before the continuum-time
limit.  Equation~\eqref{eq:matter-temporal-coupling} then gives
$Q_\tau\rightarrow0$.  The $f_s^n$ variables disappear from the
action and their sum contributes only a constant.  We recover
\begin{align}
 S_{\mathrm{pure}}
 =K\sum_{p\parallel xy}(1-\delta_{U_p,1})
 +\sum_{p\ni\tau}K_\tau([\bar U_p]),
 \label{eq:gauge-matter-to-pure-check}
\end{align}
which is exactly the class-function pure-gauge action in
Eq.~\eqref{eq:class-function-action}.  Merely setting $H_Z=0$ while keeping
$J$ finite at finite temperature still allows electric charges and is
therefore not the strict no-matter limit.

As a second check, take $G=\mathbb Z_2$ and identify group elements with
$\pm1$.  The quantum-double projectors are
$A(s)=(1+A_s^x)/2$ and $B(p)=(1+B_p^z)/2$, hence the couplings used in
Tupitsyn \emph{et al.}~\cite{TupitsynKitaevProkofevStamp2010}
are $J_x=J/2$ and $J_z=\kappa/2$.  Choose the nontrivial class coefficient in
$H_X$ to be $h_x$.  For the diagonal field choose
$\lambda_{\mathrm{sgn}}=h_z$, so that the explicit Higgs
term in Eq.~\eqref{eq:explicit-Higgs-interaction} becomes
$-h_z \Phi_\ell$ for $\Phi_\ell=\pm1$.  Since
$1-\delta_{Q,1}=(1-Q)/2$, Eq.~\eqref{eq:general-class-gauge-matter-action}
becomes, up to an additive constant,
\begin{align}
 S_{\mathbb Z_2}
 =-\sum_{b}\lambda_{\mathrm{bond}}^{\parallel,\perp}
 \mu_u\sigma_{uv}\mu_v
 -\sum_p\lambda_{\mathrm{pl}}^{\parallel,\perp}
 \prod_{b\in p}\sigma_b,
 \label{eq:Z2-gauge-Higgs-check}
\end{align}
with the four anisotropic couplings
\begin{align}
 \lambda_{\mathrm{bond}}^{\parallel}
 &=\frac{Q_\tau}{2}
 =-\frac12\ln\tanh(\Delta\tau J_x),
 &\lambda_{\mathrm{bond}}^{\perp}&=\Delta\tau h_z,\notag\\
 \lambda_{\mathrm{pl}}^{\parallel}
 &=\frac{K_\tau([-1])}{2}
 =-\frac12\ln\tanh(\Delta\tau h_x),
 &\lambda_{\mathrm{pl}}^{\perp}&=\Delta\tau J_z.
 \label{eq:Z2-coupling-benchmark}
\end{align}
These are precisely Eqs.~(8a)--(8d) of that reference.  The apparent factors
of two are entirely due to its use of Pauli stabilizers instead of the
projectors $A(s)$ and $B(p)$.  Fixing the redundant $\mu$ variables gives its
Ising model with bond and plaquette interactions.

Only now specialize the class-sum coefficients in $H_X$ to a common value,
$\lambda_{[g]}=\lambda$ for every nonidentity class.  The class dependence of the
temporal plaquettes collapses to
\begin{align}
 K_\tau([g])&=K_\tau\quad([g]\neq[1]),\qquad
 e^{-K_\tau}=\frac{e^{\abs{G}\lambda\Delta\tau}-1}
 {e^{\abs{G}\lambda\Delta\tau}+\abs{G}-1},
\end{align}
and the second term of Eq.~\eqref{eq:general-class-gauge-matter-action}
simplifies to
$K_\tau\sum_{p\ni\tau}(1-\delta_{U_p,1})$.  No other term in the
gauge--matter action is changed by this specialization.

\section{Fredenhagen--Marcu order parameters from non-invertible 1-form symmetries}
Consider the fixed-point quantum-double Hamiltonian. The Wilson loop operators labelled by irreps of the group furnish a non-invertible electric 1-form symmetry, as described in Ref.~\cite{noninvertible1form}:
\begin{align}
W^\Gamma(\mathcal{C}) = \text{tr} \Big[\mathcal{P}\prod_{\ell\in \mathcal{C}}(\bm{Z}^{\Gamma}(\ell))^{O_\ell}\Big ]\equiv  \text{tr}[\bm{W}^\Gamma(\mathcal{C})] , \label{Wilson-Loops}
\end{align}

\begin{figure}[t]
 \centering
 \includegraphics[width=0.9\textwidth]{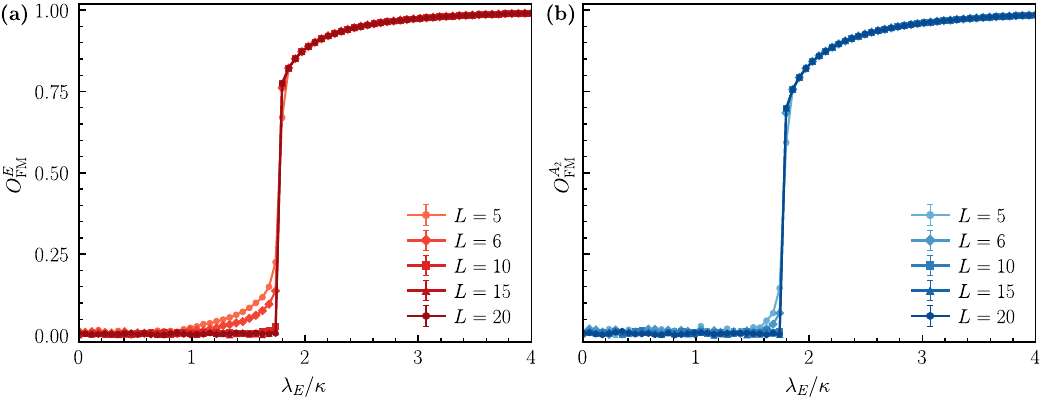}
 \caption{\textbf{Zoom out of the FM order parameter.}
 Fredenhagen--Marcu order parameters (a) $O_{\mathrm{FM}}^E$ and
 (b) $O_{\mathrm{FM}}^{A_2}$ versus $\lambda_E/\kappa$ for the $S_3$
 gauge--Higgs theory at $\kappa=1$, $J=10$, $\lambda/\kappa=0.01$, and
 $\lambda_{A_2}=0$.  Increasing $L=5,6,10,15,20$ sharpens the onset near
 $\lambda_E/\kappa\simeq 1.75$, we use $N_\tau=500,\Delta\tau=0.1$. }
 \label{fig:FM-E-A2}
\end{figure}
where we defined $O_\ell$ as the orientation of the edge $\ell$ with respect to the loop $\mathcal{C}$. If the orientations match $O_\ell=1$, otherwise, it acts as the adjoint $O_\ell=\dagger$ realizing the conjugate representation $\bm{Z}^{\bar{\Gamma}}$. In contrast to the Abelian case the electric 1-form operators do not form a group. Instead for two arbitrary irreps $\Gamma_1,\Gamma_2$ and a fixed closed loop $\mathcal{C}$ the product of Wilson loops reduces to computing the tensor product of irreps $\Gamma_1\otimes \Gamma_2$, which can be decomposed into a sum of irreps:
\begin{align}
   W^{\Gamma_1}(\mathcal{C})W^{\Gamma_2}(\mathcal{C}) = \sum_\Gamma N^{\Gamma}_{\Gamma_1\Gamma_2} W^\Gamma(\mathcal{C}).\label{RepG_Wilson}
\end{align}
The fusion coefficients $N^{\Gamma}_{\Gamma_1\Gamma_2}$ are given by the multiplicity of the irrep $\Gamma$ in the tensor-product representation $\Gamma_1\otimes \Gamma_2$. 
In analogy to the $\mathbb{Z}_2$ toric code we have
\begin{align}
\tilde{O}_\Gamma = 
\lim _{\left|L_{1 / 2}\right| \rightarrow \infty} \sqrt{\left|\tilde{C}_\Gamma\left(\left|L_{1 / 2}\right|\right)\right|}, \quad \tilde{C}_\Gamma\left(\left|L_{1 / 2}\right|\right)= \dfrac{1}{d_\Gamma}\ev{\text{tr} \Big[\mathcal{P}\prod_{\ell\in L_{1/2}}(\bm{Z}^{\Gamma}(\ell))\Big ]}.
\end{align} 
The factor $d_\Gamma$ comes from normalizing the Wilson loops so that their effect on the QD ground state is the identity.
The FM order parameter factorizes the bulk modification of the bare 1-form symmetry away from the fixed point leading to 
\begin{align}
O_{FM}^\Gamma= \lim _{\left|L_{1 / 2}\right| \rightarrow \infty} \sqrt{\left|C_\Gamma\left(\left|L_{1 / 2}\right|\right)\right|},\quad C_\Gamma\left(\left|L_{1 / 2}\right|\right) =\dfrac{1}{\sqrt{d_\Gamma}} \dfrac{\ev{\text{tr} \Big[\mathcal{P}\prod_{\ell\in L_{1/2}}(\bm{Z}^{\Gamma}(\ell))\Big ]}}{ \sqrt{\ev{\text{tr} \Big[\mathcal{P}\prod_{\ell\in L}(\bm{Z}^{\Gamma}(\ell))\Big]}}}.
\end{align}

When applying the isometry $V_{\text KW}$, we obtain a product of links with the two Higgs vertex elements at the ends, which ensures gauge invariance. We therefore evaluate in the gauge theory the observable defined by
\begin{align}
w_\Gamma^{\mathrm{open}}(P)=\frac{1}{d_\Gamma}\chi_\Gamma\!\left(f_{s'}^{-1}U_Pf_s\right),\qquad
w_\Gamma^{\mathrm{closed}}(\mathcal C)=\frac{1}{d_\Gamma}\chi_\Gamma(U_{\mathcal C}).
\end{align}
The FM estimator is formed from ensemble averages,
\begin{align}
C_\Gamma(r=L_{1 / 2})=\frac{\langle w_\Gamma^{\mathrm{open}}(P_r)\rangle}
{\sqrt{\langle w_\Gamma^{\mathrm{closed}}(\mathcal C_{2r})\rangle}},\qquad
O_{\mathrm{FM}}^\Gamma(r)=\sqrt{|C_\Gamma(r)|}.
\end{align}

Figure~\ref{fig:FM-E-A2} shows a magnified finite-size cut through the
main-text FM data.  Although the direct $A_2$ field is absent
($\lambda_{A_2}=0$), the $A_2$ response accompanies the response to the
$E$-charge field $\lambda_E$, consistently with the fusion rule
$E\otimes E=A_1\oplus A_2\oplus E$.

\end{document}